\documentclass[10pt,journal]{IEEEtran}

\usepackage{amsmath,graphicx,multirow, booktabs,tabularx,amssymb,bm,svg,tikz,algorithm,algpseudocode,comment,tcolorbox,pgfplots,mwe, color, enumitem, subfig, textcomp, adjustbox, overpic}

\usepackage{breqn}

\usepackage{adjustbox} 
\usepackage{upgreek} 
\usepackage{bbm}

\usepackage{acro}
\usetikzlibrary{spy, calc, arrows.meta}
\setlist{nosep, leftmargin=14pt}

\usepackage{xurl}
\usepackage{hyperref}

\definecolor{lightblue}{HTML}{ADD8E6}  
\definecolor{softpink}{HTML}{FFC0CB}     
\definecolor{strongpink}{HTML}{FF69B4}   
\definecolor{vividorange}{HTML}{FF7300}

\usepackage[
	style=ieee,
	dashed=false,
	maxbibnames=20,
	minbibnames=2,
	maxcitenames=1,
	mincitenames=1,
	backend=biber,
	natbib=true,
	defernumbers=false,
	sorting=none,
	useprefix=false
	]{biblatex}

\algnewcommand\algorithmicparfor{\textbf{parfor}}
\algnewcommand\algorithmicpardo{\textbf{do}}
\algnewcommand\algorithmicendparfor{\textbf{end\ parfor}}
\algrenewtext{ParFor}[1]{\algorithmicparfor\ #1\ \algorithmicpardo}
\algrenewtext{EndParFor}{\algorithmicendparfor}

\usepgfplotslibrary{groupplots}   
\pgfplotsset{compat=1.16}
\pgfplotsset{table/comment chars={\#}}
\definecolor{cLIIF}{HTML}{0072B2}
\definecolor{cCLIF}{HTML}{D55E00}
\definecolor{cMAISIa}{HTML}{009E73}
\definecolor{cMAISIb}{HTML}{56B4E9}
\definecolor{cMEDa}{HTML}{CC79A7}
\definecolor{cMEDb}{HTML}{E69F00}
\definecolor{cDDS}{HTML}{4D4D4D}
\definecolor{cDec}{HTML}{0072B2}
\definecolor{cDC}{HTML}{009E73}
\definecolor{cEnc}{HTML}{E69F00}
\definecolor{cLat}{HTML}{D55E00}
\definecolor{cDen}{HTML}{999999}
\pgfplotsset{
  paper/.style={
    font=\footnotesize,
    tick label style={font=\scriptsize},
    label style={font=\footnotesize},
    title style={font=\footnotesize\bfseries, yshift=-3pt},
    axis line style={black!60, line width=0.4pt},
    tick style={black!60, line width=0.4pt},
    major grid style={black!12, line width=0.3pt},
    minor grid style={black!6, line width=0.2pt},
    grid=major,
    legend style={draw=none, fill=white, fill opacity=0.85, text opacity=1, font=\scriptsize,
                  row sep=-1pt, inner sep=1pt},
    legend cell align=left,
    every axis plot/.append style={line width=1.0pt, mark size=1.7pt},
    unbounded coords=jump,
    xtick pos=bottom, ytick pos=left,
    scaled y ticks=false, yticklabel style={/pgf/number format/fixed},
  },
  sLIIF/.style={cLIIF, mark=*},
  sCLIF/.style={cCLIF, mark=square*},
  sMAISIa/.style={cMAISIa, mark=triangle*, mark size=2.1pt},
  sMAISIb/.style={cMAISIb, mark=triangle*, mark options={rotate=180}, mark size=2.1pt},
  sMEDa/.style={cMEDa, mark=diamond*, mark size=2.2pt},
  sMEDb/.style={cMEDb, mark=pentagon*, mark size=2.0pt},
  sDDS/.style={cDDS, mark=x, mark size=2.4pt},
  pow2x/.style={xmode=log, log basis x=2},
  discard if not/.style 2 args={x filter/.code={
      \edef\tempa{\thisrow{#1}}\edef\tempb{#2}\ifx\tempa\tempb\else\fi}},
}

\providecommand{\figcropdir}{images/Method/crops/}
\providecolor{cLIIF}{HTML}{0072B2}
\providecolor{cCLIF}{HTML}{D55E00}
\providecolor{cDen}{HTML}{999999}
\definecolor{cSlate}{HTML}{4D5B6A}       
\definecolor{cSlateFill}{HTML}{E6EAEE}
\definecolor{cDenStroke}{HTML}{6B6B6B}   
\definecolor{cZone}{HTML}{F7EADB}        
\definecolor{cZoneEdge}{HTML}{EBCDAE}
\newcommand{\figsix}{\fontsize{6}{7}\selectfont}
\newcommand{\figfive}{\fontsize{5}{6}\selectfont}

\tikzset{
  flow/.style    ={-{Stealth[length=1.6mm,width=1.2mm]},line width=0.6pt,black!70},
  dflow/.style   ={flow,dash pattern=on 1.4pt off 1pt},
  lab/.style     ={font=\scriptsize,inner sep=0pt,align=center,text=black!85},
  note/.style    ={font=\figsix,inner sep=0pt,align=center,text=black!60},
  tag/.style     ={font=\figfive,fill=white,inner sep=0.4pt,rounded corners=0.4mm},
  ptitle/.style  ={font=\footnotesize\bfseries,anchor=west,inner sep=0pt},
  ztitle/.style  ={font=\figsix\bfseries,anchor=north west,inner sep=0pt,text=black!65},
  headbox/.style ={line width=0.5pt,rounded corners=0.8mm,minimum height=4.6mm,inner sep=0.6pt,
                   align=center,font=\figsix\bfseries},
  liifhead/.style={headbox,draw=cLIIF,fill=cLIIF!13,text=cLIIF},
  clifhead/.style={headbox,draw=cCLIF,fill=cCLIF!13,text=cCLIF},
  count/.style   ={font=\scriptsize\bfseries,inner sep=0pt},
}

\newcommand{\imgcube}[5]{%
  \begin{scope}[shift={(#1,#2)}]
    \node[anchor=south west,inner sep=0] at (0,0)
         {\includegraphics[width=#3mm,height=#3mm]{\figcropdir#5_cor.png}};
    \begin{scope}[shift={(0,#3)},cm={1,0,0.5,0.36,(0,0)},transform shape]
      \node[anchor=south west,inner sep=0] at (0,0)
           {\includegraphics[width=#3mm,height=#4mm]{\figcropdir#5_ax.png}};
    \end{scope}
    \begin{scope}[shift={(#3,0)},cm={0.5,0.36,0,1,(0,0)},transform shape]
      \node[anchor=south west,inner sep=0] at (0,0)
           {\includegraphics[width=#4mm,height=#3mm]{\figcropdir#5_sag.png}};
    \end{scope}
    \draw[black!80,line width=0.35pt,line join=round] (0,0) rectangle (#3,#3)
      (0,#3) -- ++(0.5*#4,0.36*#4) -- ++(#3,0) -- ++(0,-#3) -- (#3,0)
      (#3,#3) -- ++(0.5*#4,0.36*#4);
  \end{scope}}

\newcommand{\wirecube}[5]{%
  \draw[#5,line join=round] (#1,#2) rectangle ++(#3,#3)
    (#1,#2+#3) -- ++(0.5*#4,0.36*#4) -- ++(#3,0) -- ++(0,-#3) -- (#1+#3,#2)
    (#1+#3,#2+#3) -- ++(0.5*#4,0.36*#4);}

\newcommand{\solidcube}[6]{%
  \colorlet{figbase}{#5}%
  \fill[figbase,opacity=#6] (#1,#2) rectangle ++(#3,#3);
  \fill[figbase!80!black,opacity=#6] (#1,#2+#3) -- ++(0.5*#4,0.36*#4) -- ++(#3,0) -- ++(-0.5*#4,-0.36*#4) -- cycle;
  \fill[figbase!60!black,opacity=#6] (#1+#3,#2) -- ++(0.5*#4,0.36*#4) -- ++(0,#3) -- ++(-0.5*#4,-0.36*#4) -- cycle;
  \wirecube{#1}{#2}{#3}{#4}{figbase!70!black,line width=0.5pt}}

\newcommand{\gridcube}[6]{%
  \begin{scope}[shift={(#1,#2)}]
    \colorlet{figbase}{#5}\pgfmathsetmacro\gcs{#3/#4}
    \fill[figbase!22] (0,0) rectangle (#3,#3);
    \fill[figbase!38] (0,#3) -- ++(0.5*#3,0.36*#3) -- ++(#3,0) -- (#3,#3) -- cycle;
    \fill[figbase!52] (#3,0) -- ++(0.5*#3,0.36*#3) -- ++(0,#3) -- (#3,#3) -- cycle;
    \foreach \gi in {1,...,#4}{\draw[figbase!75!black,line width=0.2pt]
        (\gi*\gcs-\gcs,0) -- (\gi*\gcs-\gcs,#3) (0,\gi*\gcs-\gcs) -- (#3,\gi*\gcs-\gcs);}
    \draw[figbase!75!black,line width=0.45pt,line join=round] (0,0) rectangle (#3,#3)
      (0,#3) -- ++(0.5*#3,0.36*#3) -- ++(#3,0) -- ++(0,-#3) -- (#3,0) (#3,#3) -- ++(0.5*#3,0.36*#3);
    \node[note,anchor=north] at (0.5*#3+0.25*#3,-0.9) {#6};
  \end{scope}}

\newcommand{\block}[8]{%
  \colorlet{figbase}{#6}%
  \fill[figbase] (#1,#2) rectangle ++(#3,#4);
  \fill[figbase!85!black] (#1,#2+#4) -- ++(0.5*#5,0.36*#5) -- ++(#3,0) -- ++(-0.5*#5,-0.36*#5) -- cycle;
  \fill[figbase!70!black] (#1+#3,#2) -- ++(0.5*#5,0.36*#5) -- ++(0,#4) -- ++(-0.5*#5,-0.36*#5) -- cycle;
  \draw[#7,line width=0.45pt,line join=round] (#1,#2) rectangle ++(#3,#4)
    (#1,#2+#4) -- ++(0.5*#5,0.36*#5) -- ++(#3,0) -- ++(0,-#4) -- (#1+#3,#2) (#1+#3,#2+#4) -- ++(0.5*#5,0.36*#5);
  \node[lab] at (#1+0.5*#3,#2+0.5*#4) {#8};}

\newcommand{\frustum}[8]{%
  \pgfmathsetmacro\fdl{#6*#4}\pgfmathsetmacro\fdr{#6*#5}\colorlet{figbase}{#7}%
  \fill[figbase] (#1,#3-0.5*#4) -- (#1,#3+0.5*#4) -- (#2,#3+0.5*#5) -- (#2,#3-0.5*#5) -- cycle;
  \fill[figbase!85!black] (#1,#3+0.5*#4) -- ++(0.5*\fdl,0.36*\fdl) -- ($(#2,#3+0.5*#5)+(0.5*\fdr,0.36*\fdr)$) -- (#2,#3+0.5*#5) -- cycle;
  \fill[figbase!70!black] (#2,#3-0.5*#5) -- ++(0.5*\fdr,0.36*\fdr) -- ++(0,#5) -- (#2,#3+0.5*#5) -- cycle;
  \draw[#8,line width=0.45pt,line join=round]
    (#1,#3-0.5*#4) -- (#1,#3+0.5*#4) -- (#2,#3+0.5*#5) -- (#2,#3-0.5*#5) -- cycle
    (#1,#3+0.5*#4) -- ++(0.5*\fdl,0.36*\fdl) -- ($(#2,#3+0.5*#5)+(0.5*\fdr,0.36*\fdr)$) -- ++(0,-#5) -- (#2,#3-0.5*#5)
    (#2,#3+0.5*#5) -- ++(0.5*\fdr,0.36*\fdr);}

\newcommand{\imgsq}[5]{%
  \node[inner sep=0pt,anchor=south west] at (#1,#2) {\includegraphics[width=#3mm,height=#3mm]{\figcropdir#4}};
  \draw[#5] (#1,#2) rectangle ++(#3,#3);}

\definecolor{cSlateHeavy}{HTML}{B9C2CC}  

\newcommand{\imgcubeop}[6]{%
  \begin{scope}[shift={(#1,#2)}]
    \fill[white] (0,0) rectangle (#3,#3);
    \node[anchor=south west,inner sep=0,opacity=#6] at (0,0)
         {\includegraphics[width=#3mm,height=#3mm]{\figcropdir#5_cor.png}};
    \begin{scope}[shift={(0,#3)},cm={1,0,0.5,0.36,(0,0)},transform shape]
      \node[anchor=south west,inner sep=0,opacity=#6] at (0,0)
           {\includegraphics[width=#3mm,height=#4mm]{\figcropdir#5_ax.png}};
    \end{scope}
    \begin{scope}[shift={(#3,0)},cm={0.5,0.36,0,1,(0,0)},transform shape]
      \node[anchor=south west,inner sep=0,opacity=#6] at (0,0)
           {\includegraphics[width=#4mm,height=#3mm]{\figcropdir#5_sag.png}};
    \end{scope}
    \draw[black!55,line width=0.35pt,dash pattern=on 0.9pt off 0.6pt] (0,0) -- (0.5*#4,0.36*#4) -- ++(#3,0) (0.5*#4,0.36*#4) -- ++(0,#3);
    \draw[black!80,line width=0.35pt,line join=round] (0,0) rectangle (#3,#3)
      (0,#3) -- ++(0.5*#4,0.36*#4) -- ++(#3,0) -- ++(0,-#3) -- (#3,0)
      (#3,#3) -- ++(0.5*#4,0.36*#4);
  \end{scope}}

\newcommand{\featcube}[5]{%
  \begin{scope}[shift={(#1,#2)}]
    \pgfmathsetmacro\gcs{#3/#4}
    \fill[cSlate!30] (0,0) rectangle (#3,#3);
    \fill[cSlate!45] (0,#3) -- ++(0.5*#3,0.36*#3) -- ++(#3,0) -- (#3,#3) -- cycle;
    \fill[cSlate!60] (#3,0) -- ++(0.5*#3,0.36*#3) -- ++(0,#3) -- (#3,#3) -- cycle;
    \node[anchor=south west,inner sep=0,opacity=0.35] at (0,0)
         {\includegraphics[width=#3mm,height=#3mm]{\figcropdir#5_cor.png}};
    \foreach \gi in {1,...,#4}{\draw[cSlate!85!black,line width=0.22pt]
        (\gi*\gcs-\gcs,0) -- (\gi*\gcs-\gcs,#3) (0,\gi*\gcs-\gcs) -- (#3,\gi*\gcs-\gcs);}
    \foreach \gi in {1,...,#4}{\draw[cSlate!85!black,line width=0.18pt]
        (\gi*\gcs-\gcs,#3) -- ++(0.5*#3,0.36*#3) (#3,\gi*\gcs-\gcs) -- ++(0.5*#3,0.36*#3);}
    \draw[cSlate!85!black,line width=0.5pt,line join=round] (0,0) rectangle (#3,#3)
      (0,#3) -- ++(0.5*#3,0.36*#3) -- ++(#3,0) -- ++(0,-#3) -- (#3,0) (#3,#3) -- ++(0.5*#3,0.36*#3);
  \end{scope}}

\newcommand{\pointsthreed}[7]{%
  \pgfmathsetseed{#6}%
  \foreach \k in {1,...,#5}{%
    \pgfmathsetmacro\pu{rnd}\pgfmathsetmacro\pv{rnd}\pgfmathsetmacro\pw{rnd}%
    \pgfmathsetmacro\pr{0.26-0.11*\pw}\pgfmathsetmacro\po{1-0.55*\pw}%
    \fill[#7,opacity=\po] ({#1+0.4+(#3-0.8)*\pu+0.5*#4*\pw},{#2+0.4+(#3-0.8)*\pv+0.36*#4*\pw}) circle (\pr);}}

\newcommand{\smallpin}[2]{%
  \draw[black!70,line width=0.45pt] (#1,#2) -- ++(0,1.05);
  \fill[cSlate,draw=black!75,line width=0.25pt] (#1,#2+1.45) circle (0.52);
  \fill[white,opacity=0.8] (#1-0.16,#2+1.61) circle (0.14);}
\newcommand{\restarticon}[4]{%
  \draw[#4,line width=0.6pt,-{Stealth[length=0.85mm,width=1.05mm]}]
       ($(#1,#2)+(100:#3)$) arc[start angle=100,end angle=-215,radius=#3];}
\newcommand{\zonepin}[5]{%
  \path[fill=cZone,draw=cZoneEdge,line width=0.5pt,rounded corners=1.5mm] (#1,#2) rectangle (#3,#4);
  \smallpin{#1+2.0}{#4-3.1}
  \node[ztitle] at (#1+3.4,#4-1.0) {#5};}
\newcommand{\zonerestart}[5]{%
  \path[fill=cZone,draw=cZoneEdge,line width=0.5pt,rounded corners=1.5mm] (#1,#2) rectangle (#3,#4);
  \restarticon{#1+2.4}{#4-2.05}{0.85}{black!70}
  \node[ztitle] at (#1+4.0,#4-1.0) {#5};}
\tikzset{comment/.style={font=\figsix\itshape,inner sep=0pt,align=left,anchor=west,text=black!65}}

\DeclareAcronym{3D}{
    short = 3-D,
    long = three-dimensional,
}

\DeclareAcronym{2D}{
    short = 2-D,
    long = two-dimensional,
}

\DeclareAcronym{CT}{
    short = CT,
    long = computed tomography,
}

\DeclareAcronym{MRI}{
    short = MRI,
    long = magnetic resonance imaging,
}

\DeclareAcronym{VAE}{
    short = VAE,
    long = variational autoencoder,
}

\DeclareAcronym{DM}{
    short = DM,
    long = diffusion model,
}

\DeclareAcronym{LDM}{
    short = LDM,
    long = latent diffusion model,
}

\DeclareAcronym{DPS}{
    short = DPS,
    long = diffusion posterior sampling,
}

\DeclareAcronym{LIIF}{
    short = LIIF,
    long = local implicit image function,
}

\DeclareAcronym{INR}{
    short = INR,
    long = implicit neural representation,
}

\DeclareAcronym{DDPM}{
    short = DDPM,
    long = denoising diffusion probabilistic model,
}

\DeclareAcronym{CNN}{
    short = CNN,
    long = convolutional neural network,
}

\DeclareAcronym{GPU}{
    short = GPU,
    long = graphics processing unit,
    short-plural = s,
    long-plural = s,
}

\DeclareAcronym{KL}{
    short = KL,
    long = Kullback-Leibler,
}

\DeclareAcronym{STD}{
	short = STD,
	long = standard deviation,
}

\DeclareAcronym{FID}{
    short = FID,
    long = Fréchet Inception Distance,
}

\DeclareAcronym{MAE}{
    short = MAE,
    long = mean absolute error,
}

\DeclareAcronym{PSNR}{
    short = PSNR,
    long = peak signal-to-noise ratio,
}

\DeclareAcronym{ROI}{
    short = ROI,
    long = region of interest,
}

\DeclareAcronym{HU}{
    short = HU,
    long = Hounsfield unit,
}

\DeclareAcronym{SNR}{
    short = SNR,
    long = signal-to-noise ratio,
}

\DeclareAcronym{SSIM}{
    short = SSIM,
    long = structural similarity index measure,
}

\DeclareAcronym{MSE}{
    short = MSE,
    long = mean squared error,
}

\DeclareAcronym{RMSE}{
    short = RMSE,
    long = root mean squared error,
}

\DeclareAcronym{AE}{
    short = AE,
    long = autoencoder,
}

\DeclareAcronym{FBP}{
	short = FBP,
	long = filtered back-projection,
}
\DeclareAcronym{FFT}{
	short = FFT,
	long = fast Fourier transform,
}
\DeclareAcronym{DDS}{
	short = DDS,
	long = decomposed diffusion sampler,
}

\DeclareAcronym{ZF}{
	short = ZF,
	long = zero-filled,
}

\DeclareAcronym{IFFT}{
	short = IFFT,
	long = inverse fast Fourier transform,
}

\DeclareAcronym{GAN}{
	short = GAN,
	long = generative adversarial network,
}

\DeclareAcronym{MLP}{
	short = MLP,
	long = multilayer perceptron,
}

\DeclareAcronym{CLIF}{
	short = CLIF,
	long = convolutional local image function,
}

\DeclareAcronym{TV}{
	short = TV,
	long = total variation,
}

\DeclareAcronym{DDIM}{
    short = DDIM,
    long = denoising diffusion implicit models,
}

\DeclareAcronym{CoReAE}{
    short = CoRe-AE,
    long = Coordinate-Rendering Autoencoder,
}

\DeclareAcronym{LPIPS}{
    short = LPIPS,
    long = learned perceptual image patch similarity,
}

\DeclareAcronym{HFEN}{
    short = HFEN,
    long = high-frequency error norm,
}

\newcommand{\boldc}{\bm{c}}

\newcommand{\boldg}{\bm{g}}
\newcommand{\boldh}{\bm{h}}

\newcommand{\bolds}{\bm{s}}

\newcommand{\boldu}{\bm{u}}

\newcommand{\boldx}{\bm{x}}
\newcommand{\boldy}{\bm{y}}
\newcommand{\boldz}{\bm{z}}

\newcommand{\boldmu}{\bm{\mu}}

\newcommand{\boldxhat}{\hat{\boldx}}
\newcommand{\xhat}{\hat{x}}

\newcommand{\boldeta}{\bm{\eta}}

\newcommand{\boldI}{\bm{I}}

\newcommand{\boldM}{\bm{M}}

\newcommand{\calD}{\mathcal{D}}

\newcommand{\calG}{\mathcal{G}}

\newcommand{\calL}{\mathcal{L}}

\newcommand{\calN}{\mathcal{N}}

\newcommand{\calT}{\mathcal{T}}

\newcommand{\calX}{\mathcal{X}}
\newcommand{\calY}{\mathcal{Y}}
\newcommand{\calZ}{\mathcal{Z}}

\newcommand{\boldcalA}{\bm{\mathcal{A}}}

\newcommand{\boldcalD}{\bm{\mathcal{D}}}
\newcommand{\boldcalE}{\bm{\mathcal{E}}}
\newcommand{\boldcalF}{\bm{\mathcal{F}}}

\newcommand{\boldSigma}{\bm{\Sigma}}
\newcommand{\boldsigma}{\bm{\sigma}}
\newcommand{\boldtheta}{\bm{\theta}}
\newcommand{\boldphi}{\bm{\phi}}
\newcommand{\boldpsi}{\bm{\psi}}

\newcommand{\boldepsilon}{\bm{\epsilon}}

\newcommand{\boldzero}{\bm{0}}

\newcommand{\R}{\mathbb{R}}

\title{Continuous 3-D Latent Diffusion for Medical Image Generation and Reconstruction}

\author{Youness Mellak, Antoine De Paepe, Dimitris Visvikis, Alexandre Bousse
    \thanks{This work was supported by Contrat de Plan \'Etat--Région (CPER) 2021-2027 IMAGIIS (INNOV-XS)}
    
    \thanks{This work was granted access to the HPC resources of IDRIS under the allocation 2026-AD011018234 made by GENCI.}
    
	\thanks{This work did not involve human subjects or animals in its research.}
	\thanks{Youness Mellak, Antoine De Paepe, Dimitris Visvikis and Alexandre Bousse are with Univ. Brest, LaTIM, Inserm, U1101, 29238~Brest, France.}
   
	\thanks{Corresponding authors: \href{mailto:bousse@univ-brest.fr}{{\texttt{bousse@univ-brest.fr}}} and \href{mailto:visvikis@univ-brest.fr}{\texttt{visvikis@univ-brest.fr}} }
}

\begin{document}

\maketitle

\begin{abstract}
	High-resolution three-dimensional (3-D) medical latent diffusion remains limited by the cost of decoding full volumes, which existing latent autoencoders (AEs) handle through overlapping sub-volume decoding. We propose a continuous 3-D latent diffusion model framework for computed tomography (CT) and magnetic resonance imaging (MRI) generation and measurement-guided reconstruction, built on a coordinate-rendering autoencoder (CoRe-AE). Its decoder applies its convolutional stage once on the compact latent grid and renders voxel intensities from spatial coordinates through a lightweight implicit head, evaluated in memory-bounded chunks. This design avoids overlapping sub-volumes and remains differentiable with respect to the latent, so that the decoder can be used inside measurement-guided reconstruction. The framework was evaluated on CT volumes of 512\textsuperscript{3} voxels and MRI volumes of 256\textsuperscript{3} voxels. At 512\textsuperscript{3}, the decoder was 4--18 times faster than the MAISI and 3D MedDiffusion AEs and required the lowest peak graphics processing unit (GPU) memory, including when its gradient was computed. CoRe-AE also achieved the highest peak signal-to-noise ratio (PSNR) and structural similarity index (SSIM) among the evaluated AEs on both modalities. The resulting frozen 3-D latent prior generated full volumes without patch seams and, without retraining, reconstructed sparse-view CT and accelerated MRI through hard data consistency, within 2.4--4.9\,dB PSNR of the pixel-domain decomposed diffusion sampler (DDS). A single volumetric latent prior thus supports both generation and reconstruction of high-resolution medical volumes on one GPU.

    Our code and pretrained weights are available at {\urlstyle{tt}\url{https://github.com/Mellak/continuous-3d-latent-diffusion}}.
	
\end{abstract}

\section{Introduction}\label{sec:intro}

High-resolution medical images such as \acs{CT} and \acs{MRI} are inherently \acs{3D}: anatomical structures extend across slices, and image quality must remain consistent in the axial, coronal, and sagittal planes. Volumetric coherence is therefore essential for both image generation and reconstruction. \Acp{DM} have become a reference approach for image generation~\citep{ho2020denoising} and have also been applied to native \ac{3D} medical image synthesis \citep{khader2023denoising}. However, clinical volumes commonly reach 512\textsuperscript{3} voxels or larger, making direct voxel-space modeling computationally demanding. At each denoising step, a voxel-space \ac{DM} evaluates a dense \ac{3D} \ac{CNN} over the entire volume and stores large intermediate activation maps. A practical framework must therefore preserve global anatomical coherence without repeatedly processing every voxel at full resolution.

\Acp{LDM} reduce this cost by moving the diffusion process from image space to a compact latent representation. An \ac{AE} first compresses the input volume, after which the generative model is trained on the resulting latent codes rather than directly on voxels \citep{Rombach_2022_CVPR}. This strategy has enabled the modeling of anatomical variability in compressed \ac{3D} brain \ac{MRI} representations~\citep{pinaya2022brain}. Subsequent methods have pursued different strategies for scaling volumetric generation. PatchDDM trains \acp{DM} on local \ac{3D} patches and applies them to larger volumes at inference~\citep{pmlr-v227-bieder24a}. MedSyn and LAND incorporate anatomical conditioning for high-resolution chest \ac{CT} synthesis \citep{medsyn2024,oliveras2026anatomically}. More recent frameworks, including MAISI, MAISI-v2, and 3D~MedDiffusion, combine latent compression, anatomical conditioning, and patch-based processing to generate \ac{CT} and \ac{MRI} volumes at resolutions up to 512\textsuperscript{3} \citep{Guo_2025_WACV,zhao2025maisi,wang2025meddiffusion}.

Despite reducing the cost of the generative model, \acp{LDM} do not eliminate the computational burden associated with high-resolution volumes. Instead, part of this burden remains in the \ac{AE}. The encoder must map a full-resolution volume to the latent space, while the decoder must reconstruct the latent representation on the original voxel grid. At 512\textsuperscript{3} resolution, a dense encoding or decoding operation may exceed the memory capacity of a single \ac{GPU}. This limitation becomes particularly restrictive when the decoder is used repeatedly within an iterative reconstruction algorithm.

Existing \acp{AE} generally address this limitation through spatial decomposition. Hierarchical and sub-volume methods combine coarse global representations with high-resolution local regions \citep{sun2022hierarchical}. Volume-compression and tensor-splitting strategies reduce the cost of dense \ac{3D} encoding and decoding \citep{Guo_2025_WACV}, while patch-volume \acp{AE} process the input and output as multiple smaller regions~\citep{wang2025meddiffusion}. Related approaches partition the generative model itself through patch-based diffusion or position-aware score blending \citep{hu2024learning,song2024diffusionblend}. Although these methods reduce peak memory, they generally require repeated evaluation of a dense network on multiple, often overlapping, sub-volumes, followed by aggregation or blending. Their runtime therefore increases with the number of processed regions, which is especially costly when decoding must be performed repeatedly.

Coordinate-based decoding provides an alternative to dense, fixed-grid volume reconstruction~\citep{shen2022nerp,feng2023imjense, liu2024scan,feng2025spatiotemporal}. \Acp{INR} represent a signal as a function of spatial coordinates, making them resolution-agnostic, differentiable, and queryable at arbitrary locations~\citep{mildenhall2021nerf,sitzmann2020implicit, muller2022instant,yu2025bilevel}. This principle was initially developed mainly for \ac{2D} image representation: \ac{LIIF} conditions coordinate queries on local latent features, thereby connecting a discrete feature grid to a continuous image representation~\citep{Chen_2021_CVPR}, while implicit decoders have been combined with \acp{LDM} to enable arbitrary-scale generation without moving the diffusion process out of latent space~\citep{Kim_2024_CVPR}, and \ac{CLIF} extended this idea with local convolutional rendering~\citep{chen2024image}. In medical imaging, coordinate-based models have subsequently been coupled with measurement operators for reconstruction from incomplete observations \citep{khorashadizadeh2024lofi,yu2025bilevel} and combined with physical models and generative priors to enforce data consistency in ill-posed inverse problems~\citep{tong2024diffinr,chu2025highly}. For the present work, two properties are particularly important: the decoder can render an image through coordinate queries rather than dense sub-volume processing, and the resulting mapping remains differentiable with respect to the latent representation.

The development of continuous representations has recently extended from \ac{2D} images and scan-specific optimization to learned volumetric medical models. In particular, MedIL introduced an implicit latent representation for heterogeneous-resolution brain \ac{MRI} and lung \ac{CT}, enabling learned \ac{3D} generation at arbitrary output resolutions~\citep{spears2025medil}. This work demonstrates the value of continuous decoding for accommodating heterogeneous image dimensions and voxel spacings. Its decoder, however, resamples features before a convolutional decoder that operates at the output resolution, so the decoding cost grows with the size of the rendered volume.
However, the use of such a decoder for efficient full-volume processing at 512\textsuperscript{3} resolution, and especially for repeated differentiation within a measurement-guided reconstruction procedure, remains largely unexplored. The question addressed in this work is therefore whether a coordinate-conditioned \ac{AE} can process full high-resolution medical volumes on a single \ac{GPU} while remaining sufficiently efficient and differentiable to serve as an interface between a learned \ac{3D} prior and a medical acquisition operator.

To address this gap, we propose \acs{CoReAE}, a \ac{3D} \ac{AE} in which a lightweight encoder maps a full-resolution volume to a compact latent grid, and a coordinate-conditioned decoder predicts voxel intensities as a continuous function of spatial coordinates. Our main contributions are the following: First, we design a memory-efficient encoder that applies dense, wide-channel layers only after spatial downsampling, so that full 512\textsuperscript{3} volumes fit on a single \ac{GPU}. Second, our decoder runs its convolutions only once, at latent resolution. It then predicts each voxel from its coordinates, chunk by chunk, without overlapping patches. We introduce two versions of this rendering step: an \ac{MLP}-based \ac{LIIF} head and a convolutional \ac{CLIF} head. The decoder remains differentiable, and its gradient on a full volume fits in the same memory budget.
Third, we train an \ac{LDM} on this latent space and use it for both unconditional generation and measurement-guided reconstruction. We generate full \ac{CT} and \ac{MRI} volumes at 512\textsuperscript{3} and 256\textsuperscript{3}, and we use the same frozen prior for sparse-view \ac{CT} and undersampled \ac{MRI} reconstruction, within the diffusion posterior-sampling framework \citep{chung2022diffusion} and with the hard-data-consistency procedure of ReSample~\citep{song2024solving}. Fourth, we analyze fidelity and computational cost in a controlled setting, with decoder ablations and with decoding and reconstruction costs compared to MAISI, 3D MedDiffusion and \acs{DDS}.

The remainder of this paper is organized as follows. Section~\ref{sec:method} presents the proposed framework, Section~\ref{sec:results} describes the experiments and results, Section~\ref{sec:discussion} discusses the findings and limitations, and Section~\ref{sec:conclusion} concludes the paper.
\section{Materials and Methods}\label{sec:method}

\subsection{Problem Formulation and Overview}\label{sec:formulation}

Let $\boldx \in \calX \triangleq \R^{ D \times H \times W}$ denote a high-resolution \ac{3D} image volume, where $D,H,W$ are the spatial dimensions and $[\boldx]_i$ denotes the image intensity at voxel $i$. We address two tasks under constrained memory. The generation problem is to learn the data distribution $p(\boldx)$, from which new realistic volumes can be sampled. The reconstruction problem is to recover $\boldx$ from incomplete measurements $\boldy \in \calY \triangleq \R^n$ acquired through the forward model
\begin{equation}\label{eq:forward_model}
	\boldy = \boldcalA\!\left(\boldx\right) + \boldeta \, ,
\end{equation}
where $\boldcalA \colon \calX \to \calY$ is an acquisition operator, instantiated in this work as a projection operator for \ac{CT} or an undersampled Fourier transform for \ac{MRI}, and $\boldeta$ denotes a centered measurement noise. The combined problem is therefore to estimate a plausible volume under the learned prior $p(\boldx)$ that is consistent with the measurements, i.e., solving
\begin{equation}\label{eq:inv_prob}
    \text{find $\boldx\sim p(\boldx)$ s.t.}\quad \boldy \approx \boldcalA\!\left(\boldx\right)\, .
\end{equation}
Solving \eqref{eq:inv_prob} is equivalent to sampling from the posterior
$\boldx \sim p(\boldx \mid \boldy) \propto p(\boldy \mid \boldx)\,p(\boldx)$, where the likelihood $p(\boldy \mid \boldx)$ is defined by the forward model \eqref{eq:forward_model}. In this work, the prior $p(\boldx)$ is learned using a \ac{DM} \citep{ho2020denoising}, and we used a  \ac{DPS} algorithm inspired by \citeauthor{zhu2023denoising}~\citep{zhu2023denoising} and \citeauthor{song2024solving}~\citep{song2024solving}.

To make both tasks tractable at clinical resolution on a single \ac{GPU}, $\boldx$ is represented by a compact latent variable $\boldz \in \calZ \triangleq \R^{C_z \times d \times h \times w}$, with latent channels $C_z$ and 
latent spatial dimensions $(d,h,w) = (D/r, H/r, W/r)$ obtained from a spatial downsampling factor $r$; the values of $C_z$ and $r$ are specified in Section~\ref{sec:encoder}. The latent is produced and decoded by a jointly trained encoder--decoder pair $(\boldcalE_{\boldphi},\boldcalD_{\boldpsi})$  with trainable parameters $\boldphi$ and $\boldpsi$. The encoder $\boldcalE_{\boldphi}$ has two components: the encoder mean $\boldmu_{\boldphi} \colon \calX \to \calZ$ and the encoder \ac{STD} $\boldsigma_{\boldphi} \colon \calX \to \calZ$, which are used to sample a latent variable $\boldz$ from an image $\boldx$, i.e.,

\begin{equation}
	\label{eq:encoder_sampling}
	\begin{aligned}
		q_{\boldphi}(\boldz \mid \boldx)
		&=
		\calN\!\left(
		    \boldz;
		    \boldmu_{\boldphi}(\boldx),
		    \boldSigma_{\boldphi}(\boldx)
		\right),
		\\
		\boldz
		&\sim q_{\boldphi}(\boldz \mid \boldx),
		\\
		\boldSigma_{\boldphi}(\boldx)
		&=
		\mathrm{diag}\!\left(
		    \boldsigma_{\boldphi}^{2}(\boldx)
		\right),
		\\
		\left(
		    \boldmu_{\boldphi}(\boldx),
		    \boldsigma_{\boldphi}(\boldx)
		\right)
		&=
		\boldcalE_{\boldphi}(\boldx).
	\end{aligned}
\end{equation}

The inverse problem \eqref{eq:inv_prob} can then be reformulated in the latent space by replacing the image prior $p(\boldx)$ with a prior $p(\boldz)$ over latent representations. Specifically, we solve
\begin{equation}\label{eq:inv_prob_latent}
    \text{find $\boldz \sim p(\boldz)$ s.t.}\quad
    \boldy \approx \boldcalA \circ \boldcalD_{\boldpsi}(\boldz)\, .
\end{equation}

Generation and reconstruction are then both posed on the latent space $\calZ$ through a single diffusion prior $p(\boldz)$ trained on encoded latents, so that $p(\boldx)$, the model's distribution over volumes, is the pushforward of $p(\boldz)$ through $\boldcalD_{\boldpsi}$.

In this work, the decoder $\boldcalD_{\boldpsi}$ is implicit: it predicts intensities at queried spatial coordinates and can therefore render a volume, or a coordinate subset, in memory-bounded chunks rather than instantiating the full $D\times H \times W$ grid at once.

The framework follows a two-stage design. The \ac{AE} pair $(\boldcalE_{\boldphi}, \boldcalD_{\boldpsi})$ is trained first and then frozen (Section~\ref{sec:ae} and Fig.~\ref{fig:method}); the diffusion prior is trained on the resulting latents (Section~\ref{sec:ldm}) and reused, without modification, as a prior that regularizes reconstruction from $\boldy$ via decoded data consistency $\boldcalA \circ \boldcalD_{\boldpsi}$ (Section~\ref{sec:recon}). The implicit decoder is evaluated in coordinate chunks, allowing the composition $\boldcalA \circ \boldcalD_{\boldpsi}$ to be differentiated with
bounded memory.

\subsection{Latent Autoencoder}\label{sec:ae}

\begin{figure*}[!t]
	\centering
\begin{tikzpicture}[x=1mm,y=1mm,font=\scriptsize]
\path[use as bounding box] (0,0) rectangle (181,84);

\smallpin{128.6}{80.0}
\node[note,anchor=west] at (130.2,81.0) {runs once per step};
\restarticon{156.2}{81.0}{0.85}{black!70}
\node[note,anchor=west] at (157.9,81.0) {repeats};

\begin{scope}[shift={(0,56)}]
\node[ptitle] at (0,25.1) {(a) Dense CNN autoencoder};
\zonerestart{0.5}{0.5}{38.5}{22.5}{$512^3$ volume: a random patch}
\zonerestart{40.5}{0.5}{104.5}{22.5}{$128^3$ patch: the whole dense network}
\imgcube{3}{3.8}{12.5}{7}{gt}
\draw[black!40,line width=0.3pt] (7.5,11.8)--(23.2,14.0) (7.5,8.675)--(23.2,6.0)
      (11.475,12.41)--(34.2,15.8) (10.625,8.675)--(31.7,6.0);
\wirecube{7.5}{8.675}{3.125}{1.7}{white,line width=0.7pt}
\imgcube{23.2}{6.0}{8.5}{5}{gtpatch}
\node[lab] at (11.2,2.4) {$\boldx$};
\node[lab] at (27.5,3.9) {$\boldx_k$};
\draw[flow] (35,10.5)--(41,10.5);
\frustum{41.2}{54.2}{10.5}{12}{4}{0.4}{cDen!55}{black!65,line width=0.6pt}
\node[lab] at (47.2,10.5) {$\boldcalE$};
\draw[flow] (55.1,10.5)--(58.4,10.5);
\gridcube{58.6}{7.5}{6}{4}{cDen}{$\boldz_k$}
\draw[flow] (67.8,10.5)--(71.3,10.5);
\frustum{71.5}{85}{10.5}{4}{12}{0.4}{cDen!55}{black!65,line width=0.6pt}
\node[lab] at (79,10.5) {$\boldcalD$};
\draw[flow] (87.6,10.5)--(90.7,10.5);
\imgcube{90.9}{6.0}{8.5}{5}{w128patch}
\node[lab] at (95.2,3.9) {$\hat{\boldx}_k$};
\node[note] at (95.2,1.7) {L1 vs $\boldx_k$};
\node[comment] at (146,16.6) {a new random patch every step};
\node[comment] at (146,10.5) {sees \textbf{1/64} of the volume};
\node[comment] at (146,4.4) {whole network repeats per patch};
\end{scope}

\begin{scope}[shift={(0,28)}]
\node[ptitle] at (0,25.1) {(b) CoRe-AE with the {\color{cLIIF}LIIF} head};
\zonepin{0.5}{0.5}{38.5}{22.5}{$512^3$ grid: light layers}
\zonepin{40.5}{0.5}{104.5}{22.5}{$128^3$ latent grid: heavy layers}
\zonerestart{106.5}{0.5}{181}{22.5}{LIIF head: random points}
\imgcube{3}{3.8}{12.5}{7}{gt}
\node[lab] at (11.2,2.4) {$\boldx$ (whole volume)};
\draw[flow] (19.5,10.5)--(23,10.5);
\frustum{23.2}{33.5}{10.5}{13}{4.5}{0.3}{white}{cSlate,line width=0.45pt}
\node[lab] at (28.1,10.5) {$\boldcalE^{\mathrm{down}}$};
\draw[flow] (35,10.5)--(41,10.5);
\block{41.2}{8}{12}{5}{3}{cSlateHeavy}{cSlate!90!black,line width=0.7pt}{$\boldcalE^{\mathrm{enc}}$}
\draw[flow] (54.9,10.5)--(58.4,10.5);
\gridcube{58.6}{7.5}{6}{4}{cSlate}{$\boldz$}
\node[note] at (63.1,3.3) {$4{\times}128^3$};
\draw[flow] (67.8,10.5)--(71.3,10.5);
\block{71.5}{8}{14}{5}{3}{cSlateHeavy}{cSlate!90!black,line width=0.7pt}{$\boldcalD^{\mathrm{feat}}$}
\node[note] at (78.5,5.6) {no upsampling};
\draw[flow] (87.2,10.5)--(90.7,10.5);
\featcube{90.9}{6.5}{8}{6}{gt}
\node[lab] at (96.9,4.6) {$\boldu$};
\node[note] at (96.9,2.4) {$48{\times}128^3$};
\draw[flow] (103.2,10.5)--(108.6,10.5);
\node[liifhead,minimum width=12mm,minimum height=5mm] (hL) at (114.8,10.5) {LIIF head};
\restarticon{122.2}{14.4}{0.85}{cLIIF}
\draw[flow] (121,10.5)--(124.5,10.5);
\imgcubeop{124.7}{4.2}{11}{6.5}{liif}{0.38}
\pointsthreed{124.7}{4.2}{11}{6.5}{42}{5}{cLIIF}
\fill[cLIIF,draw=white,line width=0.5pt] (130.3,10.8) circle (0.6);
\draw[cLIIF,line width=0.4pt] (130.3,10.8)--(146.9,10.5);
\fill[cLIIF,draw=white,line width=0.5pt] (147.8,10.5) circle (0.75);
\draw[cLIIF,line width=0.5pt] (147.8,10.5) circle (1.05);
\node[comment] at (146,16.6) {\textcolor{cLIIF}{\textbf{sample random points}} in the volume};
\node[comment] at (149.8,10.5) {one point $\to$ one value $\hat{x}(\boldc)$};
\node[comment] at (146,4.4) {decoded at \textbf{any} point};
\end{scope}

\begin{scope}[shift={(0,0)}]
\node[ptitle] at (0,25.1) {(c) CoRe-AE with the {\color{cCLIF}CLIF} head};
\zonepin{0.5}{0.5}{38.5}{22.5}{$512^3$ grid: light layers}
\zonepin{40.5}{0.5}{104.5}{22.5}{$128^3$ latent grid: heavy layers}
\zonerestart{106.5}{0.5}{181}{22.5}{CLIF head: a random patch}
\imgcube{3}{3.8}{12.5}{7}{gt}
\node[lab] at (11.2,2.4) {$\boldx$ (whole volume)};
\draw[flow] (19.5,10.5)--(23,10.5);
\frustum{23.2}{33.5}{10.5}{13}{4.5}{0.3}{white}{cSlate,line width=0.45pt}
\node[lab] at (28.1,10.5) {$\boldcalE^{\mathrm{down}}$};
\draw[flow] (35,10.5)--(41,10.5);
\block{41.2}{8}{12}{5}{3}{cSlateHeavy}{cSlate!90!black,line width=0.7pt}{$\boldcalE^{\mathrm{enc}}$}
\draw[flow] (54.9,10.5)--(58.4,10.5);
\gridcube{58.6}{7.5}{6}{4}{cSlate}{$\boldz$}
\draw[flow] (67.8,10.5)--(71.3,10.5);
\block{71.5}{8}{14}{5}{3}{cSlateHeavy}{cSlate!90!black,line width=0.7pt}{$\boldcalD^{\mathrm{feat}}$}
\draw[flow] (87.2,10.5)--(90.7,10.5);
\featcube{90.9}{6.5}{8}{6}{gt}
\node[lab] at (96.9,4.6) {$\boldu$};
\draw[flow] (103.2,10.5)--(108.6,10.5);
\node[clifhead,minimum width=12mm,minimum height=5mm] (hC) at (114.8,10.5) {CLIF head};
\restarticon{122.2}{14.4}{0.85}{cCLIF}
\draw[flow] (121,10.5)--(124.5,10.5);
\imgcubeop{124.7}{4.2}{11}{6.5}{clif}{0.38}
\solidcube{128.6}{7.9}{3}{2}{cCLIF}{0.6}
\draw[cCLIF,line width=0.4pt] (131.6,10.9)--(145.9,10.5);
\imgsq{146}{8.4}{4.2}{clif_T.png}{cCLIF,line width=0.6pt}
\node[comment] at (146,16.6) {\textcolor{cCLIF}{\textbf{sample a random patch}} $\mathcal{T}$};
\node[comment,align=left] at (151.4,10.5) {convolutions decode\\all its voxels at once};
\node[comment] at (146,4.4) {decoded \textbf{tile by tile}};
\end{scope}
\end{tikzpicture}
	\caption{One training step. A pin marks parts that run once per step; a circular arrow marks parts that repeat. (a)~A dense patch-based \ac{AE} passes a new random $128^3$ patch through the whole \ac{CNN} at every step, seeing $1/64$ of the volume. (b, c)~\ac{CoReAE} encodes the whole $512^3$ volume once, with all heavy layers on the $128^3$ latent grid; only the small head repeats, decoding the features $\boldu$ at (b)~random points (\ac{LIIF}) or (c)~a random patch (\ac{CLIF}). The $\ell_1$ loss in \eqref{eq:ae_loss} compares the predictions with $\boldx$ at the same locations. Sizes are schematic.}
	\label{fig:method}

	\vspace{4mm}
	\input{./images/Method/fig_heads}
	\caption{How each head decodes the features $\boldu$ (Fig.~\ref{fig:method}), shown on one axial slice ($1$ feature cell $= 4\times4$ output voxels). (a)~\ac{LIIF} decodes one point $\boldc$: each of its nearest cells (4 in \ac{2D}, 8 in \ac{3D}) gives $(\boldu_j,\Delta\boldc_j,v)$ to a shared \ac{MLP}, and the outputs are summed with weights $w_j=S_j/S$ \eqref{eq:liif}. (b)~\ac{CLIF} decodes one patch $\mathcal{T}$: each voxel receives the feature of its nearest cell $\boldu_{j^\star}$, its offset $\Delta\boldc$ and the cell size $v$, and a small \ac{3D} \ac{CNN} returns all voxels of the patch jointly \eqref{eq:clif}. Feature colors are illustrative.}
	\label{fig:heads}
\end{figure*}

In the following, for notational conciseness, the dependence on $(\boldphi,\boldpsi)$ is made explicit through $\boldcalE_{\boldphi}$ and $\boldcalD_{\boldpsi}$. Consequently, the dependence of the training objectives in Section~\ref{sec:ae_train} on $(\boldphi,\boldpsi)$ is expressed through the corresponding encoded and decoded outputs.

\subsubsection{Encoder}\label{sec:encoder}

The encoder $\boldcalE_{\boldphi}$ operates in two stages (Fig.~\ref{fig:method}), i.e., 
\begin{equation}\label{eq:encoder}
	\boldcalE_{\boldphi} = \boldcalE_{\boldphi_2}^{\mathrm{enc}} \circ \boldcalE_{\boldphi_1}^{\mathrm{down}} \, , \quad \, \boldphi = \left(\boldphi_1,\boldphi_2\right) \, .
\end{equation}

The first stage, $\boldcalE_{\boldphi_1}^{\mathrm{down}}$, reduces the spatial resolution by a factor of $r=4$ along each axis using a small stack of lightweight, depthwise-separable convolutional blocks interleaved with two strided convolutions:
\begin{equation}\label{eq:down}
	\boldh = \boldcalE_{\boldphi_1}^{\mathrm{down}}(\boldx) \in \R^{C_h \times d \times h \times w} \, ,
\end{equation}
where $C_h$ is the hidden channel width. Depthwise convolutions mix space and channels separately, so their cost grows linearly, rather than quadratically, with the channel count. This lets $\boldcalE_{\boldphi_1}^{\mathrm{down}}$ stay cheap in activation memory even though it operates on the largest, full-resolution feature maps.

The second stage, $\boldcalE_{\boldphi_2}^{\mathrm{enc}}$, applies standard convolutional residual blocks at full channel width on the already-reduced grid $\boldh$, which contains 64 times fewer voxels than $\boldx$. Deferring dense, wide-channel computation until after this reduction is what makes it affordable: $\boldcalE_{\boldphi_2}^{\mathrm{enc}}$ can spend most of the encoder's capacity exactly where the posterior is read from, without inflating peak activation memory. It defines the encoder mean and the encoder \ac{STD},
\begin{equation}\label{eq:posterior}
    \left(\boldmu_{\boldphi}(\boldx), \boldsigma_{\boldphi}(\boldx)\right) = \boldcalE_{\boldphi_2}^{\mathrm{enc}}(\boldh )  \in \R^{2\times C_z \times d \times h \times w} \, ,
\end{equation}
which define the factorized Gaussian posterior $q_\phi(\boldz\mid\boldx) = \calN\left(\boldz;\boldmu_{\boldphi}, \boldSigma_{\boldphi}  \right)$, with $\boldSigma_{\boldphi}$ defined in \eqref{eq:encoder_sampling}\footnote{In our implementation, the encoder \ac{STD} was implemented as  $\boldsigma_{\boldphi} = \exp ( \bolds_{\boldphi})$ where $\bolds_{\boldphi}$ is the actual second output component of $\boldcalE_{\boldphi}$.}. In this work we used $C_z=4$.

\subsubsection{Implicit Decoder}\label{sec:liif}

We regard the discrete image $\boldx \in \calX = \R^{D \times H \times W}$ as the sampling of a continuous image on the normalized domain $\Omega \triangleq [-1,1]^3$, writing, with a slight abuse of notation, $\boldx(\boldc)\in\R$ and $\boldxhat(\boldc)\in\R$ for the true and estimated intensities at a location $\boldc \in \Omega$. The voxels of $\boldx$ correspond to the uniform grid $\calG = \{\boldc_i \in \Omega \mid i = 1,\ldots,DHW\}$, which, the underlying image being continuous, may in principle be chosen at any resolution, while the latent representation $\boldz$ is defined on a coarse grid $\widetilde{\mathcal{G}}$ of size $d \times h \times w$, obtained by downsampling the image grid by a factor $r$ along each spatial axis. In this view, decoding is no longer the construction of a dense $D \times H \times W$ tensor, which is what makes conventional decoders expensive, but the evaluation of the continuous image at any queried $\boldc$, and it decomposes into two operations: describing the local image content at each cell of $\tilde{\calG}$, and combining the corresponding predictions at the query point.
We factor the decoder into a feature-decoding stage and a coordinate-decoding stage, 
\begin{equation}\label{eq:decoder} 	
	\boldcalD_{\boldpsi} 	= \boldcalD_{\boldpsi_2}^{\mathrm{head}} 	\circ \boldcalD_{\boldpsi_1}^{\mathrm{feat}}, 	\qquad 	\boldpsi = \left(\boldpsi_1,\boldpsi_2\right). 
\end{equation}

The first stage maps the compact latent grid to a richer feature grid. Since the diffusion model operates on $\boldz$, the number of latent channels $C_z$ is kept small. A stack of stride-one convolutional residual blocks therefore expands the channel dimension while preserving the latent-grid resolution,
\begin{equation}\label{eq:feat}
	\boldu = 	\boldcalD_{\boldpsi_1}^{\mathrm{feat}}(\boldz) 	\in \R^{C_u \times d \times h \times w}, 	\qquad C_u > C_z .
\end{equation}
Because these convolutions are applied on the coarse grid $\tilde{\calG}$ rather than the full image grid $\calG$, this stage remains memory efficient.

The second stage renders voxel intensities at arbitrary coordinates with a lightweight rendering head $\boldcalD_{\boldpsi_2}^{\mathrm{head}}$. We consider two heads: a \ac{MLP}-based \ac{LIIF} head and a convolutional \ac{CLIF} head (Fig.~\ref{fig:heads}).

\paragraph{\ac{LIIF} head} This head uses the \ac{LIIF} local ensemble. The vector-valued decoder
$\boldcalD_{\boldpsi_2}^{\mathrm{head}}$ is defined using a shared real-valued \ac{MLP}, $\calD_{\boldpsi_2}^{\mathrm{mlp}}$. For each query coordinate $\boldc_i$, the \ac{MLP} predicts an intensity from each neighboring latent feature, and these predictions are combined using normalized trilinear weights:
\begin{equation}
    \label{eq:liif}
    \xhat(\boldc_i)
    =
    \sum_{j \in \mathcal{N}(\boldc_i)}
    w_{i,j}\,
    \calD_{\boldpsi_2}^{\mathrm{mlp}}
    \bigl(
        \boldu_j,\,
        \Delta\boldc_{i,j},\,
        v_i
    \bigr),
    \,
    i = 1,\dots,DHW .
\end{equation}
Here, $\mathcal{N}(\boldc_i)$ is the set of eight latent-grid cells neighboring $\boldc_i$, $\boldu_j \in \R^{C_u}$ is the feature vector at the center $\tilde{\boldc}_j$ of the $j$-th cell, and $\Delta\boldc_{i,j} = \boldc_i - \tilde{\boldc}_j$ is the relative offset between the query coordinate and that cell center. Moreover, $v_i$ denotes the query-cell size, and $w_{i,j}$ is the normalized trilinear weight satisfying $\sum_{j \in \mathcal{N}(\boldc_i)} w_{i,j} = 1$. As in \ac{LIIF}~\citep{Chen_2021_CVPR}, $\Delta\boldc_{i,j}$ and $v_i$ are expressed in units of the latent-cell size. This formulation enables continuous intensity prediction on arbitrary query grids.

\paragraph{\ac{CLIF} head} The \ac{LIIF} head predicts each voxel independently, so it cannot use the local context of neighboring voxels. The \ac{CLIF} head, introduced for \ac{2D} images in~\citep{chen2024image}, replaces the \ac{MLP} by a small convolutional network applied on the output grid. For each voxel $\boldc_i$, we build an input vector $\boldg_i = \bigl(\boldu_{j^\star(i)},\, \Delta\boldc_{i,j^\star(i)},\, v_i\bigr)$, where $j^\star(i)$ is the nearest latent cell, with the same latent-cell units as above. The head then predicts
\begin{equation}
    \label{eq:clif}
    \xhat(\boldc_i)
    =
    \bigl[\boldcalD_{\boldpsi_2}^{\mathrm{conv}}\bigl(\{\boldg_k\}_{k \in \calT^+}\bigr)\bigr]_i ,
    \qquad
    \boldc_i \in \calT ,
\end{equation}
where $\boldcalD_{\boldpsi_2}^{\mathrm{conv}}$ is a stack of $3^3$ convolutional residual blocks, $\calT$ is a cubic tile of the output grid, and $\calT^+$ is the same tile extended by a halo larger than the receptive field of $\boldcalD_{\boldpsi_2}^{\mathrm{conv}}$. The network uses only channel-wise normalization, with no statistics over space. Each output voxel therefore depends only on its own neighborhood, and the tiled output is identical to the output computed on the full grid.

During training, we sample $N \ll DHW$ coordinates per iteration for the \ac{LIIF} head, and small output tiles for the \ac{CLIF} head. During inference, the full volume is rendered in memory-bounded query chunks (\ac{LIIF}) or tiles (\ac{CLIF}). Thus, decoding requires no overlapping sub-volumes or blending, remains differentiable with respect to $\boldz$, and can be used inside the reconstruction procedure of Section~\ref{sec:recon}.

\subsubsection{Training Objective}\label{sec:ae_train}

We train the \ac{AE} with a regularized reconstruction objective. At each iteration, we sample $N$ coordinates from the target volume and minimize
\begin{align}\label{eq:ae_loss}
	\calL 	= {} & \frac{1}{N} \sum_{i=1}^{N} 	\left| \hat{\boldx}(\boldc_i) - \boldx(\boldc_i) \right| \nonumber \\ 	& + \lambda_{\mathrm{KL}} 	D_{\mathrm{KL}}\!\left( 	q_\phi(\cdot \mid \boldx) 	\,\|\, \calN\left(\boldzero_{\calZ},\boldI_{\calZ}\right) 	\right),
\end{align}
where $\hat{\boldx}(\boldc_i)$ is given by \eqref{eq:liif} or \eqref{eq:clif} and $q_\phi(\cdot \mid \boldx)$ by \eqref{eq:encoder_sampling} and \eqref{eq:posterior}.

\subsection{Latent Diffusion Model}\label{sec:ldm}

With $\boldcalE_{\boldphi}$ and $\boldcalD_{\boldpsi}$ fixed, a \ac{3D} U-Net noise predictor $\boldepsilon_{\boldtheta}$ parametrized by $\boldtheta$ is trained as an unconditional \ac{DDPM} directly on latents $\boldz = \boldmu_{\boldphi}(\boldx)$ from the training set,
\begin{equation}\label{eq:ddpm_loss}
	\calL_{\mathrm{DDPM}} = \mathbb{E}_{\boldz,\bm{\epsilon},t}\left[ \left\| \boldepsilon - \boldepsilon_{\boldtheta}(\boldz_t,t) \right\|_2^2 \right] \, ,
\end{equation}
where $\boldepsilon \sim \calN(\boldzero_{\calZ},\boldI_{\calZ})$,  $\boldz_t$ is a degraded version of $\boldz$ at uniformly sampled time step $t=1,\dots,T$ and  noise schedule $\{\alpha_t\}_{t=1}^T$, i.e.,
\begin{equation}
	\boldz_t = \sqrt{\bar{\alpha}_t} \boldz+  \sqrt{1-\bar{\alpha}_t} \boldepsilon
\end{equation}
with $\bar{\alpha}_t = \prod_{t'=1}^t \alpha_{t'}$

In this setting, the \ac{DM}  operates in the lower-dimensional latent space $\calZ$  rather than on the image space $\calX$. The trained noise predictor $\boldepsilon_{\boldtheta}$ defines an implicit generative prior $p(\boldz)$ over $\calZ$, reused by both downstream tasks below.

\subsection{Measurement-Guided Reconstruction}\label{sec:recon}

For reconstruction, we reuse the frozen latent prior $p(\boldz)$ and solve the latent inverse problem~\eqref{eq:inv_prob_latent} using the hard-data-consistency procedure of ReSample~\citep{song2024solving}, extended from \ac{2D} images to \ac{3D} latent volumes.

At each reverse step, Tweedie's formula provides an estimate $\hat{\boldz}_{0|t}$ of the clean latent representation. This estimate is then corrected toward the measurements $\boldy$ in three stages.

\emph{Stage 1.} At high noise levels, $t>t_1$, no correction is applied because the decoded estimate is not yet reliable.

\emph{Stage 2.} At intermediate noise levels, $t_2<t\leq t_1$, data consistency is enforced in image space every $g$ steps. The decoded volume $\boldcalD_{\boldpsi}(\hat{\boldz}_{0|t})$ initializes a regularized least-squares problem in $\boldx$, solved with $K$ conjugate-gradient iterations. The solution is then combined with the decoded volume through a weighted correction of weight $\kappa$, and re-encoded into the latent space with the posterior mean $\boldmu_{\boldphi}$. This stage is cheap because the decoder output is treated as fixed, so no backpropagation through the decoder is needed. However, latent \acp{AE} are lossy and only approximately invertible \citep{chung2023prompt}, so re-encoding may discard part of the correction and smooth fine details.

\emph{Stage 3.} At low noise levels, $t\leq t_2$, data consistency is therefore enforced directly in latent space. Starting from $\hat{\boldz}_{0|t}$, the latent is optimized through $\boldcalA\circ\boldcalD_{\boldpsi}$ with $M$ Adam iterations of step size $\tau$. This stage is more expensive because it backpropagates through the decoder, but it refines fine structures and reduces the errors introduced by re-encoding~\citep[App.~B]{song2024solving}. The decoder is evaluated in coordinate chunks (\ac{LIIF}) or tiles (\ac{CLIF}), with gradient checkpointing on each chunk or tile, so that memory stays bounded.

After each correction, the corrected latent is mapped back to the diffusion trajectory with the $\mathtt{StochasticResample}$ step of ReSample~\citep[Prop.~2]{song2024solving}. The complete procedure is given in Algorithm~\ref{algo:recon}.

We follow the medical-image settings of ReSample~\citep{song2024solving}. We use $T=100$ deterministic \ac{DDIM} steps ($\eta=0$). Stage~1 covers the first 10 steps ($t_1=90$). Stage~2 applies the image-space correction every $g=5$ steps, with $\rho=10^{-3}\|\boldcalA\|_2^2$, $K=10$ conjugate-gradient iterations, tolerance $\epsilon_{\mathrm{tol}}=10^{-4}$, and $\kappa=0.9$. Stage~3 covers the last 10 steps ($t_2=10$), with $M=5$ Adam iterations of step size $\tau=5\times10^{-3}$. Stochastic resampling uses $\gamma=40$ with the adaptive variance schedule of ReSample.

\begin{algorithm}[t]
    \scriptsize
    \caption{Latent-Implicit \ac{3D} Reconstruction with
    ReSample~\citep{song2024solving}}
    \label{algo:recon}
    \begin{algorithmic}[1]
        \Require
        $\boldy$,
        $\boldcalA(\cdot)$,
        $\boldmu_{\boldphi}(\cdot)$,
        $\boldcalD_{\boldpsi}(\cdot)$,
        $\boldepsilon_{\boldtheta}(\cdot,t)$,
        $\{\bar{\alpha}_t\}_{t=0}^{T}$
        \Require
        $\eta$, $\{\delta_t\}$;
        thresholds $t_1>t_2$;
        guidance period $g$;
        $\rho$, $\kappa$, $K$, $\epsilon_{\mathrm{tol}}$;
        $\tau$, $M$;
        $\gamma$
        \State $\boldz_T\sim\calN(\boldzero,\boldI)$
        \For{$t=T,\dots,1$}
            \State
            $\boldepsilon_1\sim\calN(\boldzero,\boldI)$,
            \quad
            $\hat{\boldepsilon}_t
            =
            \boldepsilon_{\boldtheta}(\boldz_t,t)$
            \State
            $\hat{\boldz}_{0|t}
            =
            \tfrac{1}{\sqrt{\bar{\alpha}_t}}
            \bigl(
                \boldz_t
                -
                \sqrt{1-\bar{\alpha}_t}\,
                \hat{\boldepsilon}_t
            \bigr)$
            \Comment{Tweedie's formula}
            \State
            $\boldz'_{t-1}
            =
            \sqrt{\bar{\alpha}_{t-1}}\,
            \hat{\boldz}_{0|t}
            +
            \sqrt{
                1-\bar{\alpha}_{t-1}-\eta^2\delta_t^2
            }\,
            \hat{\boldepsilon}_t
            +
            \eta\delta_t\,\boldepsilon_1$
            \Comment{\acs{DDIM} step}
            \If{$t>t_1$ \textbf{or} ($t>t_2$ \textbf{and} $(t_1-t)\bmod g\neq0$)}
                \State
                $\boldz_{t-1}=\boldz'_{t-1}$
                \Comment{Stage 1, and unguided steps of Stage 2}
            \Else
                \If{$t>t_2$}
                    \Comment{Stage 2: image-space correction, every $g$ steps}
                    \State
                    $\boldx^\star
                    \approx
                    \operatorname*{arg\,min}_{\boldx}
                    \;
                    \tfrac{1}{2}
                    \left\|
                        \boldy-\boldcalA(\boldx)
                    \right\|_2^2
                    +
                    \tfrac{\rho}{2}
                    \left\|
                        \boldx
                        -
                        \boldcalD_{\boldpsi}
                        \bigl(\hat{\boldz}_{0|t}\bigr)
                    \right\|_2^2$
                    \Comment{CG, $K$ iter., tol.\ $\epsilon_{\mathrm{tol}}$}
                    \State
                    $\hat{\boldx}_0
                    =
                    \boldcalD_{\boldpsi}\bigl(\hat{\boldz}_{0|t}\bigr)
                    +
                    \kappa\,\bigl(\boldx^\star-\boldcalD_{\boldpsi}\bigl(\hat{\boldz}_{0|t}\bigr)\bigr)$
                    \Comment{Weighted correction}
                    \State
                    $\boldz^{\mathrm{dc}}_{0|t}
                    =
                    \boldmu_{\boldphi}
                    \bigl(\hat{\boldx}_0\bigr)$
                    \Comment{Re-encode with the posterior mean}
                \Else
                    \Comment{Stage 3: latent-space correction}
                    \State  
                    $\boldz^{\mathrm{dc}}_{0|t}
                    \approx
                    \operatorname*{arg\,min}_{\boldz}
                    \;
                    \tfrac{1}{2}
                    \left\|
                        \boldy
                        -
                        \boldcalA
                        \bigl(
                            \boldcalD_{\boldpsi}(\boldz)
                        \bigr)
                    \right\|_2^2$
                    \Comment{Adam, $M$ iter., step $\tau$, init.\ $\hat{\boldz}_{0|t}$}
                \EndIf
                \State
                $\boldz_{t-1}
                =
                \mathtt{StochasticResample}
                \bigl(
                    \boldz^{\mathrm{dc}}_{0|t},
                    \boldz'_{t-1},
                    \gamma
                \bigr)$
            \EndIf
        \EndFor
        \State
        \Return
        $\hat{\boldx}
        =
        \boldcalD_{\boldpsi}(\boldz_0)$
    \end{algorithmic}
\end{algorithm}

\section{Experiments and Results}\label{sec:results}

We evaluate the proposed framework in three stages. First, we assess the latent \ac{AE}, since it defines the representation used by the \ac{LDM}. Second, we evaluate unconditional volume generation and compare the generated distributions with existing \ac{3D} medical generative models. Finally, we study measurement-guided reconstruction against classical baselines and a slice-wise pretrained \ac{2D} \ac{DM}-prior baseline.

\subsection{Experimental Details}\label{sec:exp_details}

We evaluate the proposed framework on two volumetric medical imaging modalities. For \ac{CT}, we use 700 volumes from LIDC-IDRI~\citep{armato2011lidc} and the CT-LYMPH-NODES collection~\citep{roth2015ctlymphnodes}, resampled or cropped to a fixed spatial size of 512\textsuperscript{3}. For \ac{MRI}, we use 540 T1-weighted volumes from the IXI dataset~\citep{ixi}, processed to 256\textsuperscript{3}.

For \ac{CT}, intensities are clipped to $[-1000,1500]$\,HU and linearly rescaled to $[-1,1]$. Volumes with different axial extents are cropped or zero-padded to obtain a fixed 512\textsuperscript{3} grid. For \ac{MRI}, each volume is normalized using the 1st and 99th intensity percentiles, clipped to this range, and rescaled to $[-1,1]$ before resizing or cropping to 256\textsuperscript{3}. All splits are performed at the subject level.

The \ac{AE} is trained separately for each modality with a downsampling factor $r=4$, $C_z=4$ latent channels, and $N=200{,}000$ randomly sampled coordinate queries per iteration. This gives latent grids of size 4\texttimes{}128\textsuperscript{3} for \ac{CT} and $4\times64^3$ for \ac{MRI}. After training, the \ac{AE} is frozen and a separate unconditional \ac{3D} \ac{LDM} is trained for each modality. Both the \ac{AE} and the \ac{3D} U-Net diffusion model are trained on a single 48\,GB \ac{GPU}, without model parallelism, gradient checkpointing, or patch-wise training.

\subsection{Autoencoding Performance}\label{sec:ae_results}

The \ac{AE} sets an upper bound on downstream quality, since the diffusion prior operates entirely in latent space and details lost during encoding cannot be recovered during generation or reconstruction. We therefore evaluate the \ac{AE} first, in terms of fidelity (\acs{PSNR}, \acs{SSIM}, \ac{LPIPS}, \ac{HFEN}, \ac{FID}) and computational cost (runtime and peak \ac{GPU} memory).

We compare \ac{CoReAE} with the \acp{AE} used in MAISI~\citep{zhao2025maisi} and 3D~MedDiffusion~\citep{wang2025meddiffusion}. \ac{CoReAE} encodes the volume once into a compact latent grid and decodes voxel intensities through coordinate queries, with chunks (\ac{LIIF}) or tiles (\ac{CLIF}) used only to bound memory. In contrast, MAISI uses overlapping sliding-window autoencoding with window sizes 128\textsuperscript{3} and 256\textsuperscript{3}, denoted MAISI-W128 and MAISI-W256. For 3D MedDiffusion we evaluate its $4\times$ and $8\times$ patch-volume \acp{AE}, decoding the latent with overlapping 128\textsuperscript{3} patches. Both baselines blend overlapping predictions using Gaussian-weighted averaging with 50\% patch overlap (a stride of half the patch size). Fidelity metrics are reported as mean~$\pm$~\ac{STD} over held-out volumes.

\begin{table}[htbp]
	\centering
	\caption{
		Autoencoding performance on held-out \ac{CT} and \ac{MRI} volumes. Mean~$\pm$~\ac{STD} over volumes. Bold: best value for each metric and dataset.
	}
	\label{tab:ae_results}
	\scriptsize
	\setlength{\tabcolsep}{4pt}
	\renewcommand{\arraystretch}{1.05}
	\resizebox{\linewidth}{!}{%
	\begin{tabular}{c l c c c c c}
		\toprule
		\textbf{Dataset} & \textbf{Method}
		& \textbf{\acs{PSNR} $\uparrow$}
		& \textbf{\acs{SSIM} $\uparrow$}
		& \textbf{\acs{LPIPS} $\downarrow$}
		& \textbf{\acs{HFEN} $\downarrow$}
		& \textbf{\acs{FID}$\downarrow$} \\
		\midrule

		\multirow{6}{*}{\rotatebox[origin=c]{90}{\textbf{\ac{CT} $512^3$}}}
		& \acs{CoReAE} (\acs{LIIF}) & \textbf{42.32 $\pm$ 2.42} & \textbf{0.9835 $\pm$ 0.0127} & 0.0419 $\pm$ 0.0220 & 0.1407 $\pm$ 0.0323 & 0.671 \\
		& \acs{CoReAE} (\acs{CLIF}) & 40.11 $\pm$ 1.99 & 0.9781 $\pm$ 0.0127 & 0.0291 $\pm$ 0.0155 & 0.1959 $\pm$ 0.0296 & 0.094 \\
		& MAISI-W128 & 39.85 $\pm$ 1.69 & 0.9721 $\pm$ 0.0076 & \textbf{0.0135 $\pm$ 0.0024} & \textbf{0.1216 $\pm$ 0.0126} & 0.159 \\
		& MAISI-W256 & 38.32 $\pm$ 2.07 & 0.9749 $\pm$ 0.0041 & 0.0193 $\pm$ 0.0055 & 0.1453 $\pm$ 0.0177 & \textbf{0.091} \\
		& 3D MedDiffusion 4$\times$ & 39.76 $\pm$ 1.09 & 0.9300 $\pm$ 0.0248 & 0.0199 $\pm$ 0.0073 & 0.1338 $\pm$ 0.0180 & 0.107 \\
		& 3D MedDiffusion 8$\times$ & 37.16 $\pm$ 0.46 & 0.9509 $\pm$ 0.0079 & 0.0310 $\pm$ 0.0095 & 0.3092 $\pm$ 0.0303 & 0.133 \\

		\midrule

		\multirow{6}{*}{\rotatebox[origin=c]{90}{\textbf{\ac{MRI} $256^3$}}}
		& \acs{CoReAE} (\acs{LIIF}) & \textbf{33.77 $\pm$ 0.74} & \textbf{0.9615 $\pm$ 0.0063} & 0.0715 $\pm$ 0.0107 & \textbf{0.1305 $\pm$ 0.0069} & 3.280 \\
		& \acs{CoReAE} (\acs{CLIF}) & 31.88 $\pm$ 0.68 & 0.9373 $\pm$ 0.0090 & 0.0432 $\pm$ 0.0053 & 0.1826 $\pm$ 0.0085 & 1.450 \\
		& MAISI-W128 & 31.80 $\pm$ 0.78 & 0.9361 $\pm$ 0.0124 & \textbf{0.0317 $\pm$ 0.0048} & 0.2068 $\pm$ 0.0150 & 0.395 \\
		& MAISI-W256 & 30.96 $\pm$ 0.75 & 0.9293 $\pm$ 0.0140 & 0.0332 $\pm$ 0.0047 & 0.2273 $\pm$ 0.0158 & \textbf{0.246} \\
		& 3D MedDiffusion 4$\times$ & 30.79 $\pm$ 0.71 & 0.9241 $\pm$ 0.0138 & 0.0402 $\pm$ 0.0051 & 0.2437 $\pm$ 0.0161 & 0.694 \\
		& 3D MedDiffusion 8$\times$ & 25.29 $\pm$ 0.54 & 0.7418 $\pm$ 0.0325 & 0.1064 $\pm$ 0.0125 & 0.6743 $\pm$ 0.0339 & 2.202 \\

		\bottomrule
	\end{tabular}%
	}
\end{table}

On \ac{CT} (Table~\ref{tab:ae_results}), \ac{CoReAE} (\ac{LIIF}) gives the highest \ac{PSNR} and \ac{SSIM} (42.32~dB, 0.9835), about 2.5~dB above the best baseline (MAISI-W128, 39.85~dB). \ac{CoReAE} (\ac{CLIF}) loses 2.2~dB compared with \ac{LIIF} but stays above all baselines in \ac{PSNR}. In exchange, it improves the perceptual metrics: \ac{LPIPS} goes from 0.042 to 0.029, and \ac{FID} from 0.671 to 0.094, close to the best value (0.091, MAISI-W256). MAISI-W128 keeps the best \ac{LPIPS} and \ac{HFEN}. The two heads therefore show a clear trade-off: the \ac{LIIF} head favors voxel-wise fidelity, and the \ac{CLIF} head favors perceptual quality.

\begin{figure*}[htbp]
    \centering
    \includegraphics[width=\textwidth]{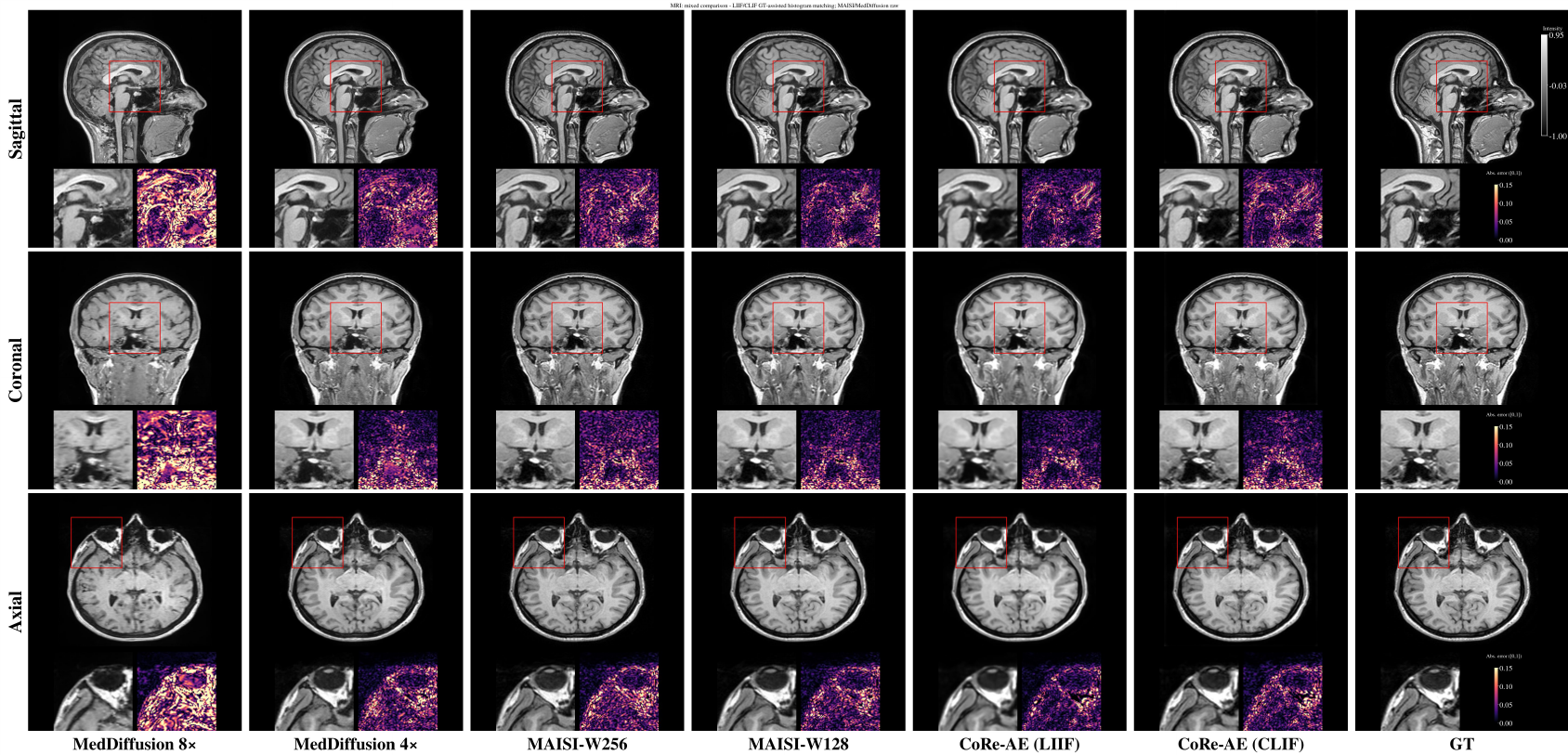}
    \caption{
        Qualitative comparison of \ac{MRI} autoencoding on a representative IXI volume. Rows show central sagittal, coronal, and axial slices. For each method, the top panel shows the decoded slice and the selected region of interest (red box), and the bottom panels show its magnification and the absolute error map, with intensities normalized to $[-1,1]$. Columns compare 3D MedDiffusion ($8\times$ and $4\times$), MAISI (window $256^3$ and $128^3$), the proposed \ac{AE} with the \ac{LIIF} and CLIF heads, and the ground truth (GT).
    }
    \label{fig:ae_mri_qualitative}
\end{figure*}

\begin{figure*}[htbp]
\centering
\includegraphics[width=\textwidth]{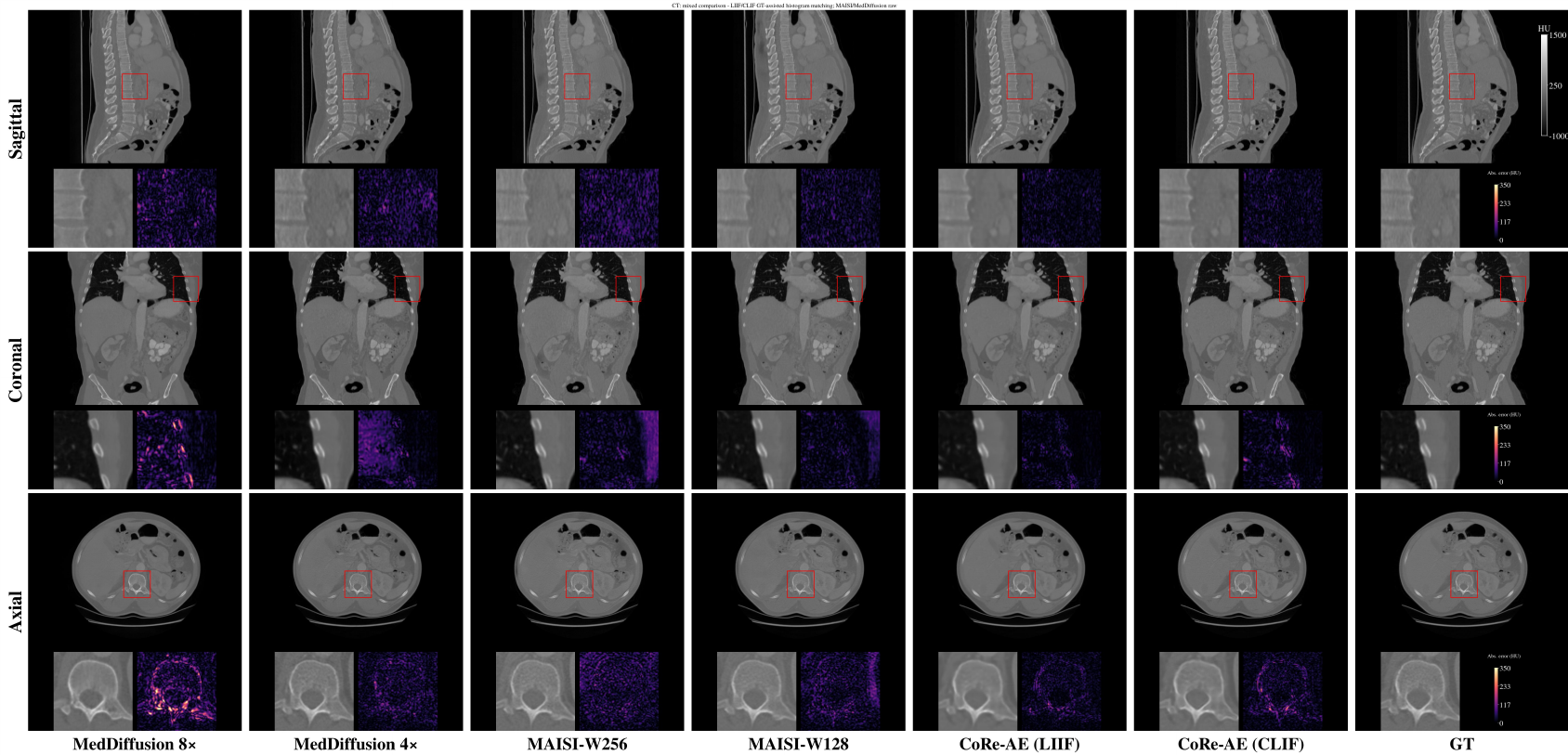}
\caption{
Qualitative comparison of \ac{CT} autoencoding on a representative held-out volume. Rows show central sagittal, coronal, and axial slices. For each method, the top panel shows the decoded slice and the selected region of interest (red box), and the bottom panels show its magnification and the absolute error map in HU. Columns compare 3D MedDiffusion ($8\times$ and $4\times$), MAISI (window $256^3$ and $128^3$), the proposed \ac{AE} with the \ac{LIIF} and CLIF heads, and the ground truth (GT). Decoder outputs are converted to HU using the range $[-1000,1500]$.
}
\label{fig:ae_ct_qualitative}
\end{figure*}

The \ac{MRI} results follow the same pattern. \ac{CoReAE} (\ac{LIIF}) gives the best \ac{PSNR}, \ac{SSIM} and \ac{HFEN} (33.77~dB, 0.9615, 0.131), about 2~dB above MAISI-W128. \ac{CoReAE} (\ac{CLIF}) is close to MAISI-W128 in \ac{PSNR} (31.88 vs.\ 31.80~dB) and reduces the \ac{FID} of the \ac{LIIF} head from 3.28 to 1.45. MAISI keeps the best \ac{LPIPS} and \ac{FID}. The qualitative comparisons in Figs.~\ref{fig:ae_mri_qualitative} and~\ref{fig:ae_ct_qualitative} agree with these numbers: \ac{CoReAE} preserves anatomy in the axial, coronal, and sagittal views, with no visible patch seams, and the remaining error is located on edges and fine textures.

\begin{table*}[htbp]
	\centering
	\caption{Computational cost of the \acp{AE} for one 512\textsuperscript{3} \ac{CT} volume (batch size 1, single 48\,GB \ac{GPU}). Units: number of decoded chunks, tiles, windows or patches. Volume processed: voxels processed by the decoder, including overlaps and halos, relative to the volume size. Memory: peak allocated \ac{GPU} memory. Decoding + gradient: decoding and gradient with respect to the latent. OOM: out of memory. The measurement protocol is given in Supplementary Section~\ref{sec:s_protocol}.}
	\label{tab:cost512}
	\scriptsize
	\setlength{\tabcolsep}{4pt}
	\renewcommand{\arraystretch}{1.05}
	\resizebox{\linewidth}{!}{%
	\begin{tabular}{l c c c c c c c c c}
		\toprule
		& & & & \multicolumn{2}{c}{\textbf{Autoencoding}} & \multicolumn{2}{c}{\textbf{Decoding}} & \multicolumn{2}{c}{\textbf{Decoding + gradient}} \\
		\cmidrule(lr){5-6}\cmidrule(lr){7-8}\cmidrule(lr){9-10}
		\textbf{Method} & \textbf{Params (M)} & \textbf{Units} & \textbf{Vol.\ processed} & \textbf{GFLOPs} & \textbf{Time (s)} & \textbf{Time (s)} & \textbf{Mem.\ (GiB)} & \textbf{Time (s)} & \textbf{Mem.\ (GiB)} \\
		\midrule
		\acs{CoReAE} (\acs{LIIF}) & 1.5 & 512 chunks & 1.00$\times$ & \textbf{91\,645} & \textbf{8.2} & \textbf{7.8} & \textbf{1.7} & \textbf{28.5} & 9.6 \\
		\acs{CoReAE} (\acs{CLIF}) & 2.0 & 216 tiles & 1.39$\times$ & 207\,309 & 9.1 & 8.7 & 2.1 & 30.8 & \textbf{9.4} \\
		MAISI-W128 & 20.9 & 343 windows & 5.36$\times$ & 4\,166\,453 & 242.8 & 142.7 & 3.3 & 596.6 & 15.1 \\
		MAISI-W256 & 20.9 & 27 windows & 3.38$\times$ & 2\,387\,787 & 115.0 & 69.9 & 18.1 & OOM & $>$48 \\
		3D MedDiffusion 4$\times$ & 26.0 & 343 patches & 5.36$\times$ & 1\,107\,631 & 41.6 & 37.4 & 5.2 & 72.9 & 12.7 \\
		3D MedDiffusion 8$\times$ & 52.3 & 343 patches & 5.36$\times$ & 1\,088\,333 & 38.2 & 34.0 & 4.0 & 71.5 & 12.7 \\
		\bottomrule
	\end{tabular}%
	}
\end{table*}

Table~\ref{tab:cost512} gives the cost of each \ac{AE} for a 512\textsuperscript{3} \ac{CT} volume. \ac{CoReAE} decodes the volume in 7.8~s (\ac{LIIF}) and 8.7~s (\ac{CLIF}), against 34.0 to 142.7~s for the baselines, i.e., 4 to 18 times faster. It also has the lowest decoding memory (1.7 and 2.1~GiB). The same holds when the gradient with respect to the latent is computed, as required in reconstruction: \ac{CoReAE} needs about 30~s and less than 10~GiB, while MAISI-W256 runs out of memory on a 48\,GB \ac{GPU}.

The cost of the baselines follows from their decoding strategy. A dense full-resolution decode is infeasible under a practical memory budget, so the volume is decoded in overlapping sub-volumes, the full \ac{3D} decoder is re-evaluated on each, and the outputs are blended. The baselines therefore process 3.4 to 5.4 times the volume, and their runtime grows with the number of sub-volumes. This forces a trade-off between memory and speed: MAISI-W256 is faster than MAISI-W128 (69.9 vs.\ 142.7~s) but needs 18.1~GiB instead of 3.3~GiB. \ac{CoReAE} avoids this trade-off. Its convolutions run once on the compact latent grid, and only the light rendering head runs at full resolution, so it processes about 1 time the volume. The runtime follows the computation (GFLOPs), not the number of parameters (Supplementary Fig.~\ref{fig:s_model_cost}), and the memory of \ac{CoReAE} stays almost constant when the output size grows (Supplementary Fig.~\ref{fig:s_resolution}).

\ac{CoReAE} thus combines the highest voxel-wise fidelity with a much lower cost than the baselines. This efficiency is not merely a convenience. In guided reconstruction, the decoder is embedded in an iterative loop and is evaluated many times per volume as the prior is combined with the measurement model. Its per-call cost therefore multiplies across iterations and dominates the total reconstruction time. A decoder that decodes a volume in seconds rather than minutes, and whose gradient fits in memory, is what makes reconstruction with this prior practical (Section~\ref{sec:recon}).

\subsection{Unconditional Volume Generation}
We assess unconditional generation with a 2.5-D \ac{FID} protocol. Each volume is sliced along the axial (XY), coronal (YZ), and sagittal (ZX) planes; per-slice features are extracted with a RadImageNet-pretrained ResNet-50~\citep{mei2022radimagenet}, and \ac{FID} is computed per plane and averaged. Preprocessing is identical across methods, and we use 40 unseen volumes per dataset from LIDC-IDRI \acs{CT} and IXI \acs{MRI} T1.

For \acs{MRI}, IXI is anonymized with a face mask that we could apply to our samples but not to the MAISI samples, so the reference and the two methods do not share a common spatial support. We therefore report the \acs{MRI} scores as indicative only and draw no ranking from them. 3D MedDiffusion is excluded from \acs{MRI} altogether, as its generator synthesizes only the brain parenchyma and omits the skull and surrounding anatomy present in IXI, precluding a meaningful whole-head comparison.

\begin{table}[htbp]
	\centering
	\caption{Unconditional generation quality: 2.5-D \ac{FID} per plane and averaged (lower is better). Bold: best value for each plane and dataset. \ac{MRI} scores are indicative only (see text).}
	\label{tab:generation_fid}
	\scriptsize
	\setlength{\tabcolsep}{5pt}
	\renewcommand{\arraystretch}{1.05}
	\begin{tabular}{clcccc}
		\toprule
		\textbf{Dataset} & \textbf{Method}
		& \textbf{FID$_{\mathrm{XY}}$ $\downarrow$}
		& \textbf{FID$_{\mathrm{YZ}}$ $\downarrow$}
		& \textbf{FID$_{\mathrm{ZX}}$ $\downarrow$}
		& \textbf{FID$_{\mathrm{avg}}$ $\downarrow$} \\
		\midrule
		\multirow{4}{*}{\rotatebox[origin=c]{90}{\textbf{CT}}}
		& \acs{CoReAE} (\acs{LIIF}) & \textbf{2.202} & \textbf{2.805} & \textbf{2.560} & \textbf{2.522} \\
		& \acs{CoReAE} (\acs{CLIF}) & 2.496 & 2.924 & 2.646 & 2.689 \\
		& MAISI & 3.788 & 4.751 & 3.947 & 4.162 \\
		& 3D MedDiffusion & 2.375 & 4.682 & 3.253 & 3.437 \\
		\midrule
		\multirow{3}{*}{\rotatebox[origin=c]{90}{\textbf{MRI}}}
		& \acs{CoReAE} (\acs{LIIF}) & \textbf{3.579} & \textbf{3.550} & \textbf{3.210} & \textbf{3.446} \\
		& \acs{CoReAE} (\acs{CLIF}) & 3.658 & 3.610 & 3.250 & 3.506 \\
		& MAISI & 5.769 & 6.047 & 6.555 & 6.124 \\
		\bottomrule
	\end{tabular}
\end{table}

\begin{figure*}[!t]
\centering
\includegraphics[width=\textwidth]{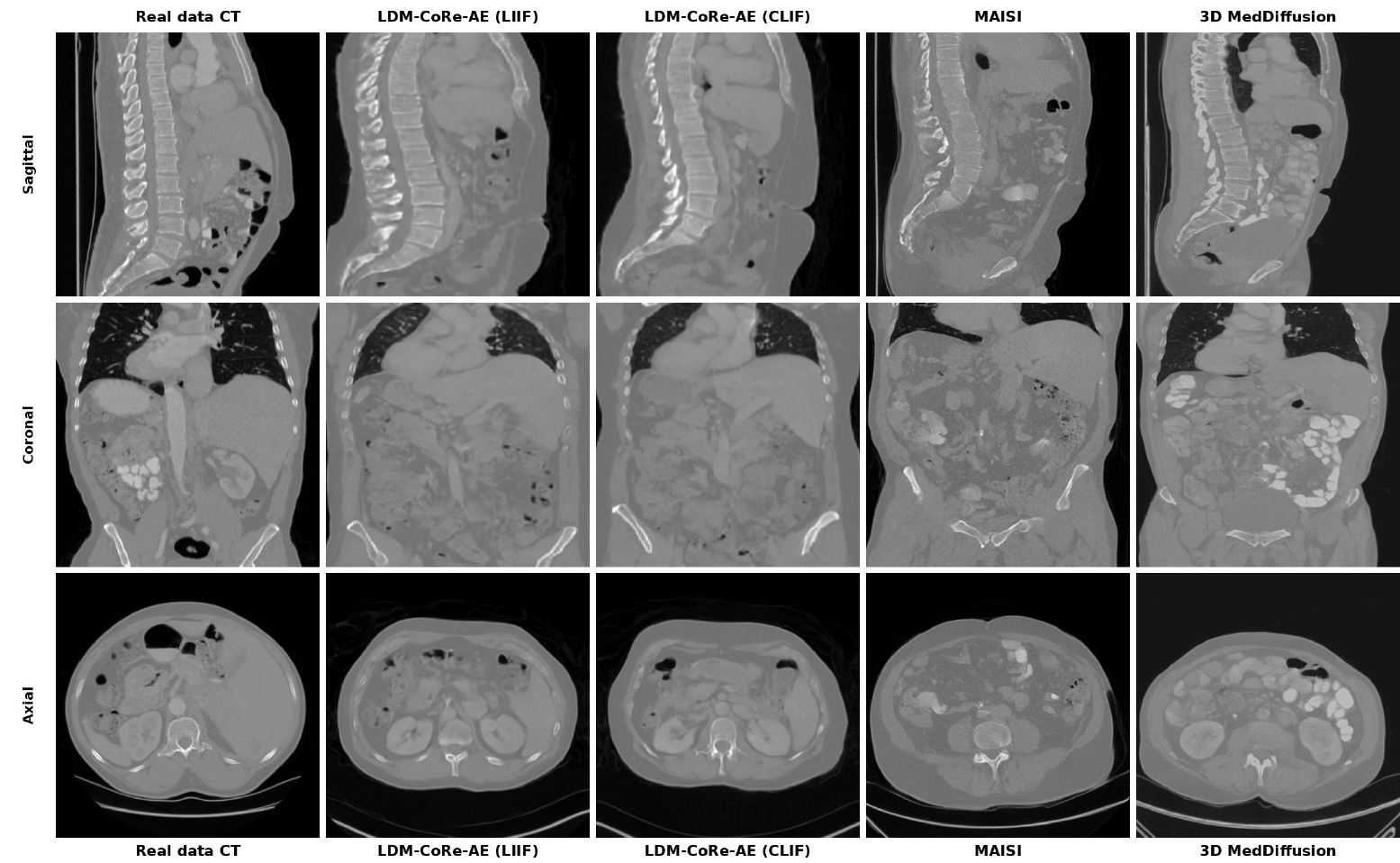}

\caption{
Qualitative comparison of unconditional \ac{CT} volume generation.
Rows show sagittal, coronal, and abdominal axial views with comparable
torso framing. Columns show an unpaired real \ac{CT} volume,
LDM-CoRe-AE (LIIF), LDM-CoRe-AE (CLIF), MAISI, and 3D MedDiffusion.
The first generated volume from each proposed model is shown.
}
\label{fig:generation_ct_qualitative}
\end{figure*}

On the matched-support \acs{CT} benchmark (Table~\ref{tab:generation_fid}), \ac{CoReAE} gives the lowest \ac{FID} on every plane and on average, with 2.522 for the \ac{LIIF} head and 2.689 for the \ac{CLIF} head, against 3.437 for 3D MedDiffusion and 4.162 for MAISI. 3D MedDiffusion is close to our method on axial slices (2.375 vs.\ 2.202), but its \ac{FID} increases on the coronal and sagittal planes (4.682 and 3.253). Our scores stay similar across the three planes. The two heads give close results, so the choice of the rendering head has little effect on generation quality. On IXI \acs{MRI}, our method also reports lower \ac{FID} on every plane and on average (3.446 vs.\ 6.124); we attribute part of this gap to the unmatched masking support rather than to sample quality, and treat it as indicative only. The qualitative comparison in Fig.~\ref{fig:generation_ct_qualitative} is consistent with these scores: all methods produce anatomically coherent volumes across the three views, with our samples showing no patch seams or slice discontinuities. Overall, our compact representation generates full volumes with lower \ac{FID} than strong diffusion baselines, at a fraction of their decoding cost (Section~\ref{sec:ae_results}).

\subsection{Image Reconstruction}\label{sec:recon_results}

We evaluate the frozen latent prior $p(\boldz)$ as a regularizer for the inverse problem~\eqref{eq:inv_prob_latent} on sparse-view \ac{CT} and accelerated single-coil \ac{MRI}. Data consistency is imposed through $\boldcalA\circ\boldcalD_{\boldpsi}$, with the decoder rendering memory-bounded slabs of adjacent slices (Section~\ref{sec:recon}). For \ac{CT}, the parallel-beam projection operator is implemented using TorchRadon~\citep{torch_radon}, with 30 or 60 noisy views. For \ac{MRI}, we use $\boldcalA(\boldx)=\boldM\boldcalF(\boldx)$, where $\boldcalF$ is the slice-wise two-dimensional Fourier transform and $\boldM$ is a variable-density Cartesian mask. The central $8\%$ of $k$-space is fully sampled, while the remaining lines are undersampled to obtain acceleration factors $R=4$ and $R=8$.

Among approaches that extend pretrained \ac{2D} diffusion priors to volumetric reconstruction, DiffusionMBIR introduces inter-slice regularization~\citep{chung2023solving}, whereas TPDM combines priors trained on perpendicular planes~\citep{lee2023improving}. We use \ac{DDS}~\citep{chung2024decomposed} as a strong representative of this family because it couples slice-wise denoising with conjugate-gradient data consistency and has demonstrated improved reconstruction quality and lower sampling cost than earlier volumetric diffusion solvers.

Accordingly, the quantitative comparison is made against \ac{DDS}, while \ac{FBP} and \ac{ZF}-\ac{IFFT} are included as analytic and zero-filled references for \ac{CT} and \ac{MRI}, respectively. Results are evaluated in \ac{HU} for \ac{CT} and as magnitude images for \ac{MRI}, on 10 held-out volumes per modality. In addition to \ac{PSNR} and \ac{SSIM}, we report metrics that are sensitive to fine details and texture: \ac{LPIPS}, \ac{HFEN}, the error of the image gradients, the Fourier ring correlation in the upper half of the frequency band (FRC$_{\mathrm{HF}}$), the ratio between the reconstructed and ground-truth energy in this band ($P_{\mathrm{HF}}$, ideal value 1), and the relative errors of the gray-level co-occurrence contrast and homogeneity inside the body ($\Delta$Contrast, $\Delta$Homog.).

\begin{figure*}[htbp]
\centering
\includegraphics[width=0.85\textwidth]{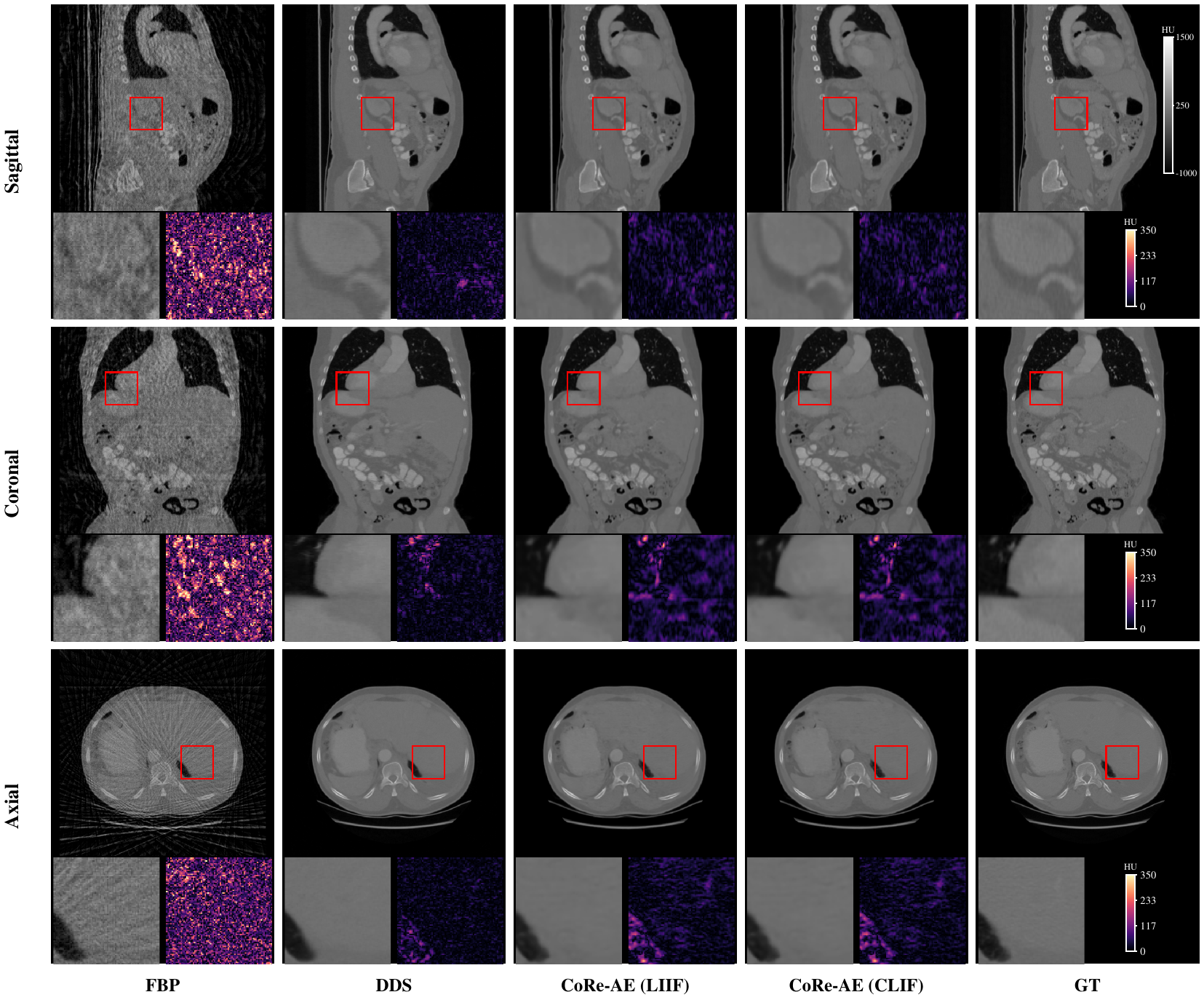}
\caption{
Qualitative comparison of sparse-view \ac{CT} reconstruction from 30 noisy projections. Rows show sagittal, coronal, and axial slices. For each method, the top panel shows the reconstruction and the selected region of interest, and the bottom panels show its magnification and the absolute error map in HU. Columns compare \ac{FBP}, \ac{DDS}, \acs{CoReAE} (\acs{LIIF}), \acs{CoReAE} (\acs{CLIF}), and the ground truth (GT). All images use the same HU window.
}
\label{fig:reconstruction_ct_30views_qualitative}
\end{figure*}

\begin{figure*}[htbp]
\centering
\includegraphics[width=0.85\textwidth]{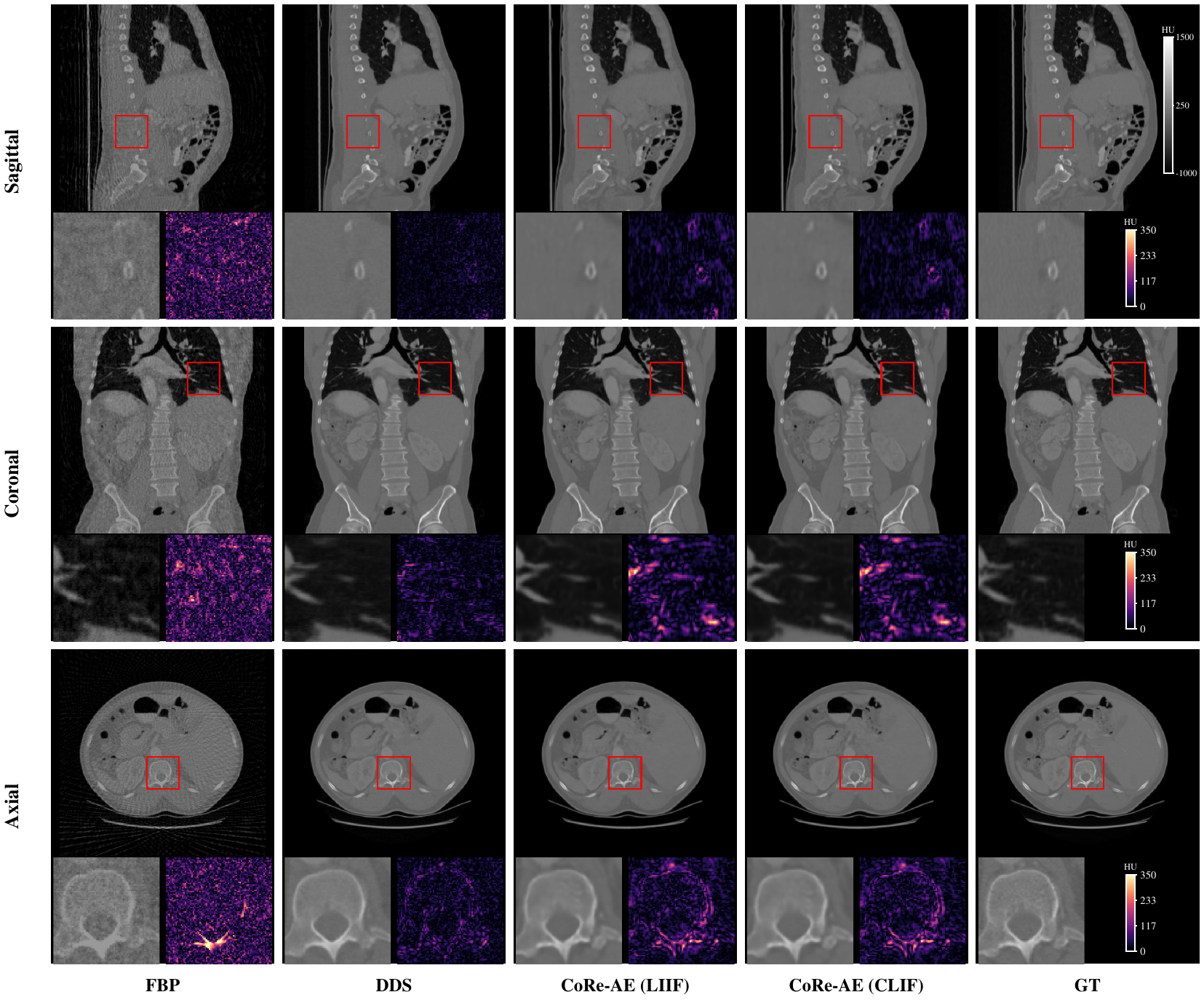}
\caption{
Qualitative comparison of sparse-view \ac{CT} reconstruction from 60 noisy projections. Rows show sagittal, coronal, and axial slices. For each method, the top panel shows the reconstruction and the selected region of interest, and the bottom panels show its magnification and the absolute error map in HU. Columns compare \ac{FBP}, \ac{DDS}, \acs{CoReAE} (\acs{LIIF}), \acs{CoReAE} (\acs{CLIF}), and the ground truth (GT). All images use the same HU window.
}
\label{fig:reconstruction_ct_60views_qualitative}
\end{figure*}

\begin{figure*}[htbp]
\centering
\includegraphics[width=0.85\textwidth]{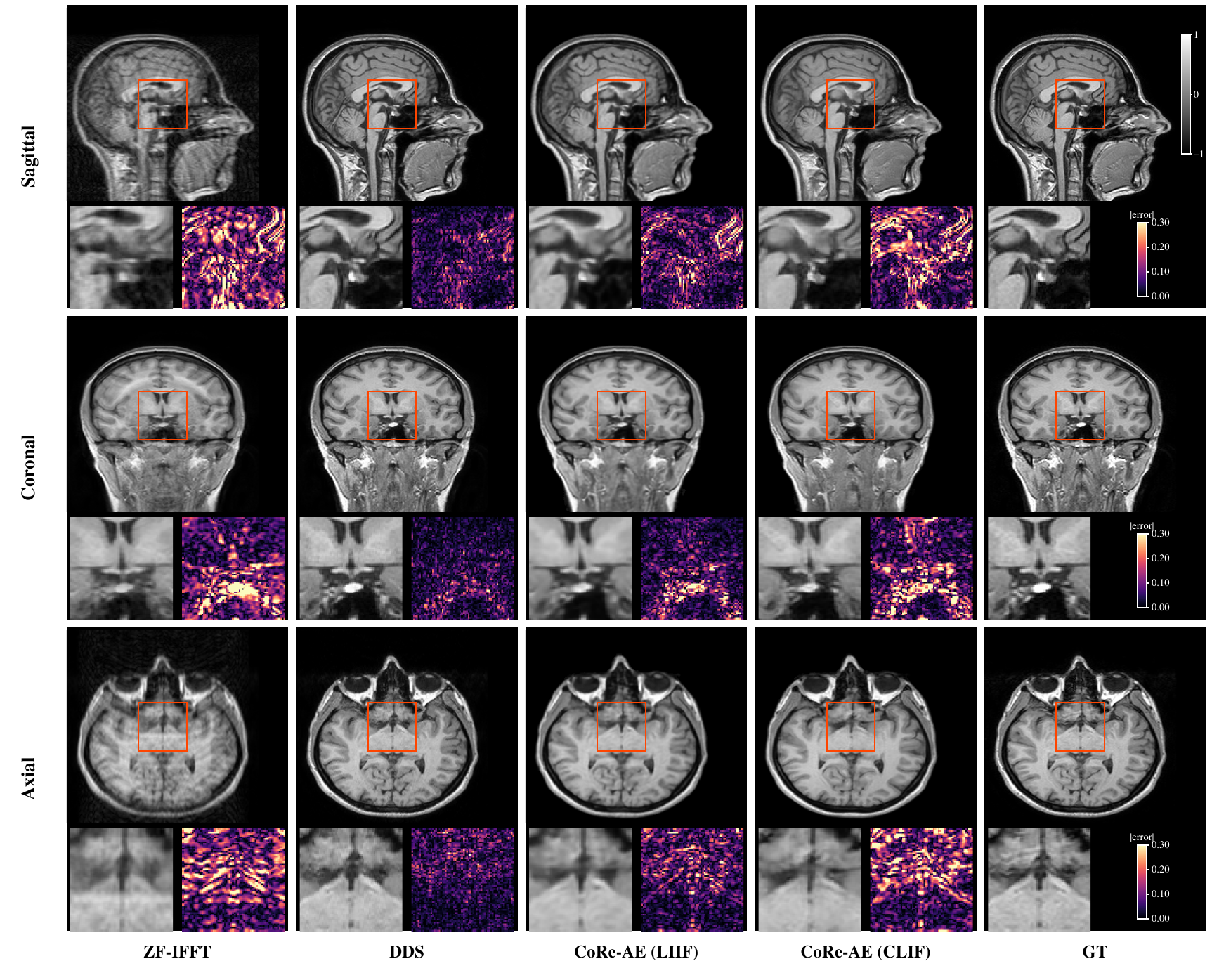}
\caption{
Qualitative comparison of single-coil Cartesian \ac{MRI} reconstruction at $R=4$. Rows show sagittal, coronal, and axial slices. For each method, the top panel shows the reconstruction and the selected region of interest, and the bottom panels show its magnification and the absolute error map. Columns compare \ac{ZF}, \ac{DDS}, \acs{CoReAE} (\acs{LIIF}), \acs{CoReAE} (CLIF), and the ground truth (GT). All images use the same normalized intensity window.
}
\label{fig:reconstruction_mri_x4_qualitative}
\end{figure*}

\begin{figure*}[htbp]
\centering
\includegraphics[width=0.85\textwidth]{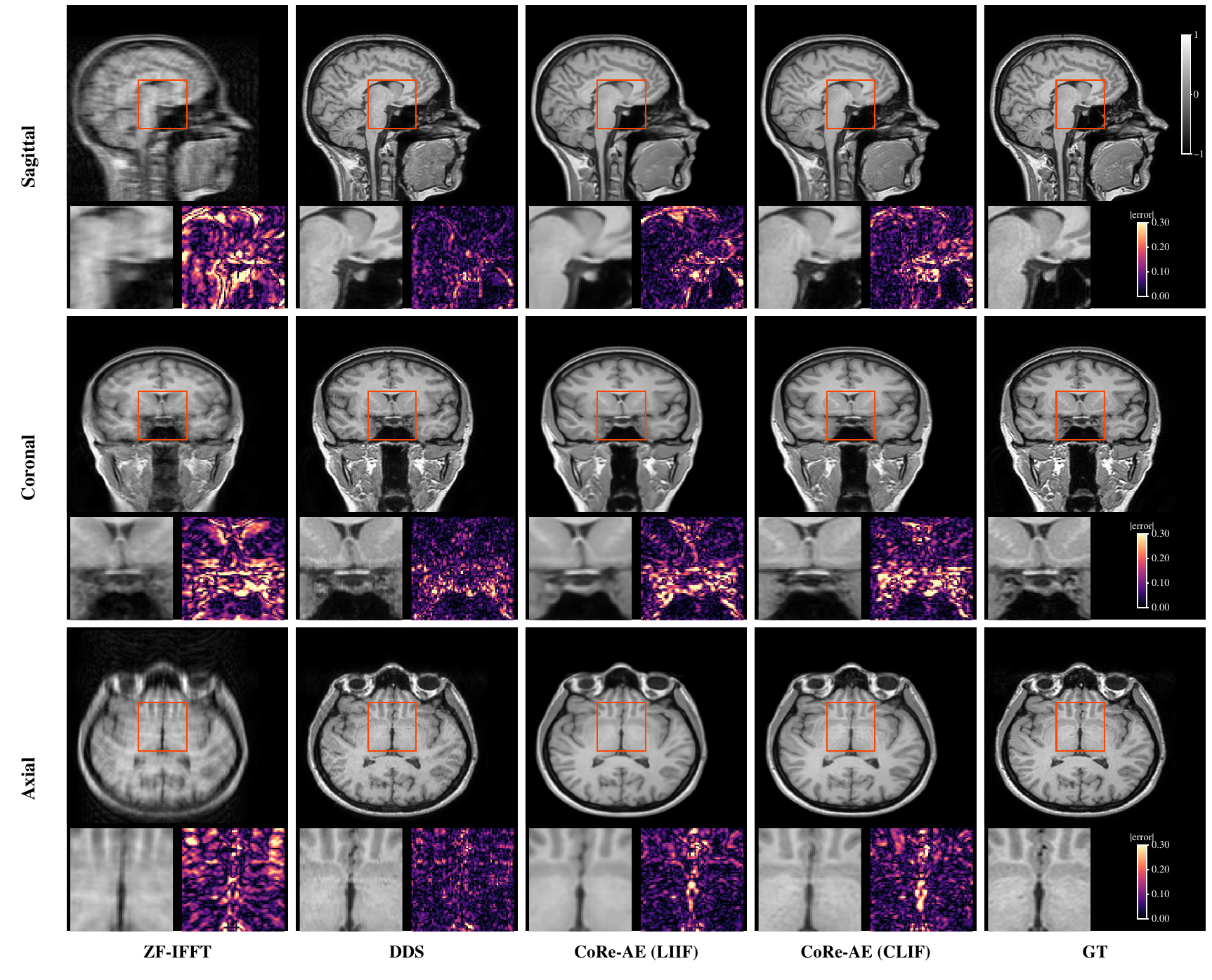}
\caption{
Qualitative comparison of single-coil Cartesian \ac{MRI} reconstruction at $R=8$. Rows show sagittal, coronal, and axial slices. For each method, the top panel shows the reconstruction and the selected region of interest, and the bottom panels show its magnification and the absolute error map. Columns compare \ac{ZF}, \ac{DDS}, \acs{CoReAE} (\acs{LIIF}), \acs{CoReAE} (CLIF), and the ground truth (GT). All images use the same normalized intensity window.
}
\label{fig:reconstruction_mri_x8_qualitative}
\end{figure*}

\begin{table*}[htbp]
	\centering
	\caption{Sparse-view \acs{CT} reconstruction (30 and 60 projections) on 10 held-out volumes.}
	\label{tab:ct-rec-metrics}
	\scriptsize
	\setlength{\tabcolsep}{3pt}
	\renewcommand{\arraystretch}{1.05}
	\resizebox{\linewidth}{!}{%
	\begin{tabular}{c l c c c c c c c c c}
		\toprule
		\textbf{Views} & \textbf{Method}
		& \textbf{\acs{PSNR} $\uparrow$} & \textbf{\acs{SSIM} $\uparrow$} & \textbf{\acs{LPIPS} $\downarrow$} & \textbf{\acs{HFEN} $\downarrow$}
		& \textbf{Grad.\ err.\ $\downarrow$} & \textbf{FRC$_{\mathrm{HF}}$ $\uparrow$} & \textbf{$P_{\mathrm{HF}}$ ($\to1$)}
		& \textbf{$\Delta$Contrast (\%) $\downarrow$} & \textbf{$\Delta$Homog.\ (\%) $\downarrow$} \\
		\midrule
		\multirow{4}{*}{30}
		& \acs{FBP} & 10.08 $\pm$ 0.37 & 0.009 $\pm$ 0.052 & 0.739 $\pm$ 0.077 & 2.642 $\pm$ 1.613 & 3.023 $\pm$ 1.747 & 0.101 $\pm$ 0.044 & 28.01 $\pm$ 3.24 & 548.8 $\pm$ 409.9 & 64.0 $\pm$ 1.4 \\
		& \acs{DDS} & \textbf{41.53 $\pm$ 3.98} & \textbf{0.964 $\pm$ 0.018} & 0.050 $\pm$ 0.039 & \textbf{0.203 $\pm$ 0.055} & \textbf{0.227 $\pm$ 0.061} & 0.964 $\pm$ 0.023 & \textbf{1.02 $\pm$ 0.04} & \textbf{6.2 $\pm$ 6.2} & \textbf{3.4 $\pm$ 1.9} \\
		& \acs{CoReAE} (\acs{LIIF}) & 38.26 $\pm$ 2.52 & 0.962 $\pm$ 0.019 & \textbf{0.043 $\pm$ 0.022} & 0.254 $\pm$ 0.048 & 0.255 $\pm$ 0.046 & \textbf{0.972 $\pm$ 0.014} & 0.90 $\pm$ 0.06 & 10.8 $\pm$ 9.4 & 6.8 $\pm$ 1.9 \\
		& \acs{CoReAE} (\acs{CLIF}) & 38.35 $\pm$ 2.52 & 0.963 $\pm$ 0.019 & \textbf{0.043 $\pm$ 0.022} & 0.251 $\pm$ 0.048 & 0.252 $\pm$ 0.046 & \textbf{0.972 $\pm$ 0.014} & 0.90 $\pm$ 0.06 & 10.3 $\pm$ 9.4 & 6.7 $\pm$ 1.9 \\
		\cmidrule(lr){1-11}
		\multirow{4}{*}{60}
		& \acs{FBP} & 11.71 $\pm$ 0.83 & 0.048 $\pm$ 0.076 & 0.644 $\pm$ 0.081 & 1.952 $\pm$ 1.335 & 2.320 $\pm$ 1.422 & 0.140 $\pm$ 0.059 & 23.34 $\pm$ 3.02 & 364.9 $\pm$ 288.5 & 57.6 $\pm$ 1.6 \\
		& \acs{DDS} & \textbf{43.67 $\pm$ 3.52} & \textbf{0.974 $\pm$ 0.012} & \textbf{0.038 $\pm$ 0.029} & \textbf{0.150 $\pm$ 0.036} & \textbf{0.193 $\pm$ 0.045} & 0.968 $\pm$ 0.017 & \textbf{1.02 $\pm$ 0.03} & \textbf{5.8 $\pm$ 7.4} & \textbf{5.0 $\pm$ 2.9} \\
		& \acs{CoReAE} (\acs{LIIF}) & 38.74 $\pm$ 2.52 & 0.966 $\pm$ 0.017 & 0.041 $\pm$ 0.022 & 0.237 $\pm$ 0.047 & 0.241 $\pm$ 0.046 & 0.973 $\pm$ 0.014 & 0.91 $\pm$ 0.06 & 8.8 $\pm$ 9.4 & 6.5 $\pm$ 2.0 \\
		& \acs{CoReAE} (\acs{CLIF}) & 38.82 $\pm$ 2.52 & 0.966 $\pm$ 0.017 & 0.041 $\pm$ 0.022 & 0.234 $\pm$ 0.048 & 0.239 $\pm$ 0.046 & \textbf{0.974 $\pm$ 0.014} & 0.92 $\pm$ 0.06 & 8.6 $\pm$ 9.4 & 6.4 $\pm$ 2.0 \\
		\bottomrule
	\end{tabular}%
	}
\end{table*}

\begin{table*}[htbp]
	\centering
	\caption{Accelerated \acs{MRI} reconstruction ($\times4$, $\times8$) on 10 held-out IXI volumes ($256^3$).}
	\label{tab:mri-rec-metrics}
	\scriptsize
	\setlength{\tabcolsep}{3pt}
	\renewcommand{\arraystretch}{1.05}
	\resizebox{\linewidth}{!}{%
	\begin{tabular}{c l c c c c c c c c c}
		\toprule
		\textbf{Accel.} & \textbf{Method}
		& \textbf{\acs{PSNR} $\uparrow$} & \textbf{\acs{SSIM} $\uparrow$} & \textbf{\acs{LPIPS} $\downarrow$} & \textbf{\acs{HFEN} $\downarrow$}
		& \textbf{Grad.\ err.\ $\downarrow$} & \textbf{FRC$_{\mathrm{HF}}$ $\uparrow$} & \textbf{$P_{\mathrm{HF}}$ ($\to1$)}
		& \textbf{$\Delta$Contrast (\%) $\downarrow$} & \textbf{$\Delta$Homog.\ (\%) $\downarrow$} \\
		\midrule
		\multirow{4}{*}{$\times4$}
		& \acs{ZF} & 24.06 $\pm$ 3.18 & 0.660 $\pm$ 0.079 & 0.246 $\pm$ 0.086 & 0.618 $\pm$ 0.294 & 0.663 $\pm$ 0.257 & 0.516 $\pm$ 0.161 & 0.45 $\pm$ 0.15 & 63.1 $\pm$ 8.5 & 20.5 $\pm$ 8.9 \\
		& \acs{DDS} & \textbf{36.82 $\pm$ 4.11} & \textbf{0.961 $\pm$ 0.021} & \textbf{0.015 $\pm$ 0.008} & \textbf{0.171 $\pm$ 0.031} & \textbf{0.266 $\pm$ 0.044} & \textbf{0.905 $\pm$ 0.035} & \textbf{0.93 $\pm$ 0.05} & \textbf{6.2 $\pm$ 4.7} & \textbf{3.3 $\pm$ 2.6} \\
		& \acs{CoReAE} (\acs{LIIF}) & 33.49 $\pm$ 3.80 & 0.930 $\pm$ 0.033 & 0.061 $\pm$ 0.026 & 0.204 $\pm$ 0.039 & 0.413 $\pm$ 0.053 & 0.852 $\pm$ 0.053 & 0.88 $\pm$ 0.13 & 20.5 $\pm$ 5.9 & 13.4 $\pm$ 7.4 \\
		& \acs{CoReAE} (\acs{CLIF}) & 33.87 $\pm$ 3.61 & 0.861 $\pm$ 0.064 & 0.046 $\pm$ 0.020 & 0.209 $\pm$ 0.049 & 0.528 $\pm$ 0.067 & 0.848 $\pm$ 0.075 & 0.90 $\pm$ 0.09 & 16.8 $\pm$ 6.7 & 8.3 $\pm$ 6.1 \\
		\cmidrule(lr){1-11}
		\multirow{4}{*}{$\times8$}
		& \acs{ZF} & 22.43 $\pm$ 3.35 & 0.620 $\pm$ 0.098 & 0.289 $\pm$ 0.089 & 0.746 $\pm$ 0.228 & 0.762 $\pm$ 0.193 & 0.438 $\pm$ 0.147 & 0.28 $\pm$ 0.13 & 72.8 $\pm$ 9.0 & 35.7 $\pm$ 13.0 \\
		& \acs{DDS} & \textbf{33.62 $\pm$ 4.17} & \textbf{0.921 $\pm$ 0.046} & \textbf{0.037 $\pm$ 0.021} & \textbf{0.275 $\pm$ 0.047} & \textbf{0.364 $\pm$ 0.056} & \textbf{0.842 $\pm$ 0.056} & \textbf{0.92 $\pm$ 0.08} & \textbf{10.1 $\pm$ 7.0} & \textbf{5.5 $\pm$ 3.5} \\
		& \acs{CoReAE} (\acs{LIIF}) & 31.23 $\pm$ 3.48 & 0.863 $\pm$ 0.062 & 0.079 $\pm$ 0.033 & 0.311 $\pm$ 0.049 & 0.505 $\pm$ 0.061 & 0.808 $\pm$ 0.060 & 0.86 $\pm$ 0.20 & 22.6 $\pm$ 7.1 & 15.6 $\pm$ 9.9 \\
		& \acs{CoReAE} (\acs{CLIF}) & 31.11 $\pm$ 3.62 & 0.850 $\pm$ 0.068 & 0.049 $\pm$ 0.021 & 0.343 $\pm$ 0.053 & 0.557 $\pm$ 0.070 & 0.799 $\pm$ 0.077 & 0.87 $\pm$ 0.09 & 17.9 $\pm$ 7.0 & 9.1 $\pm$ 6.5 \\
		\bottomrule
	\end{tabular}%
	}
\end{table*}

The latent prior reconstructs good-quality volumes from severely undersampled measurements (Tables~\ref{tab:ct-rec-metrics} and~\ref{tab:mri-rec-metrics}). On \ac{CT}, it reaches about 38.3~dB \ac{PSNR} at 30 views and 38.8~dB at 60 views, with both heads. On \ac{MRI}, it reaches 33.5--33.9~dB at $\times4$ and 31.1--31.2~dB at $\times8$, far above the zero-filled reference. \ac{DDS} scores highest in \ac{PSNR} and \ac{SSIM} in every setting, as expected, since it runs a full-resolution diffusion model on the volume itself while our prior works in the compact latent space and reconstructs through the decoder. The \ac{PSNR} gap to \ac{DDS} is 3.2 and 4.9~dB on \ac{CT} (30 and 60 views), and 3.0 and 2.4~dB on \ac{MRI} ($\times4$ and $\times8$).

On \ac{CT}, the two heads give almost the same results. The \ac{SSIM} stays close to \ac{DDS} (0.962--0.966 vs.\ 0.964--0.974), and \ac{LPIPS} is even lower than \ac{DDS} at 30 views (0.043 vs.\ 0.050). The FRC$_{\mathrm{HF}}$ is also as high as for \ac{DDS} (0.972--0.974 vs.\ 0.964--0.968), so the fine details that are recovered are at the right place. However, our reconstructions keep only 90--92\% of the high-frequency energy of the ground truth ($P_{\mathrm{HF}}$), and their texture contrast error is higher (8.6--10.8\% vs.\ about 6\%). The loss is therefore mainly in the amplitude of fine details: the texture is smoother. Our \ac{PSNR} gains only 0.5~dB from 30 to 60 views, against 2.1~dB for \ac{DDS}, which indicates that the latent prior, rather than the number of measurements, limits our result.

On \ac{MRI}, the two heads show the same trade-off as in autoencoding (Section~\ref{sec:ae_results}). The \ac{LIIF} head gives a higher \ac{SSIM} and a lower gradient error. The \ac{CLIF} head gives a lower \ac{LPIPS} (0.046 vs.\ 0.061 at $\times4$) and a better texture, with lower contrast and homogeneity errors (16.8 vs.\ 20.5\% and 8.3 vs.\ 13.4\% at $\times4$).

The qualitative comparisons (Figs.~\ref{fig:reconstruction_ct_30views_qualitative}--\ref{fig:reconstruction_mri_x8_qualitative}) agree with these numbers. Our reconstructions preserve the anatomy, without the streak and aliasing artifacts of the analytic reconstructions, but fine textures are smoother than with \ac{DDS}. A task-based evaluation with pretrained segmentation networks on the same test volumes (Supplementary Section~\ref{sec:s_task}) shows that large organs, bones and brain structures are largely preserved, with \ac{CT} Dice scores similar to \ac{DDS} for several structures. However, our \ac{CT} reconstructions have larger boundary errors (mean HD95 of 41.7 vs.\ 17.8~mm for \ac{LIIF} at 30 views), the pancreas is not recovered at either number of views, and the in-organ \ac{HU} bias and \ac{MAE} are higher than with \ac{DDS} (about 22--24 vs.\ 1--2~\ac{HU}, and 40--42 vs.\ 22--26~\ac{HU}). The distinction from \ac{DDS} is the prior itself: \ac{DDS} guides reconstruction with a \ac{2D} slice-wise \ac{DM}, whereas ours is a native \ac{3D} generative prior applied through the same decoder used for generation.

In terms of cost, one denoising step of our method is about $75\times$ cheaper than one step of \ac{DDS}, because it calls a single \ac{3D} latent U-Net instead of a \ac{2D} U-Net on every slice. With image-space correction only, a \ac{CT} 512\textsuperscript{3} reconstruction takes 194--197~s, against 1155~s for \ac{DDS}. The latent-space stage, however, needs one decoder pass with gradient per iteration, and with the full schedule it makes the reconstruction slower than \ac{DDS} (about 1500~s). The full cost analysis is given in Supplementary Section~\ref{sec:s_recon}.

\section{Discussion}\label{sec:discussion}

The results place the main contribution of this work in the \ac{AE}, rather than in the diffusion process. For matched latent resolution, denoiser architecture, precision, and sampling schedule, the diffusion-stage cost is comparable to that of a conventional \ac{3D} \ac{LDM}. \ac{CoReAE} instead changes the cost of representing and decoding high-resolution volumes, while keeping the highest voxel-wise fidelity among the evaluated \acp{AE}. In our single-\ac{GPU} setting, it provides a common volumetric representation that can be used for both generation and reconstruction, without introducing a cheaper diffusion sampler.

The autoencoding experiments show that the decoding advantage comes from where computation is repeated. Patch-based approaches run the complete decoder on every sub-volume. They recompute the same intermediate activations and, when patches overlap, blend redundant predictions. At 512\textsuperscript{3}, they process 3.4 to 5.4 times the volume (Table~\ref{tab:cost512}). Our decoder runs its convolutional stage only once, on the latent grid, and reuses these features for all output voxels. Only the feature lookup and the small rendering head are repeated, chunk by chunk or tile by tile. The gain therefore comes from avoiding repeated decoder passes, not from a smaller model (Supplementary Fig.~\ref{fig:s_model_cost}).

This factorization also establishes the main limitation of the method. The decoder preserves large-scale anatomy but smooths some fine structures and high-frequency texture. With the \ac{LIIF} head, each voxel is predicted independently, and the training objective provides no explicit supervision over coherent decoded neighborhoods. This head gives the best \ac{PSNR}, but not the best perceptual scores (Table~\ref{tab:ae_results}). The \ac{CLIF} head renders local tiles and therefore uses the context of neighboring voxels. It improves \ac{LPIPS} and \ac{FID}, at the cost of a lower \ac{PSNR}, but MAISI still keeps the best \ac{LPIPS}. The two heads thus show a trade-off between voxel-wise fidelity and perceptual quality. Because the \ac{AE} is frozen before diffusion training, its representational range sets a fidelity ceiling.
Generation and reconstruction therefore inherit the same limitation.

For generation, the method produces qualitatively coherent full volumes without visible patch seams, and its \ac{CT} \ac{FID} is the lowest of the compared methods (Table~\ref{tab:generation_fid}). Its scores are also similar across the three planes, while 3D MedDiffusion degrades on the coronal and sagittal planes. These results should still be read with care: the 2.5-D \ac{FID} relies on a \ac{2D} feature extractor, uses 40 volumes per dataset, and does not directly measure fine texture, which remains smoothed by the frozen \ac{AE}. The results therefore demonstrate efficient full-volume synthesis of good quality rather than a complete assessment of distribution matching. The \ac{MRI} scores should be interpreted cautiously because the face-masked IXI references and unmasked MAISI samples do not share the same spatial support, while 3D MedDiffusion is excluded because it generates only brain parenchyma. Since voxel- or wavelet-domain priors may remain tractable at 256\textsuperscript{3} resolution~\citep{friedrich2024wdm, de2025adaptive}, the decoder-level benefit is clearer in the 512\textsuperscript{3} \ac{CT} setting.

The reconstruction results expose the \ac{AE} limitation more strongly. \ac{DDS}~\citep{chung2024decomposed} achieves higher \ac{PSNR} and \ac{SSIM} in every evaluated setting, by 3.2--4.9~dB on \ac{CT} and 2.4--3.0~dB on \ac{MRI}. Its data-consistency updates act directly on image pixels, without passing through an \ac{AE}, and its \ac{2D} denoiser benefits from a large number of slice-level training instances. Combined with axial \ac{TV} regularization, this yields a strong volumetric reconstruction baseline~\citep{du2026improving}. Our method, in contrast, optimizes in the latent space, which has several limitations. First, it works through the composite mapping $\boldcalA\circ\boldcalD_{\boldpsi}$, which is nonlinear and generally nonconvex even when $\boldcalA$ is linear~\citep{rout2023solving,song2024solving}. Second, the reconstruction cannot be better than what the frozen \ac{AE} can represent. Image-space corrections must be re-encoded through a lossy encoder, which can attenuate recovered details~\citep{chung2023prompt,song2024solving}, and latent-space corrections remain restricted to the signals that the decoder can produce. This ceiling is visible in our results: our \ac{PSNR} gains only 0.5~dB from 30 to 60 \ac{CT} views, against 2.1~dB for \ac{DDS}. It is also visible in the task-based evaluation: large structures are segmented reliably, but boundary errors are larger than with \ac{DDS}, small soft-tissue organs such as the pancreas are not recovered, and the \ac{HU} errors inside organs are higher (Supplementary Section~\ref{sec:s_task}). Third, the guidance depends on several parameters that need careful tuning. Finally, the latent-space stage is costly, as each iteration requires a decoder pass with gradient. These difficulties are not specific to our method: a recent benchmark shows that \ac{CT} reconstruction remains challenging for diffusion priors in general~\citep{shi2026dm4ct}. The reconstruction experiments should therefore be read as a demonstration that a frozen \ac{3D} latent prior trained for generation can also serve for reconstruction, not as evidence of superiority over pixel-domain methods.

Taken together, the results show that \ac{CoReAE} reduces the cost of decoding without losing voxel-wise fidelity, and provides a good generative prior that can also be used for inverse problems. This comes at the price of smoother fine textures, and of a lower accuracy than pixel-domain methods on sparse-view \ac{CT} and accelerated \ac{MRI}. The proposed factorization is most useful when repeated dense volumetric decoding is the dominant computational constraint, but its benefit diminishes when image-space processing remains tractable or fine-detail preservation is the primary objective. The conclusions are also limited by the use of two public datasets, simulated sparse-view \ac{CT}, and simulated single-coil \ac{MRI}. The study does not include external clinical validation, multi-coil acquisitions, realistic acquisition imperfections, reader studies, or explicit volumetric-consistency metrics. Finally, we did not evaluate memorization. Diffusion models can reproduce their training images~\citep{carlini2023extracting,somepalli2023diffusion}, and this has also been shown for \ac{3D} latent diffusion models trained on medical data, in particular when the training set is small~\citep{dar2023investigating,akbar2025beware,dar2025unconditional} and \ac{FID} does not measure this risk.

Future work should address decoder fidelity, reconstruction, diffusion cost, and memorization. Bit-plane decomposition offers a route toward near-lossless implicit representations by predicting binary bit-planes instead of a single continuous intensity~\citep{han2025towards}. Adapting this idea to a latent-conditioned \ac{3D} medical decoder could raise the autoencoding fidelity ceiling without returning to repeated dense decoding. For reconstruction, the image-space correction relies on re-encoding, while the \ac{AE} is not exactly invertible.
Guidance algorithms that account for this mismatch, together with a cheaper latent-space stage and a systematic tuning of the guidance in \ac{3D}, should reduce the gap to pixel-domain methods. In addition, the coordinate-based decoder can render a volume at any resolution, but this property was not exploited here, as we trained and evaluated each model at a single fixed resolution to maximize accuracy. Following MAISI-v2~\citep{zhao2025maisi}, the model could be trained on volumes of different sizes and voxel spacings to generate volumes at multiple resolutions. At the same time, more compact latent representations should be explored to reduce the cost of diffusion itself. \Acp{AE} with spatial downsampling factors of up to \texttimes{}32 substantially reduce the number of latent tokens processed during denoising~\citep{xie2024sana,chen2025sana}, suggesting that compact or anisotropic \ac{3D} latent grids could lower both training and sampling costs for high-resolution medical volumes. The central challenge is to obtain this additional compression without further degrading fine structures or widening the reconstruction gap. Finally, a memorization analysis of the trained models should be carried out~\citep{dar2025unconditional,akbar2025beware}, while larger and more heterogeneous training cohorts remain a complementary direction.

\section{Conclusion}\label{sec:conclusion}

We introduced a continuous \ac{3D} \ac{LDM} for high-resolution medical volume generation and reconstruction. Its central component, \ac{CoReAE}, combines a compact volumetric latent representation with a coordinate-conditioned decoder, allowing full \ac{CT} and \ac{MRI} volumes to be rendered without patch-wise decoding or fixed output grids. With either the \ac{LIIF} or the \ac{CLIF} head, \ac{CoReAE} achieved the highest \ac{PSNR} and \ac{SSIM} among the evaluated \acp{AE}, while decoding a 512\textsuperscript{3} \ac{CT} volume 4 to 18 times faster and with the lowest \ac{GPU} memory. This design enables 512\textsuperscript{3} \ac{CT} generation on a single \ac{GPU} and produces coherent volumes without visible patch seams, with the lowest \ac{CT} \ac{FID} among the compared methods.

The same frozen \ac{3D} diffusion prior was also applied to sparse-view \ac{CT} and accelerated \ac{MRI} through hard data consistency, without task-specific retraining. The lossy \ac{AE} limits fine-detail recovery, and the reconstructions remain 2.4 to 4.9~dB below the pixel-domain \ac{DDS} solver. Downstream segmentation preserves large anatomical structures, but shows larger boundary errors, failures on small organs, and a lower intensity fidelity than with \ac{DDS}. The results nevertheless show that one volumetric latent prior can support both unconditional generation and inverse-problem reconstruction. Overall, the framework offers a practical route to continuous high-resolution medical volume modeling, with fine-texture preservation and reconstruction accuracy remaining the main targets for further improvement.

\section*{Acknowledgment}

All authors declare that they have no known conflicts of interest in terms of competing financial interests or personal relationships that could have an influence or are relevant to the work reported in this paper.

\AtNextBibliography{\footnotesize} 
\printbibliography

@article{sun2022hierarchical,
  title={Hierarchical amortized {GAN} for {3D} high resolution medical image synthesis},
  author={Sun, Li and Chen, Junxiang and Xu, Yanwu and Gong, Mingming and Yu, Ke and Batmanghelich, Kayhan},
  journal={IEEE Journal of Biomedical and Health Informatics},
  volume={26},
  number={8},
  pages={3966--3975},
  year={2022},
  publisher={IEEE},
  doi={10.1109/JBHI.2022.3172976}
}

@InProceedings{Guo_2025_WACV,
  author = {Guo, Pengfei and Zhao, Can and Yang, Dong and Xu, Ziyue and Nath, Vishwesh and Tang, Yucheng and Simon, Benjamin and Belue, Mason and Harmon, Stephanie and Turkbey, Baris and Xu, Daguang},
  title = {MAISI: Medical {AI} for Synthetic Imaging},
  booktitle = {Proceedings of the Winter Conference on Applications of Computer Vision (WACV)},
  month = {2},
  year = {2025},
  pages = {4430-4441}
}

@article{wang2025meddiffusion,
  author = {Wang, Haoshen and Liu, Zhentao and Sun, Kaicong and Wang, Xiaodong and Shen, Dinggang and Cui, Zhiming},
  title = {{3D} MedDiffusion: A {3D} Medical Latent Diffusion Model for Controllable and High-quality Medical Image Generation},
  journal = {IEEE Transactions on Medical Imaging},
  year = {2025},
  volume = {44},
  number = {12},
  pages = {4960--4972},
  doi = {10.1109/TMI.2025.3585372}
}

@inproceedings{hu2024learning,
  title = {Learning Image Priors Through Patch-Based Diffusion Models for Solving Inverse Problems},
  author = {Hu, Jason and Song, Bowen and Xu, Xiaojian and Shen, Liyue and Fessler, Jeffrey A.},
  booktitle = {Advances in Neural Information Processing Systems},
  volume = {37},
  year = {2024}
}

@inproceedings{song2024diffusionblend,
  title = {DiffusionBlend: Learning {3D} Image Prior through Position-aware Diffusion Score Blending for {3D} Computed Tomography Reconstruction},
  author = {Song, Bowen and Hu, Jason and Luo, Zhaoxu and Fessler, Jeffrey A. and Shen, Liyue},
  booktitle = {Advances in Neural Information Processing Systems},
  volume = {37},
  year = {2024}
}

@InProceedings{Chen_2021_CVPR,
  author = {Chen, Yinbo and Liu, Sifei and Wang, Xiaolong},
  title = {Learning Continuous Image Representation With Local Implicit Image Function},
  booktitle = {Proceedings of the IEEE/CVF Conference on Computer Vision and Pattern Recognition (CVPR)},
  month = {6},
  year = {2021},
  pages = {8628-8638}
}

@article{khorashadizadeh2024lofi,
  title={{LoFi}: Neural Local Fields for Scalable Image Reconstruction},
  author={Khorashadizadeh, AmirEhsan and Liaudat, Tob{\'i}as I. and Liu, Tianlin and McEwen, Jason D. and Dokmani{\'c}, Ivan},
  journal={arXiv preprint arXiv:2411.04995},
  year={2024},
  doi={10.48550/arXiv.2411.04995}
}

@article{tong2024diffinr,
  title={{Diff}-{INR}: Generative Regularization for Electrical Impedance Tomography},
  author={Tong, Bowen and Wang, Junwu and Liu, Dong},
  journal={arXiv preprint arXiv:2409.04494},
  year={2024},
  doi={10.48550/arXiv.2409.04494}
}

@article{chu2025highly,
  title={Highly Accelerated {MRI} via Implicit Neural Representation Guided Posterior Sampling of Diffusion Models},
  author={Chu, Jiayue and Du, Chenhe and Lin, Xiyue and Zhang, Xiaoqun and Wang, Lihui and Zhang, Yuyao and Wei, Hongjiang},
  journal={Medical Image Analysis},
  volume={100},
  pages={103398},
  year={2025},
  doi={10.1016/j.media.2024.103398}
}

@InProceedings{Rombach_2022_CVPR,
  author    = {Rombach, Robin and Blattmann, Andreas and Lorenz, Dominik and Esser, Patrick and Ommer, Bj\"orn},
  title     = {High-Resolution Image Synthesis With Latent Diffusion Models},
  booktitle = {Proceedings of the IEEE/CVF Conference on Computer Vision and Pattern Recognition (CVPR)},
  month     = {6},
  year      = {2022},
  pages     = {10684-10695}
}

@article{khader2023denoising,
  title   = {Denoising diffusion probabilistic models for {3D} medical image generation},
  author  = {Khader, Firas and Mueller-Franzes, Gustav and Tayebi Arasteh, Soroosh and Han, Tianyu and Haarburger, Christoph and Schulze-Hagen, Maximilian and Schad, Philipp and Engelhardt, Sandy and Baessler, Bettina and Foersch, Sebastian and Stegmaier, Johannes and Kuhl, Christiane and Nebelung, Sven and Kather, Jakob Nikolas and Truhn, Daniel},
  journal = {Scientific Reports},
  volume  = {13},
  number  = {1},
  pages   = {7303},
  year    = {2023},
  doi     = {10.1038/s41598-023-34341-2}
}

@InProceedings{pinaya2022brain,
  title     = {Brain Imaging Generation with Latent Diffusion Models},
  author    = {Pinaya, Walter H. L. and Tudosiu, Petru-Daniel and Dafflon, Jessica and da Costa, Pedro F. and Fernandez, Virginia and Nachev, Parashkev and Ourselin, Sebastien and Cardoso, M. Jorge},
  booktitle = {Deep Generative Models},
  pages     = {117--126},
  year      = {2022},
  publisher = {Springer Nature Switzerland},
  doi       = {10.1007/978-3-031-18576-2_12}
}

@InProceedings{pmlr-v227-bieder24a,
  title     = {Memory-Efficient {3D} Denoising Diffusion Models for Medical Image Processing},
  author    = {Bieder, Florentin and Wolleb, Julia and Durrer, Alicia and Sandkuehler, Robin and Cattin, Philippe C.},
  booktitle = {Medical Imaging with Deep Learning},
  pages     = {552--567},
  year      = {2024},
  editor    = {Oguz, Ipek and Noble, Jack and Li, Xiaoxiao and Styner, Martin and Baumgartner, Christian and Rusu, Mirabela and Heinmann, Tobias and Kontos, Despina and Landman, Bennett and Dawant, Benoit},
  volume    = {227},
  series    = {Proceedings of Machine Learning Research},
  month     = {7},
  publisher = {PMLR},
  url       = {https://proceedings.mlr.press/v227/bieder24a.html}
}

@ARTICLE{medsyn2024,
  author  = {Xu, Yanwu and Sun, Li and Peng, Wei and Jia, Shuyue and Morrison, Katelyn and Perer, Adam and Zandifar, Afrooz and Visweswaran, Shyam and Eslami, Motahhare and Batmanghelich, Kayhan},
  journal = {IEEE Transactions on Medical Imaging},
  title   = {MedSyn: Text-guided Anatomy-aware Synthesis of High-Fidelity 3D CT Images},
  year    = {2024},
  doi     = {10.1109/TMI.2024.3415032}
}

@article{oliveras2026anatomically,
  title   = {Anatomically guided latent diffusion for high-resolution {3D} chest {CT} synthesis},
  author  = {Oliveras, Anna and Mar{\'i}, Roger and Redondo, Rafael and Guardi{\`a}, Oriol and Ugwu, Cynthia Ifeyinwa and Tost, Ana and Nagarajan, Bhalaji and Migliorelli, Carolina and Ribas, Vicent and Radeva, Petia},
  journal = {Scientific Reports},
  year    = {2026},
  doi     = {10.1038/s41598-026-51634-4}
}

@misc{zhao2025maisi,
  title         = {{MAISI}-v2: Accelerated {3D} High-Resolution Medical Image Synthesis with Rectified Flow and Region-specific Contrastive Loss},
  author        = {Zhao, Can and Guo, Pengfei and Yang, Dong and Tang, Yucheng and He, Yufan and Simon, Benjamin and Belue, Mason and Harmon, Stephanie and Turkbey, Baris and Xu, Daguang},
  year          = {2025},
  eprint        = {2508.05772},
  archivePrefix = {arXiv},
  primaryClass  = {eess.IV}
}

@inproceedings{song2024solving,
  title={Solving inverse problems with latent diffusion models via hard data consistency},
  author={Song, Bowen and Kwon, Soo Min and Zhang, Zecheng and Hu, Xinyu and Qu, Qing and Shen, Liyue},
  booktitle={International Conference on Learning Representations},
  volume={2024},
  pages={7624--7654},
  year={2024}
}

@inproceedings{zhu2023denoising,
  title={Denoising diffusion models for plug-and-play image restoration},
  author={Zhu, Yuanzhi and Zhang, Kai and Liang, Jingyun and Cao, Jiezhang and Wen, Bihan and Timofte, Radu and Van Gool, Luc},
  booktitle={Proceedings of the IEEE/CVF conference on computer vision and pattern recognition},
  pages={1219--1229},
  year={2023}
}

@article{armato2011lidc,
  title   = {The Lung Image Database Consortium (LIDC) and Image Database Resource Initiative ({IDRI}): A completed reference database of lung nodules on {CT} scans},
  author  = {Armato, Samuel G. and McLennan, Geoffrey and Bidaut, Luc and McNitt-Gray, Michael F. and Meyer, Charles R. and Reeves, Anthony P. and Zhao, Binsheng and Aberle, Denise R. and Henschke, Claudia I. and Hoffman, Eric A. and Kazerooni, Ella A. and MacMahon, Heber and Van Beek, Edwin J. R. and Yankelevitz, David and Biancardi, Alberto M. and Bland, Peter H. and Brown, Matthew S. and Engelmann, Roger M. and Laderach, Geoffrey E. and Max, Denise and Pais, Richard C. and Qing, David P. Y. and Roberts, Rachael Y. and Smith, Anthony R. and Starkey, Adam and Batrah, Poonam and Caligiuri, Philip and Farooqi, Amina and Gladish, Gregory W. and Jude, Catherine M. and Munden, Reginald F. and Petkovska, Iva and Quint, Leslie E. and Schwartz, Lawrence H. and Sundaram, Baskaran and Dodd, Lori E. and Fenimore, Charles and Gur, David and Petrick, Nicholas and Freymann, John and Kirby, Justin and Hughes, Brandon and Casteele, Andre Vande and Gupte, Sachin and Sallam, M. and Heath, Michael D. and Kuhn, Michael H. and Dharaiya, Ekta and Burns, Richard and Fryd, Daniel S. and Salganicoff, Marcos and Anand, Vinay and Shreter, Uri and Vastagh, Stephen and Croft, Barbara Y.},
  journal = {Medical Physics},
  volume  = {38},
  number  = {2},
  pages   = {915--931},
  year    = {2011},
  doi     = {10.1118/1.3528204}
}

@misc{roth2015ctlymphnodes,
  title        = {A new {2.5D} representation for lymph node detection in {CT}},
  author       = {Roth, Holger and Lu, Le and Seff, Ari and Cherry, Kevin M. and Hoffman, Joanne and Wang, Shijun and Liu, Jiamin and Turkbey, Evrim B. and Summers, Ronald M.},
  year         = {2015},
  howpublished = {The Cancer Imaging Archive},
  note         = {Data set},
  doi          = {10.7937/K9/TCIA.2015.AQIIDCNM}
}

@misc{ixi,
  title        = {IXI Dataset: Information eXtraction from Images},
  author       = {{IXI}},
  year         = {2006},
  howpublished = {EPSRC GR/S21533/02},
  note         = {Accessed: 2026-06-21}
}

@article{mei2022radimagenet,
  title={RadImageNet: an open radiologic deep learning research dataset for effective transfer learning},
  author={Mei, Xueyan and Liu, Zelong and Robson, Philip M and Marinelli, Brett and Huang, Mingqian and Doshi, Amish and Jacobi, Adam and Cao, Chendi and Link, Katherine E and Yang, Thomas and others},
  journal={Radiology: Artificial Intelligence},
  volume={4},
  number={5},
  pages={e210315},
  year={2022},
  publisher={Radiological Society of North America}
}

@article{chung2022diffusion,
  title={Diffusion posterior sampling for general noisy inverse problems},
  author={Chung, Hyungjin and Kim, Jeongsol and Mccann, Michael T and Klasky, Marc L and Ye, Jong Chul},
  journal={arXiv preprint arXiv:2209.14687},
  year={2022}
}

@article{ho2020denoising,
  title={Denoising diffusion probabilistic models},
  author={Ho, Jonathan and Jain, Ajay and Abbeel, Pieter},
  journal={Advances in neural information processing systems},
  volume={33},
  pages={6840--6851},
  year={2020}
}

@inproceedings{chen2024image,
  title={Image neural field diffusion models},
  author={Chen, Yinbo and Wang, Oliver and Zhang, Richard and Shechtman, Eli and Wang, Xiaolong and Gharbi, Michael},
  booktitle={Proceedings of the IEEE/CVF Conference on Computer Vision and Pattern Recognition},
  pages={8007--8017},
  year={2024}
}

@article{de2025adaptive,
  title={Adaptive Diffusion Models for Sparse-View Motion-Corrected Head Cone-Beam {CT}},
  author={De Paepe, Antoine and Bousse, Alexandre and Phung-Ngoc, Cl{\'e}mentine and Mellak, Youness and Visvikis, Dimitris},
  journal={IEEE Transactions on Radiation and Plasma Medical Sciences},
  year={2025},
  publisher={IEEE}
}

@article{du2026improving,
  title={Improving {2D} Diffusion Models for {3D} Medical Imaging with Inter-Slice Consistent Stochasticity},
  author={Du, Chenhe and Wu, Qing and Tian, Xuanyu and Yu, Jingyi and Wei, Hongjiang and Zhang, Yuyao},
  journal={arXiv preprint arXiv:2602.04162},
  year={2026}
}

@inproceedings{han2025towards,
  title={Towards Lossless Implicit Neural Representation via Bit Plane Decomposition},
  author={Han, Woo Kyoung and Lee, Byeonghun and Cho, Hyunmin and Im, Sunghoon and Jin, Kyong Hwan},
  booktitle={Proceedings of the Computer Vision and Pattern Recognition Conference},
  pages={2269--2278},
  year={2025}
}

@inproceedings{chung2024decomposed,
  title={Decomposed diffusion sampler for accelerating large-scale inverse problems},
  author={Chung, Hyungjin and Lee, Suhyeon and Ye, Jong Chul},
  booktitle={International conference on learning representations},
  volume={2024},
  pages={38922--38949},
  year={2024}
}

@article{spears2025medil,

  title={MedIL: implicit latent spaces for generating heterogeneous medical images at arbitrary resolutions},

  author={Spears, Tyler and Zhu, Shen and Jin, Yinzhu and Shrivastava, Aman and Fletcher, P Thomas},

  journal={arXiv preprint arXiv:2504.09322},

  year={2025}

}

@article{chung2023prompt,
  title={Prompt-tuning latent diffusion models for inverse problems},
  author={Chung, Hyungjin and Ye, Jong Chul and Milanfar, Peyman and Delbracio, Mauricio},
  journal={arXiv preprint arXiv:2310.01110},
  year={2023}
}

@article{torch_radon,
Author = {Matteo Ronchetti},
Title = {TorchRadon: Fast Differentiable Routines for Computed Tomography},
Year = {2020},
Eprint = {arXiv:2009.14788},
journal={arXiv preprint arXiv:2009.14788},
}

@inproceedings{chung2023solving,
  title={Solving 3d inverse problems using pre-trained 2d diffusion models},
  author={Chung, Hyungjin and Ryu, Dohoon and McCann, Michael T and Klasky, Marc L and Ye, Jong Chul},
  booktitle={Proceedings of the IEEE/CVF conference on computer vision and pattern recognition},
  pages={22542--22551},
  year={2023}
}

@inproceedings{lee2023improving,
  title={Improving {3D} imaging with pre-trained perpendicular {2D} diffusion models},
  author={Lee, Suhyeon and Chung, Hyungjin and Park, Minyoung and Park, Jonghyuk and Ryu, Wi-Sun and Ye, Jong Chul},
  booktitle={Proceedings of the IEEE/CVF international conference on computer vision},
  pages={10710--10720},
  year={2023}
}

@article{shen2022nerp,
  title={{NeRP}: implicit neural representation learning with prior embedding for sparsely sampled image reconstruction},
  author={Shen, Liyue and Pauly, John and Xing, Lei},
  journal={IEEE transactions on neural networks and learning systems},
  volume={35},
  number={1},
  pages={770--782},
  year={2022},
  publisher={IEEE}
}

@article{feng2023imjense,
  title={{IMJENSE}: scan-specific implicit representation for joint coil sensitivity and image estimation in parallel {MRI}},
  author={Feng, Ruimin and Wu, Qing and Feng, Jie and She, Huajun and Liu, Chunlei and Zhang, Yuyao and Wei, Hongjiang},
  journal={IEEE Transactions on Medical Imaging},
  volume={43},
  number={4},
  pages={1539--1553},
  year={2023},
  publisher={IEEE}
}

@article{liu2024scan,
  title={Scan-specific unsupervised highly accelerated non-cartesian {CEST} imaging using implicit neural representation and explicit sparse prior},
  author={Liu, Bei and She, Huajun and Du, Yiping P},
  journal={IEEE Transactions on Biomedical Engineering},
  volume={71},
  number={10},
  pages={3032--3045},
  year={2024},
  publisher={IEEE}
}

@article{feng2025spatiotemporal,
  title={Spatiotemporal implicit neural representation for unsupervised dynamic {MRI} reconstruction},
  author={Feng, Jie and Feng, Ruimin and Wu, Qing and Shen, Xin and Chen, Lixuan and Li, Xin and Feng, Li and Chen, Jingjia and Zhang, Zhiyong and Liu, Chunlei and others},
  journal={IEEE Transactions on Medical Imaging},
  volume={44},
  number={5},
  pages={2143--2156},
  year={2025},
  publisher={IEEE}
}

@article{mildenhall2021nerf,
  title={{Nerf}: Representing scenes as neural radiance fields for view synthesis},
  author={Mildenhall, Ben and Srinivasan, Pratul P and Tancik, Matthew and Barron, Jonathan T and Ramamoorthi, Ravi and Ng, Ren},
  journal={Communications of the ACM},
  volume={65},
  number={1},
  pages={99--106},
  year={2021},
  publisher={ACM New York, NY, USA}
}

@article{sitzmann2020implicit,
  title={Implicit neural representations with periodic activation functions},
  author={Sitzmann, Vincent and Martel, Julien and Bergman, Alexander and Lindell, David and Wetzstein, Gordon},
  journal={Advances in neural information processing systems},
  volume={33},
  pages={7462--7473},
  year={2020}
}

@article{muller2022instant,
  title={Instant neural graphics primitives with a multiresolution hash encoding},
  author={M{\"u}ller, Thomas and Evans, Alex and Schied, Christoph and Keller, Alexander},
  journal={ACM transactions on graphics (TOG)},
  volume={41},
  number={4},
  pages={1--15},
  year={2022},
  publisher={ACM New York, NY, USA}
}

@article{yu2025bilevel,
  title={Bilevel optimized implicit neural representation for scan-specific accelerated {MRI} reconstruction},
  author={Yu, Hongze and Fessler, Jeffrey A and Jiang, Yun},
  journal={arXiv preprint arXiv:2502.21292},
  year={2025}
}

@article{chen2025sana,
  title={Sana-video: Efficient video generation with block linear diffusion transformer},
  author={Chen, Junsong and Zhao, Yuyang and Yu, Jincheng and Chu, Ruihang and Chen, Junyu and Yang, Shuai and Wang, Xianbang and Pan, Yicheng and Zhou, Daquan and Ling, Huan and others},
  journal={arXiv preprint arXiv:2509.24695},
  year={2025}
}

@article{xie2024sana,
  title={Sana: Efficient high-resolution image synthesis with linear diffusion transformers},
  author={Xie, Enze and Chen, Junsong and Chen, Junyu and Cai, Han and Tang, Haotian and Lin, Yujun and Zhang, Zhekai and Li, Muyang and Zhu, Ligeng and Lu, Yao and others},
  journal={arXiv preprint arXiv:2410.10629},
  year={2024}
}

@article{rout2023solving,
  title={Solving linear inverse problems provably via posterior sampling with latent diffusion models},
  author={Rout, Litu and Raoof, Negin and Daras, Giannis and Caramanis, Constantine and Dimakis, Alex and Shakkottai, Sanjay},
  journal={Advances in Neural Information Processing Systems},
  volume={36},
  pages={49960--49990},
  year={2023}
}

@InProceedings{Kim_2024_CVPR,
  author = {Kim, Jinseok and Kim, Tae-Kyun},
  title = {Arbitrary-Scale Image Generation and Upsampling using Latent Diffusion Model and Implicit Neural Decoder},
  booktitle = {Proceedings of the IEEE/CVF Conference on Computer Vision and Pattern Recognition (CVPR)},
  month = {6},
  year = {2024},
  pages = {9202-9211}
}

@inproceedings{friedrich2024wdm,
  title={Wdm: 3d wavelet diffusion models for high-resolution medical image synthesis},
  author={Friedrich, Paul and Wolleb, Julia and Bieder, Florentin and Durrer, Alicia and Cattin, Philippe C},
  booktitle={MICCAI workshop on deep generative models},
  pages={11--21},
  year={2024},
  organization={Springer}
}

@inproceedings{shi2026dm4ct,
  title     = {{DM4CT}: Benchmarking Diffusion Models for Computed Tomography Reconstruction},
  author    = {Shi, Jiayang and Pelt, Daniël M. and Batenburg, K. Joost},
  booktitle = {International Conference on Learning Representations (ICLR)},
  year      = {2026}
}

@inproceedings{carlini2023extracting,
  title     = {Extracting Training Data from Diffusion Models},
  author    = {Carlini, Nicholas and Hayes, Jamie and Nasr, Milad and Jagielski, Matthew and Sehwag, Vikash and Tram{\`e}r, Florian and Balle, Borja and Ippolito, Daphne and Wallace, Eric},
  booktitle = {32nd USENIX Security Symposium},
  pages     = {5253--5270},
  year      = {2023}
}

@inproceedings{somepalli2023diffusion,
  title     = {Diffusion Art or Digital Forgery? {I}nvestigating Data Replication in Diffusion Models},
  author    = {Somepalli, Gowthami and Singla, Vasu and Goldblum, Micah and Geiping, Jonas and Goldstein, Tom},
  booktitle = {Proceedings of the IEEE/CVF Conference on Computer Vision and Pattern Recognition (CVPR)},
  pages     = {6048--6058},
  year      = {2023}
}

@article{dar2023investigating,
  title   = {Investigating Data Memorization in {3D} Latent Diffusion Models for Medical Image Synthesis},
  author  = {Dar, Salman Ul Hassan and Ghanaat, Arman and Kahmann, Jannik and Ayx, Isabelle and Papavassiliu, Theano and Schoenberg, Stefan O. and Engelhardt, Sandy},
  journal = {arXiv preprint arXiv:2307.01148},
  year    = {2023}
}

@article{akbar2025beware,
  title   = {Beware of diffusion models for synthesizing medical images---a comparison with {GANs} in terms of memorizing brain {MRI} and chest x-ray images},
  author  = {Akbar, Muhammad Usman and Wang, Wuhao and Eklund, Anders},
  journal = {Machine Learning: Science and Technology},
  volume  = {6},
  pages   = {015022},
  year    = {2025},
  doi     = {10.1088/2632-2153/ad9a3a}
}

@article{dar2025unconditional,
  title   = {Unconditional latent diffusion models memorize patient imaging data},
  author  = {Dar, Salman Ul Hassan and Seyfarth, Marvin and Ayx, Isabelle and Papavassiliu, Theano and Schoenberg, Stefan O. and Siepmann, Robert Malte and Laqua, Fabian Christopher and Kahmann, Jannik and Frey, Norbert and Ba{\ss}ler, Bettina and Foersch, Sebastian and Truhn, Daniel and Kather, Jakob Nikolas and Engelhardt, Sandy},
  journal = {Nature Biomedical Engineering},
  volume  = {10},
  number  = {3},
  pages   = {458--472},
  year    = {2025},
  doi     = {10.1038/s41551-025-01468-8}
}

\cleardoublepage
\setcounter{section}{0}\setcounter{table}{0}\setcounter{figure}{0}\setcounter{equation}{0}
\renewcommand{\thesection}{S\arabic{section}}
\renewcommand{\thetable}{S\arabic{table}}
\renewcommand{\thefigure}{S\arabic{figure}}
\renewcommand{\theequation}{S\arabic{equation}}
\twocolumn[%
  \begin{center}
    {\normalfont\fontsize{24}{28}\selectfont
     Supplementary Material\par}
    \vspace{3em}
  \end{center}
]

This document provides supplementary material for the main manuscript: implementation details (Section~\ref{sec:s_impl}), the protocol used to measure runtime and memory (Section~\ref{sec:s_protocol}), the decoder cost as a function of the output size and of the model size (Section~\ref{sec:s_scaling}), the effect of the chunk and tile size (Section~\ref{sec:s_chunk}), the cost of the reconstruction (Section~\ref{sec:s_recon}), and a task-based evaluation of the reconstructions (Section~\ref{sec:s_task}).
\section{Implementation details}\label{sec:s_impl}

Table~\ref{tab:s_impl} lists the architecture and training settings of \ac{CoReAE} and of the \ac{3D} latent diffusion model. The same settings are used for \ac{CT} and \ac{MRI}, except for the volume size.

\begin{table*}[htbp]
    \centering
    \caption{Implementation details of \ac{CoReAE} and of the \ac{3D} \ac{LDM}.}
    \label{tab:s_impl}
    \small
    \renewcommand{\arraystretch}{1.1}
    \begin{tabular}{>{\raggedright\arraybackslash}p{3.6cm}>{\raggedright\arraybackslash}p{12.4cm}}
        \toprule
        \multicolumn{2}{l}{\textbf{\acs{CoReAE}}}\\
        Encoder & \ac{3D} \ac{CNN}, $4\times$ downsampling, 4 latent channels (\ac{CT} $512^3\rightarrow4\times128^3$, \ac{MRI} $256^3\rightarrow4\times64^3$) \\
        Feature decoder & 4 residual blocks, 48 channels, on the latent grid (no upsampling) \\
        \acs{LIIF} head & \ac{MLP} with 3 layers of width 128; local ensemble; relative coordinates and cell size in latent-cell units \\
        \acs{CLIF} head & 2 residual blocks, 64 channels, $3^3$ kernels, halo of 6 voxels, channel-wise normalization only \\
        Parameters & 1.5\,M (\ac{LIIF}), 2.0\,M (\ac{CLIF}) \\
        Rendering & $2^{18}$ queries per chunk (\ac{LIIF}), $96^3$ tiles (\ac{CLIF}); the output does not depend on this choice (Table~\ref{tab:s_invariance}) \\
        Training & AdamW, learning rate $2\times10^{-4}$, batch size 1, $\lambda_{\mathrm{KL}}=10^{-4}$, $N=200{,}000$ coordinate queries per iteration, 300 epochs \\
        \midrule
        \multicolumn{2}{l}{\textbf{\acs{3D} latent diffusion model}}\\
        \acs{3D} U-Net & 64 base channels, multipliers $(1,2,2,4,4,4)$, 2 residual blocks per level, attention at resolutions 16 and 32; 86.5\,M parameters \\
        Diffusion & $\epsilon$-prediction, $T=1000$ training steps, linear schedule $\beta_t\in[10^{-4},0.02]$; 100 \ac{DDIM} sampling steps \\
        Training & AdamW, learning rate $10^{-4}$, batch size 1 with gradient accumulation over 2 steps, gradient clipping at 1.0, EMA 0.9999, 500k steps \\
        Latent normalization & per-channel mean and standard deviation computed on the training latents \\
        \midrule
        \multicolumn{2}{l}{\textbf{Data}}\\
        \acs{CT} & 700 volumes (LIDC-IDRI and CT-LYMPH-NODES), $512^3$, clipped to $[-1000,1500]$\,HU and rescaled to $[-1,1]$ \\
        \acs{MRI} & 540 T1-weighted IXI volumes, $256^3$, 1st--99th percentile normalization, rescaled to $[-1,1]$ \\
        \bottomrule
    \end{tabular}
\end{table*}

\section{Measurement protocol}\label{sec:s_protocol}

All runtime and memory results were obtained on a single NVIDIA RTX~6000 \ac{GPU} (48\,GB), batch size 1, and the same software environment for all methods (PyTorch~2.8, CUDA~12.8). We time elementary units with CUDA events (median of 10 repeats after 3 warm-up runs): one \ac{LIIF} query chunk, one \ac{CLIF} tile, one MAISI window, one 3D MedDiffusion patch, one U-Net call, and one data-consistency iteration. The cost of a full volume or of a full reconstruction is then obtained as the number of units multiplied by the unit cost, plus a fixed cost. Memory is the peak allocated \ac{GPU} memory.

Every formula was validated against real full-volume runs and real reconstruction steps (Fig.~\ref{fig:s_validation}). Prediction errors are below 5\% for \ac{LIIF}, MAISI, 3D MedDiffusion and \ac{DDS}, and below 15\% for \ac{CLIF}, whose tiles have a larger Python overhead. Values at $768^3$ and $1024^3$ are predictions from measured units; all other sizes were measured directly.

\begin{figure*}[htbp]
    \centering
    \resizebox{0.8\textwidth}{!}{
\begin{tikzpicture}
\begin{groupplot}[
  group style={group size=2 by 1, horizontal sep=1.55cm},
  paper, xmode=log, ymode=log, width=0.45\linewidth, height=0.45\linewidth,
]
\nextgroupplot[title={(a) Time}, xlabel={Measured (s)}, ylabel={Predicted from unit costs (s)},
  xmin=0.01, xmax=1000, ymin=0.01, ymax=1000, legend pos=south east]
\node[font=\scriptsize, anchor=north west, fill=white, inner sep=1pt] at (rel axis cs:0.03,0.97) {grey band: $\pm10\%$};
\fill[black!9] (axis cs:0.01,0.009) -- (axis cs:1000,900) -- (axis cs:1000,1100) -- (axis cs:0.01,0.011) -- cycle;
\addplot[black!60, line width=0.6pt, mark=none, forget plot, domain=0.01:1000, samples=2] {x};
\addplot[black!80, only marks, mark=*, mark options={fill=black!80, fill opacity=0.8}] table[x=measured_s, y=predicted_s] {images/Supplementary/data/validation_decoder.dat};
\addlegendentry{Full-volume decoding}
\addplot[black!80, only marks, mark=triangle*, mark options={fill=white}] table[x=measured_s, y=predicted_s] {images/Supplementary/data/validation_steps.dat};
\addlegendentry{Reconstruction steps}
\nextgroupplot[title={(b) Peak memory}, xlabel={Measured (GiB)}, ylabel={Predicted (GiB)},
  xmin=0.2, xmax=60, ymin=0.2, ymax=60, xtick={0.3,1,3,10,30}, xticklabels={0.3,1,3,10,30},
  ytick={0.3,1,3,10,30}, yticklabels={0.3,1,3,10,30}]
\fill[black!9] (axis cs:0.2,0.18) -- (axis cs:60,54) -- (axis cs:60,66) -- (axis cs:0.2,0.22) -- cycle;
\addplot[black!60, line width=0.6pt, mark=none, domain=0.2:60, samples=2] {x};
\addplot[black!80, only marks, mark=*, mark options={fill=black!80, fill opacity=0.8}] table[x=measured_gib, y=predicted_gib] {images/Supplementary/data/validation_decoder.dat};
\node[font=\scriptsize, anchor=north west, fill=white, inner sep=1pt] at (rel axis cs:0.03,0.97) {grey band: $\pm10\%$};
\end{groupplot}
\end{tikzpicture}
 }
    \caption{Validation of the unit-cost formulas: predicted vs.\ measured time (a) and peak memory (b), for full-volume decoding and for reconstruction steps. The solid line is the identity and the grey band is $\pm10\%$.}
    \label{fig:s_validation}
\end{figure*}

\section{Decoder cost vs.\ output size and model size}\label{sec:s_scaling}

Fig.~\ref{fig:s_resolution} extends the $512^3$ cost comparison of the main manuscript to output sizes from $128^3$ to $1024^3$. The decoders of \ac{CoReAE} run their convolutions on the fixed latent grid, so only the output volume itself grows with the output size. Their memory therefore stays almost flat: the decoding memory of \ac{CLIF} goes from 1.8\,GiB at $128^3$ to 2.1\,GiB at $512^3$. The baselines run their convolutions at the output resolution in overlapping windows or patches, so their cost grows with the number of windows and with the overlap. MAISI with $256^3$ windows runs out of memory when the gradient with respect to the latent is computed, at every output size.

\begin{figure*}[htbp]
    \centering
    \resizebox{0.95\textwidth}{!}{
\begin{tikzpicture}
\begin{groupplot}[
  group style={group size=2 by 2, horizontal sep=1.55cm, vertical sep=1.95cm},
  paper, pow2x, ymode=log, width=0.49\linewidth, height=4.6cm,
  xmin=110, xmax=1180, xtick={128,256,512,768,1024}, xticklabels={$128^3$,$256^3$,$512^3$,$768^3$,$1024^3$},
  xlabel={Output volume size}, xticklabel style={font=\tiny},
]
\newcommand{\predicted}{\fill[black!5] ({axis cs:640,1}|-{rel axis cs:0,0}) rectangle ({rel axis cs:1,1});
  \node[font=\tiny, text=black!55, anchor=north] at ({axis cs:880,1}|-{rel axis cs:0,1}) {predicted};}
\newcommand{\gpu}{\addplot[black!55, densely dashed, line width=0.6pt, mark=none, domain=110:1180, samples=2, forget plot] {48};
  \node[font=\tiny, text=black!55, anchor=south west] at (axis cs:118,48) {48\,GB GPU};}
\nextgroupplot[title={(a) Decoding: latent $\rightarrow$ volume}, ylabel={Time per volume (s, log)}, ymin=0.05, ymax=20000,
  legend to name=declegend, legend columns=3, legend style={/tikz/every even column/.append style={column sep=6pt}}]
\predicted
\addplot[sLIIF]  table[x=size, y=time_fwd] {images/Supplementary/data/dec_res_liif.dat};      \addlegendentry{Ours LIIF ($2^{18}$ queries/chunk)}
\addplot[sCLIF]  table[x=size, y=time_fwd] {images/Supplementary/data/dec_res_clif.dat};      \addlegendentry{Ours CLIF ($96^3$ tiles)}
\addplot[sMAISIa] table[x=size, y=time_fwd] {images/Supplementary/data/dec_res_maisi128.dat}; \addlegendentry{MAISI ($128^3$ windows)}
\addplot[sMAISIb] table[x=size, y=time_fwd] {images/Supplementary/data/dec_res_maisi256.dat}; \addlegendentry{MAISI ($256^3$ windows)}
\addplot[sMEDa]  table[x=size, y=time_fwd] {images/Supplementary/data/dec_res_med4.dat};      \addlegendentry{3D MedDiffusion 4$\times$}
\addplot[sMEDb]  table[x=size, y=time_fwd] {images/Supplementary/data/dec_res_med8.dat};      \addlegendentry{3D MedDiffusion 8$\times$}
\nextgroupplot[title={(b) Decoding + gradient to the latent}, ylabel={Time per volume (s, log)}, ymin=0.05, ymax=60000]
\predicted
\addplot[sLIIF]  table[x=size, y=time_fb] {images/Supplementary/data/dec_res_liif.dat};
\addplot[sCLIF]  table[x=size, y=time_fb] {images/Supplementary/data/dec_res_clif.dat};
\addplot[sMAISIa] table[x=size, y=time_fb] {images/Supplementary/data/dec_res_maisi128.dat};
\addplot[sMEDa]  table[x=size, y=time_fb] {images/Supplementary/data/dec_res_med4.dat};
\addplot[sMEDb]  table[x=size, y=time_fb] {images/Supplementary/data/dec_res_med8.dat};
\node[font=\tiny, text=cMAISIb, anchor=north west] at (rel axis cs:0.02,0.98) {MAISI ($256^3$ windows): out of memory};
\nextgroupplot[title={(c) Decoding: peak GPU memory}, ylabel={Peak allocated memory (GiB)}, ymin=1, ymax=80,
  ytick={1,2,5,10,20,48}, yticklabels={1,2,5,10,20,48}]
\predicted \gpu
\addplot[sLIIF]  table[x=size, y=mem_fwd] {images/Supplementary/data/dec_res_liif.dat};
\addplot[sCLIF]  table[x=size, y=mem_fwd] {images/Supplementary/data/dec_res_clif.dat};
\addplot[sMAISIa] table[x=size, y=mem_fwd] {images/Supplementary/data/dec_res_maisi128.dat};
\addplot[sMAISIb] table[x=size, y=mem_fwd] {images/Supplementary/data/dec_res_maisi256.dat};
\addplot[sMEDa]  table[x=size, y=mem_fwd] {images/Supplementary/data/dec_res_med4.dat};
\addplot[sMEDb]  table[x=size, y=mem_fwd] {images/Supplementary/data/dec_res_med8.dat};
\nextgroupplot[title={(d) Decoding + gradient: peak GPU memory}, ylabel={Peak allocated memory (GiB)}, ymin=5, ymax=80,
  ytick={5,10,20,48}, yticklabels={5,10,20,48}]
\predicted \gpu
\addplot[sLIIF]  table[x=size, y=mem_fb] {images/Supplementary/data/dec_res_liif.dat};
\addplot[sCLIF]  table[x=size, y=mem_fb] {images/Supplementary/data/dec_res_clif.dat};
\addplot[sMAISIa] table[x=size, y=mem_fb] {images/Supplementary/data/dec_res_maisi128.dat};
\addplot[sMEDa]  table[x=size, y=mem_fb] {images/Supplementary/data/dec_res_med4.dat};
\addplot[sMEDb]  table[x=size, y=mem_fb] {images/Supplementary/data/dec_res_med8.dat};
\node[font=\tiny, text=cMAISIb, anchor=south west, fill=white, inner sep=1pt] at (rel axis cs:0.02,0.62) {MAISI ($256^3$ windows): out of memory};
\end{groupplot}
\node[anchor=south] at ($(group c1r1.north)!0.5!(group c2r1.north)+(0,0.55cm)$) {\pgfplotslegendfromname{declegend}};
\end{tikzpicture}}
    \caption{Decoder cost vs.\ output volume size (\ac{CT}). (a)~Decoding time; (b)~time of decoding plus gradient with respect to the latent; (c, d)~corresponding peak \ac{GPU} memory. The grey area ($768^3$, $1024^3$) shows values predicted from measured unit costs. The dashed line is the 48\,GB \ac{GPU} limit.}
    \label{fig:s_resolution}
\end{figure*}

Fig.~\ref{fig:s_model_cost} relates the autoencoding time at $512^3$ to the computation per volume and to the number of parameters. Time follows the computation (Spearman $\rho=1.00$), not the model size ($\rho=0.46$). For a convolutional \ac{AE}, each parameter is reused at every position where it is applied, so the cost depends on the number of voxels processed, including overlaps and halos. MAISI-W128 and MAISI-W256 use the same network (20.9\,M parameters), but W128 processes $5.36\times$ the volume against $3.38\times$ for W256, and is therefore about twice as slow. \ac{CoReAE} has both few parameters and no overlap, which explains its low cost.

\begin{figure*}[htbp]
    \centering
    \resizebox{0.75\textwidth}{!}{
\definecolor{bLIIF}{HTML}{1F4E79}
\definecolor{bCLIF}{HTML}{9E3D5C}
\definecolor{bM128}{HTML}{6F95B8}
\definecolor{bM256}{HTML}{A7BCD0}
\definecolor{bMD4}{HTML}{C48A9C}
\definecolor{bMD8}{HTML}{8C8C94}
\definecolor{bInk}{HTML}{2B2B2B}
\begin{tikzpicture}
\begin{axis}[
  paper, width=0.78\linewidth, height=0.56\linewidth,
  xmode=log, ymode=log, xmin=4.5e4, xmax=1.2e7, ymin=4, ymax=700,
  xlabel={Computation per $512^3$ volume (GFLOPs, encoder + decoder, log scale)},
  ylabel={Autoencoding time at $512^3$ (s, log scale)},
  axis line style={bInk!70}, tick style={bInk!70}, major grid style={bInk!9},
  clip=false, name=bub,
]
\addplot[bInk!45, densely dashed, line width=0.7pt, domain=4.5e4:7.6e6, samples=2] {8.1819*x/91645.2};
\node[font=\scriptsize, text=bInk!60, rotate=33, anchor=south] at (axis cs:3.2e5,30) {time $\propto$ FLOPs};
\addplot[only marks, mark=*, mark size=12.66pt, draw=bMD8!80!black, fill=bMD8, fill opacity=0.78, discard if not={label}{MedDiff-8x}] table[x=gflops, y=time_s] {images/Supplementary/data/model_cost.dat};
\addplot[only marks, mark=*, mark size=8.92pt, draw=bMD4!80!black, fill=bMD4, fill opacity=0.78, discard if not={label}{MedDiff-4x}] table[x=gflops, y=time_s] {images/Supplementary/data/model_cost.dat};
\addplot[only marks, mark=*, mark size=8.01pt, draw=bM128!80!black, fill=bM128, fill opacity=0.78, discard if not={label}{MAISI-W128}] table[x=gflops, y=time_s] {images/Supplementary/data/model_cost.dat};
\addplot[only marks, mark=*, mark size=8.01pt, draw=bM256!80!black, fill=bM256, fill opacity=0.78, discard if not={label}{MAISI-W256}] table[x=gflops, y=time_s] {images/Supplementary/data/model_cost.dat};
\addplot[only marks, mark=*, mark size=2.48pt, draw=bCLIF!80!black, fill=bCLIF, fill opacity=0.78, discard if not={label}{Ours-CLIF}] table[x=gflops, y=time_s] {images/Supplementary/data/model_cost.dat};
\addplot[only marks, mark=*, mark size=2.15pt, draw=bLIIF!80!black, fill=bLIIF, fill opacity=0.78, discard if not={label}{Ours-LIIF}] table[x=gflops, y=time_s] {images/Supplementary/data/model_cost.dat};
\node[font=\footnotesize\bfseries, text=bLIIF, anchor=north west] at (axis cs:9.6e4,7.3) {Ours LIIF};
\node[font=\scriptsize, text=bLIIF, anchor=north west] at (axis cs:9.6e4,5.9) {1.5\,M params};
\node[font=\footnotesize\bfseries, text=bCLIF, anchor=south west] at (axis cs:2.3e5,10.3) {Ours CLIF};
\node[font=\scriptsize, text=bCLIF, anchor=south west] at (axis cs:2.3e5,8.5) {2.0\,M params};
\node[font=\footnotesize, text=bM128!75!black, anchor=east] at (axis cs:3.05e6,242.8) {MAISI ($128^3$ windows), 20.9\,M};
\node[font=\footnotesize, text=bM256!55!black, anchor=west] at (axis cs:3.2e6,115) {MAISI ($256^3$ windows), 20.9\,M};
\node[font=\footnotesize, text=bMD4!75!black, anchor=south east] at (axis cs:8.6e5,48) {3D MedDiffusion 4$\times$, 26.0\,M};
\node[font=\footnotesize, text=bMD8!70!black, anchor=west] at (axis cs:1.75e6,36) {3D MedDiffusion 8$\times$, 52.3\,M};
\node[font=\footnotesize, anchor=north west, align=left, fill=white, fill opacity=0.9, text opacity=1, inner sep=3pt]
  at (rel axis cs:0.02,0.98) {Spearman $\rho$ (time, FLOPs) $=1.00$\\Spearman $\rho$ (time, parameters) $=0.46$};
\end{axis}
\coordinate (key) at ($(bub.south east)+(-1.25cm,0.45cm)$);
\foreach \p/\dy in {2/0, 20/0.55, 50/1.3} {
  \pgfmathsetmacro{\r}{1.75*sqrt(\p)}
  \fill[bInk!12, draw=bInk!45] ($(key)+(0,\dy)$) circle (\r pt);
  \node[font=\scriptsize, text=bInk!80, anchor=west] at ($(key)+(0.45,\dy)$) {\p\,M};
}
\node[font=\scriptsize\bfseries, text=bInk!80, anchor=south] at ($(key)+(0.25,1.68)$) {Parameters};
\end{tikzpicture}}
    \caption{Autoencoding time at $512^3$ vs.\ computation per volume (GFLOPs, encoder and decoder, including window overlaps and halos). The bubble area is proportional to the number of parameters. The dashed line shows constant throughput.}
    \label{fig:s_model_cost}
\end{figure*}

\section{Effect of the chunk and tile size}\label{sec:s_chunk}

The \ac{LIIF} head renders the volume in chunks of $N$ coordinate queries, and the \ac{CLIF} head in cubic tiles with a halo. Table~\ref{tab:s_chunk} and Fig.~\ref{fig:s_chunk} show that the chunk and tile size control a trade-off between runtime and memory. Runtime decreases with the chunk size up to about $2^{18}$--$2^{20}$ queries; smaller chunks are dominated by per-chunk overhead. Working memory grows linearly with the chunk or tile size.

\begin{table*}[htbp]
    \centering
    \caption{\ac{CT} $512^3$ decoding vs.\ chunk size (\ac{LIIF}) and tile size (\ac{CLIF}, halo 6). Time (s) and peak memory (GiB) of decoding (Fwd) and of decoding plus gradient with respect to the latent (Fb).}
    \label{tab:s_chunk}
    \small
    \setlength{\tabcolsep}{4pt}
    \begin{tabular}{rrcccc@{\hspace{16pt}}rrcccc}
        \toprule
        \multicolumn{6}{c}{\textbf{\acs{LIIF}}} & \multicolumn{6}{c}{\textbf{\acs{CLIF}}}\\
        \cmidrule(lr){1-6}\cmidrule(lr){7-12}
        Queries/chunk & Chunks & Fwd (s) & Fb (s) & Fwd (GiB) & Fb (GiB) & Tile & Tiles & Fwd (s) & Fb (s) & Fwd (GiB) & Fb (GiB)\\
        \midrule
        $2^{14}$ & 8192 & 82.3 & 245.2 & 1.5 & 8.5 & $32^3$ & 4096 & 12.1 & 39.0 & 0.7 & 4.8\\
        $2^{16}$ & 2048 & 20.6 & 65.5 & 1.6 & 8.7 & $48^3$ & 1331 & 7.6 & 27.7 & 0.8 & 5.3\\
        $2^{18}$ & 512 & 7.8 & 28.5 & 1.7 & 9.6 & $64^3$ & 512 & 7.4 & 26.8 & 1.1 & 6.2\\
        $2^{20}$ & 128 & 7.6 & 26.6 & 2.4 & 14.6 & $96^3$ & 216 & 8.7 & 30.8 & 2.1 & 9.4\\
        $2^{22}$ & 32 & 7.3 & 25.5 & 5.1 & 34.7 & $128^3$ & 64 & 8.4 & 30.0 & 3.8 & 15.3\\
        & & & & & & $192^3$ & 27 & 8.4 & 29.5 & 10.5 & 37.9\\
        \bottomrule
    \end{tabular}
\end{table*}

\begin{figure*}[htbp]
    \centering
    \resizebox{0.95\textwidth}{!}{
\begin{tikzpicture}
\begin{groupplot}[
  group style={group size=2 by 2, horizontal sep=1.55cm, vertical sep=1.95cm},
  paper, pow2x, width=0.49\linewidth, height=4.6cm,
]
\nextgroupplot[title={(a) LIIF: time vs.\ queries per chunk}, ymode=log, ylabel={Time per volume (s)},
  xlabel={Queries per chunk $N$}, xtick={16384,65536,262144,1048576,4194304},
  xticklabels={$2^{14}$,$2^{16}$,$2^{18}$,$2^{20}$,$2^{22}$}, ymin=0.5, ymax=400,
  legend to name=sweeplegend, legend columns=4, legend style={/tikz/every even column/.append style={column sep=6pt}}]
\addplot[cLIIF, mark=*] table[x=setting, y=time_fwd] {images/Supplementary/data/sweep_ct_liif.dat}; \addlegendentry{CT $512^3$, decode}
\addplot[cLIIF, mark=*, densely dashed, mark options={solid, fill=white}] table[x=setting, y=time_fb] {images/Supplementary/data/sweep_ct_liif.dat}; \addlegendentry{CT $512^3$, decode + backward}
\addplot[cCLIF, mark=square*] table[x=setting, y=time_fwd] {images/Supplementary/data/sweep_mri_liif.dat}; \addlegendentry{MRI $256^3$, decode}
\addplot[cCLIF, mark=square*, densely dashed, mark options={solid, fill=white}] table[x=setting, y=time_fb] {images/Supplementary/data/sweep_mri_liif.dat}; \addlegendentry{MRI $256^3$, decode + backward}
\nextgroupplot[title={(b) LIIF: memory vs.\ queries per chunk}, ylabel={Peak memory (GiB)},
  xlabel={Queries per chunk $N$}, xtick={16384,65536,262144,1048576,4194304},
  xticklabels={$2^{14}$,$2^{16}$,$2^{18}$,$2^{20}$,$2^{22}$}, ymin=0, ymax=38]
\addplot[cLIIF, mark=*] table[x=setting, y=mem_fwd] {images/Supplementary/data/sweep_ct_liif.dat};
\addplot[cLIIF, mark=*, densely dashed, mark options={solid, fill=white}] table[x=setting, y=mem_fb] {images/Supplementary/data/sweep_ct_liif.dat};
\addplot[cCLIF, mark=square*] table[x=setting, y=mem_fwd] {images/Supplementary/data/sweep_mri_liif.dat};
\addplot[cCLIF, mark=square*, densely dashed, mark options={solid, fill=white}] table[x=setting, y=mem_fb] {images/Supplementary/data/sweep_mri_liif.dat};
\nextgroupplot[title={(c) CLIF: time vs.\ tile size}, ymode=log, ylabel={Time per volume (s)},
  xlabel={Tile side $T$ (halo 6)}, xtick={32,48,64,96,128,192}, xticklabels={32,48,64,96,128,192}, ymin=0.5, ymax=60]
\addplot[cLIIF, mark=*] table[x=setting, y=time_fwd] {images/Supplementary/data/sweep_ct_clif.dat};
\addplot[cLIIF, mark=*, densely dashed, mark options={solid, fill=white}] table[x=setting, y=time_fb] {images/Supplementary/data/sweep_ct_clif.dat};
\addplot[cCLIF, mark=square*] table[x=setting, y=time_fwd] {images/Supplementary/data/sweep_mri_clif.dat};
\addplot[cCLIF, mark=square*, densely dashed, mark options={solid, fill=white}] table[x=setting, y=time_fb] {images/Supplementary/data/sweep_mri_clif.dat};
\nextgroupplot[title={(d) CLIF: memory vs.\ tile size}, ylabel={Peak memory (GiB)},
  xlabel={Tile side $T$ (halo 6)}, xtick={32,48,64,96,128,192}, xticklabels={32,48,64,96,128,192}, ymin=0, ymax=40]
\addplot[cLIIF, mark=*] table[x=setting, y=mem_fwd] {images/Supplementary/data/sweep_ct_clif.dat};
\addplot[cLIIF, mark=*, densely dashed, mark options={solid, fill=white}] table[x=setting, y=mem_fb] {images/Supplementary/data/sweep_ct_clif.dat};
\addplot[cCLIF, mark=square*] table[x=setting, y=mem_fwd] {images/Supplementary/data/sweep_mri_clif.dat};
\addplot[cCLIF, mark=square*, densely dashed, mark options={solid, fill=white}] table[x=setting, y=mem_fb] {images/Supplementary/data/sweep_mri_clif.dat};
\end{groupplot}
\node[anchor=south] at ($(group c1r1.north)!0.5!(group c2r1.north)+(0,0.55cm)$) {\pgfplotslegendfromname{sweeplegend}};
\end{tikzpicture}}
    \caption{Runtime and peak memory vs.\ chunk size (\ac{LIIF}, a--b) and tile size (\ac{CLIF}, c--d), for \ac{CT} $512^3$ and \ac{MRI} $256^3$. Solid lines: decoding; dashed lines: decoding plus gradient with respect to the latent.}
    \label{fig:s_chunk}
\end{figure*}

The chunk and tile size do not change the result. In FP32, the decoded volume and the gradient with respect to the latent are identical for every chunk, tile and halo size tested (Table~\ref{tab:s_invariance}). The only numerical difference comes from BF16 arithmetic, independently of the chunk size.

\begin{table*}[htbp]
    \centering
    \caption{Numerical invariance at native resolution, against an FP32 reference rendered with the largest chunk or tile. Max $|\Delta|$ is computed on the $[-1,1]$ intensity range. Gradient: gradient with respect to the latent of an $\ell_2$ loss.}
    \label{tab:s_invariance}
    \small
    \begin{tabular}{llcccc}
        \toprule
        Model & Change & Max $|\Delta|$ & \acs{PSNR} (dB) & Grad.\ rel.\ error & Grad.\ cosine\\
        \midrule
        \acs{CT} \acs{LIIF} & chunk $2^{16}$, $2^{18}$ (FP32) & 0 & $\infty$ & $<10^{-6}$ & 1.000000\\
        \acs{CT} \acs{CLIF} & tile $64^3$, $96^3$; halo 8 (FP32) & 0 & $\infty$ & $<10^{-6}$ & 1.000000\\
        \acs{MRI} \acs{CLIF} & tile $64^3$, $96^3$; halo 8 (FP32) & 0 & $\infty$ & $<10^{-6}$ & 1.000000\\
        \acs{CT} \acs{LIIF} & BF16 & 0.022 & 64.7 & 0.33\% & 0.999995\\
        \acs{CT} \acs{CLIF} & BF16 & 0.011 & 59.3 & 0.39\% & 0.999992\\
        \acs{MRI} \acs{CLIF} & BF16 & 0.017 & 59.4 & 0.35\% & 0.999994\\
        \bottomrule
    \end{tabular}
\end{table*}

\section{Reconstruction cost}\label{sec:s_recon}

We measured every elementary operation of one reconstruction step for \ac{CoReAE} and for \ac{DDS}: the \ac{3D} latent U-Net, the decoder (forward, and forward plus gradient), the encoder, the Radon forward and adjoint operators with one conjugate-gradient iteration (\ac{CT}, 60 views), the masked Fourier operator (\ac{MRI}), and the \ac{2D} U-Net and data-consistency step of \ac{DDS}. The cost of a step is the sum of its units, and the cost of a whole reconstruction is the schedule multiplied by the step costs. Each composed step was also run for real and agreed with the sum of its units within 4\% (\ac{LIIF}, \ac{MRI}, \ac{DDS}) and 15\% (\ac{CLIF}).

\begin{table*}[htbp]
    \centering
    \caption{Cost of one reconstruction step (s). Latent iteration: one Adam iteration of Stage~3, i.e., one decoder forward pass plus gradient and one application of the forward operator.}
    \label{tab:s_recon_step}
    \small
    \begin{tabular}{llcccc}
        \toprule
        Case & Method & Unguided step & Image-space step & Latent iteration & \acs{DDS} step\\
        \midrule
        \multirow{3}{*}{\acs{CT} $512^3$} & \acs{CoReAE} (\acs{LIIF}) & 0.21 & 9.0 & 26.6 & --\\
        & \acs{CoReAE} (\acs{CLIF}) & 0.21 & 8.8 & 26.8 & --\\
        & \acs{DDS} & -- & -- & -- & 23.1\\
        \midrule
        \multirow{3}{*}{\acs{CT} $256^3$} & \acs{CoReAE} (\acs{LIIF}) & 0.02 & 1.5 & 3.6 & --\\
        & \acs{CoReAE} (\acs{CLIF}) & 0.02 & 1.4 & 3.4 & --\\
        & \acs{DDS} & -- & -- & -- & 2.4\\
        \midrule
        \multirow{2}{*}{\acs{MRI} $256^3$} & \acs{CoReAE} (\acs{LIIF}) & 0.02 & 1.06 & 3.1 & --\\
        & \acs{CoReAE} (\acs{CLIF}) & 0.02 & 1.04 & 3.4 & --\\
        \bottomrule
    \end{tabular}
\end{table*}

\begin{table*}[htbp]
    \centering
    \caption{Breakdown of one \ac{CT} $512^3$ reconstruction step (s) and peak \ac{GPU} memory (GiB).}
    \label{tab:s_recon_breakdown}
    \small
    \begin{tabular}{lccc}
        \toprule
        Component & \acs{CoReAE} (\acs{LIIF}) & \acs{CoReAE} (\acs{CLIF}) & \acs{DDS}\\
        \midrule
        Denoiser (\acs{3D} latent U-Net / \acs{2D} U-Net on 512 slices) & 0.21 & 0.21 & 15.7\\
        Decoder forward & 7.6 & 7.4 & --\\
        Data consistency (10 CG iterations / 5 ADMM-CG iterations) & 0.78 & 0.78 & 7.4\\
        Encoder & 0.42 & 0.42 & --\\
        Decoder forward and gradient + Radon (one latent iteration) & 26.6 & 26.8 & --\\
        \midrule
        Peak memory: unguided / image-space / latent iteration & 4.3 / 7.6 / 12.4 & 4.3 / 7.4 / 12.1 & 20.0\\
        \bottomrule
    \end{tabular}
\end{table*}

\begin{table*}[htbp]
    \centering
    \caption{Whole reconstruction time (s) and number of network calls. Full schedule (Algorithm~1 of the main manuscript): 100 \ac{DDIM} steps, i.e., 74 unguided, 16 image-space and 10 latent-space steps with 5 iterations each. Image-space only: 81 unguided and 19 image-space steps. \ac{DDS}: 50 steps.}
    \label{tab:s_recon_total}
    \small
    \begin{tabular}{lccccc}
        \toprule
        & \multicolumn{2}{c}{Full schedule} & \multicolumn{2}{c}{Image-space only} & \\
        \cmidrule(lr){2-3}\cmidrule(lr){4-5}
        Case & \acs{LIIF} & \acs{CLIF} & \acs{LIIF} & \acs{CLIF} & \acs{DDS}\\
        \midrule
        \acs{CT} $512^3$ & 1494 & 1500 & \textbf{197} & \textbf{194} & 1155\\
        \acs{CT} $256^3$ & 208 & 195 & \textbf{32} & \textbf{30} & 119\\
        \acs{MRI} $256^3$ & 175 & 188 & 23 & 23 & --\\
        \midrule
        Denoiser calls & \multicolumn{2}{c}{100 (\acs{3D} latent)} & \multicolumn{2}{c}{100 (\acs{3D} latent)} & 800 (\acs{2D}, 32 slices)\\
        Decoder calls (forward / with gradient) & \multicolumn{2}{c}{16 / 50} & \multicolumn{2}{c}{19 / 0} & --\\
        \bottomrule
    \end{tabular}
\end{table*}

\begin{figure*}[htbp]
    \centering
    \resizebox{0.95\textwidth}{!}{
\begin{tikzpicture}
\begin{groupplot}[
  group style={group size=3 by 1, horizontal sep=1.25cm},
  paper, height=4.8cm, grid=none, ymajorgrids,
  xticklabel style={font=\scriptsize, align=center}, enlarge x limits=0.15,
]
\newcommand{\stack}[1]{%
  \addplot[fill=cDen, draw=none] table[x=x, y=denoiser] {images/Supplementary/data/recon_steps_#1.dat};
  \addplot[fill=cDec, draw=none] table[x=x, y=decode] {images/Supplementary/data/recon_steps_#1.dat};
  \addplot[fill=cDC, draw=none] table[x=x, y=dc] {images/Supplementary/data/recon_steps_#1.dat};
  \addplot[fill=cEnc, draw=none] table[x=x, y=encode] {images/Supplementary/data/recon_steps_#1.dat};
  \addplot[fill=cLat, draw=none] table[x=x, y=latent_opt] {images/Supplementary/data/recon_steps_#1.dat};}
\newcommand{\capped}[2]{\node[font=\tiny, anchor=south, fill=white, inner sep=1pt] at (axis cs:#1,\pgfkeysvalueof{/pgfplots/ymax}) {#2\,s\,$\uparrow$};}
\nextgroupplot[title={(a) One step, CT $512^3$}, width=0.35\linewidth, ybar stacked, /pgf/bar width=8pt,
  ylabel={Seconds per step}, ymin=0, ymax=30, restrict y to domain*=0:30, clip=false,
  xtick={0,...,4}, xticklabels={LIIF\\pixel, LIIF\\latent, CLIF\\pixel, CLIF\\latent, DDS},
  legend to name=steplegend, legend columns=5, area legend,
  legend style={/tikz/every even column/.append style={column sep=5pt}}]
\stack{ct512}
\capped{1}{133}\capped{3}{134}
\legend{Denoiser, Decoder, Data consistency, Encoder, Latent guidance (5 decoder fwd+bwd)}
\nextgroupplot[title={(b) One step, MRI $256^3$}, width=0.30\linewidth, ybar stacked, /pgf/bar width=8pt,
  ymin=0, ymax=4, restrict y to domain*=0:4, clip=false,
  xtick={0,...,3}, xticklabels={LIIF\\pixel, LIIF\\latent, CLIF\\pixel, CLIF\\latent}]
\stack{mri256}
\capped{1}{15.6}\capped{3}{17.0}
\nextgroupplot[title={(c) Whole CT reconstruction}, width=0.36\linewidth, ybar=0.4pt, /pgf/bar width=4.5pt,
  ymode=log, log origin=infty, ymin=10, ymax=3000, ylabel={Total time (s)}, xtick={0,1}, xticklabels={$256^3$, $512^3$},
  enlarge x limits=0.5, legend to name=totlegend, legend columns=5, area legend,
  legend style={/tikz/every even column/.append style={column sep=5pt}}]
\addplot[fill=cLIIF, draw=none] table[x expr=\coordindex, y=liif_pixel] {images/Supplementary/data/recon_totals_ct.dat};
\addplot[fill=cCLIF, draw=none] table[x expr=\coordindex, y=clif_pixel] {images/Supplementary/data/recon_totals_ct.dat};
\addplot[fill=cLIIF!40, draw=none] table[x expr=\coordindex, y=liif_full] {images/Supplementary/data/recon_totals_ct.dat};
\addplot[fill=cCLIF!40, draw=none] table[x expr=\coordindex, y=clif_full] {images/Supplementary/data/recon_totals_ct.dat};
\addplot[fill=cDDS, draw=none] table[x expr=\coordindex, y=dds] {images/Supplementary/data/recon_totals_ct.dat};
\legend{LIIF pixel-only, CLIF pixel-only, LIIF full schedule, CLIF full schedule, DDS}
\end{groupplot}
\node[anchor=south] at ($(group c1r1.north)!0.5!(group c3r1.north)+(0,1.15cm)$) {\pgfplotslegendfromname{steplegend}};
\node[anchor=south] at ($(group c1r1.north)!0.5!(group c3r1.north)+(0,0.70cm)$) {\pgfplotslegendfromname{totlegend}};
\end{tikzpicture}}
    \caption{Reconstruction cost. (a, b)~Time of one step, broken down by component, for \ac{CT} $512^3$ and \ac{MRI} $256^3$; latent-space steps exceed the axis and their totals are printed on top. (c)~Whole \ac{CT} reconstruction time.}
    \label{fig:s_recon_steps}
\end{figure*}

One denoising step of \ac{CoReAE} is about $75\times$ cheaper than one step of \ac{DDS} (0.21\,s for one \ac{3D} latent U-Net call vs.\ 15.7\,s for \ac{2D} U-Net calls over 512 slices), and needs $1.6$--$5\times$ less memory. With image-space correction only, a \ac{CT} $512^3$ reconstruction takes 194--197\,s vs.\ 1155\,s for \ac{DDS}. The latent-space stage needs one decoder pass with gradient per iteration; with the full schedule (50 such passes), it dominates the runtime and makes the reconstruction slower than \ac{DDS} (about 1500\,s vs.\ 1155\,s).

\section{Task-based evaluation of the reconstructions}\label{sec:s_task}

We applied pretrained public segmentation networks to the ground truth and to each reconstruction of the same volume, and compared the segmentations: TotalSegmentator~2.18 (\texttt{total} task) for \ac{CT}, with the window $[-1000,1500]$\,HU and $1\times1\times1.5$\,mm spacing, and SynthSeg (FreeSurfer~7.4.1, \texttt{--parc --robust}) for \ac{MRI}. We report the Dice score, the 95th-percentile Hausdorff distance (HD95), and the \ac{HU} error inside each ground-truth organ.

Task-based evaluation used the same 10~\ac{CT} and 10~\ac{MRI} test volumes as the reconstruction experiments. Structure-specific scores were averaged across volumes. For \ac{CT}, only volumes containing the structure in the reference field of view were included; for \ac{MRI}, all 10 volumes were included. Overall mean Dice was then calculated as the unweighted average of the structure-specific means over 15~\ac{CT} or 17~\ac{MRI} structures. Structures present in the reference but missed in a reconstruction received a Dice score of zero and were excluded from HD95, which is undefined for empty segmentations. HD95 therefore reflects detected structures only, with potentially different contributing structures across methods. Intensity accuracy was assessed using both absolute organ-average \ac{HU} bias and in-organ \ac{MAE}, since positive and negative errors can cancel in the bias.

\begin{table*}[htbp]
    \centering
    \caption{\ac{CT} task-based evaluation (TotalSegmentator): Dice per structure and means over 15 structures. Each structure is averaged over the test volumes that contain it in the reference field of view; overall means are unweighted over structures. The brain is excluded because only a small part lies in the field of view. Bold: best value for each number of views. \ac{FBP} (mean Dice 0.06 and 0.21 at 30 and 60 views) is omitted. HD95 is averaged over the structures detected by each method, so complete detection failures are not included in it.}
    \label{tab:s_ct_dice}
    \small
    \begin{tabular}{lccc@{\hspace{16pt}}ccc}
        \toprule
        & \multicolumn{3}{c}{30 views} & \multicolumn{3}{c}{60 views}\\
        \cmidrule(lr){2-4}\cmidrule(lr){5-7}
        Structure & \acs{DDS} & \acs{LIIF} & \acs{CLIF} & \acs{DDS} & \acs{LIIF} & \acs{CLIF}\\
        \midrule
        aorta           & 0.86 & \textbf{0.87} & \textbf{0.87} & \textbf{0.93} & 0.88 & 0.89\\
        heart           & \textbf{0.91} & \textbf{0.91} & \textbf{0.91} & \textbf{0.96} & 0.93 & 0.93\\
        liver           & \textbf{0.97} & 0.94 & 0.93 & \textbf{0.98} & 0.96 & 0.96\\
        lungs           & \textbf{0.99} & \textbf{0.99} & \textbf{0.99} & \textbf{1.00} & 0.99 & 0.99\\
        kidneys         & 0.60 & \textbf{0.63} & 0.61 & \textbf{0.74} & 0.70 & 0.59\\
        spleen          & 0.47 & \textbf{0.68} & 0.67 & \textbf{0.79} & 0.75 & 0.75\\
        stomach         & 0.68 & 0.68 & \textbf{0.72} & \textbf{0.86} & 0.73 & 0.72\\
        gallbladder     & \textbf{0.88} & 0.81 & 0.35 & \textbf{0.94} & 0.80 & 0.87\\
        pancreas        & 0.00 & 0.00 & 0.00 & \textbf{0.50} & 0.00 & 0.00\\
        urinary bladder & \textbf{0.96} & 0.94 & 0.94 & \textbf{0.97} & 0.95 & 0.95\\
        vertebrae       & 0.66 & 0.78 & \textbf{0.79} & 0.78 & \textbf{0.81} & 0.80\\
        ribs            & \textbf{0.82} & 0.76 & 0.77 & \textbf{0.86} & 0.77 & 0.77\\
        hips            & \textbf{0.96} & 0.95 & 0.95 & \textbf{0.97} & 0.95 & 0.95\\
        femurs          & \textbf{0.97} & 0.96 & \textbf{0.97} & \textbf{0.98} & 0.97 & 0.97\\
        skull           & 0.75 & \textbf{0.81} & 0.76 & \textbf{0.87} & 0.78 & 0.80\\
        \midrule
        Mean Dice                         & 0.77 & \textbf{0.78} & 0.75 & \textbf{0.88} & 0.80 & 0.80\\
        Mean HD95 (mm)                    & \textbf{17.8} & 41.7 & 42.0 & \textbf{5.9} & 38.5 & 15.4\\
        Mean $|$\acs{HU} bias$|$ in organ & \textbf{2.4} & 23.9 & 23.7 & \textbf{1.3} & 22.4 & 22.2\\
        \acs{MAE} in organ (\acs{HU})     & \textbf{26.4} & 42.1 & 42.1 & \textbf{21.9} & 40.4 & 40.3\\
        \bottomrule
    \end{tabular}
\end{table*}

\begin{table*}[htbp]
    \centering
    \caption{\ac{MRI} task-based evaluation (SynthSeg): mean over 17 brain structures. Bold: best value for each acceleration factor. \acs{ZF} (mean Dice 0.76 and 0.68 at $\times4$ and $\times8$) is omitted.}
    \label{tab:s_mri_dice}
    \small
    \begin{tabular}{lccc@{\hspace{16pt}}ccc}
        \toprule
        & \multicolumn{3}{c}{$\times4$} & \multicolumn{3}{c}{$\times8$}\\
        \cmidrule(lr){2-4}\cmidrule(lr){5-7}
        & \acs{DDS} & \acs{LIIF} & \acs{CLIF} & \acs{DDS} & \acs{LIIF} & \acs{CLIF}\\
        \midrule
        Mean Dice      & \textbf{0.96} & 0.94 & 0.91 & \textbf{0.94} & 0.92 & 0.91\\
        Mean HD95 (mm) & \textbf{1.0} & \textbf{1.0} & 1.3 & \textbf{1.1} & 1.2 & 1.3\\
        \bottomrule
    \end{tabular}
\end{table*}

On \ac{CT}, the lungs, liver, urinary bladder, hips and femurs keep a Dice score of at least 0.93 for all methods. At 30 views, \ac{CoReAE} (\ac{LIIF}) reaches a mean Dice similar to \ac{DDS} (0.78 vs.\ 0.77), and a higher Dice on several structures, such as the spleen, vertebrae and skull. Similar Dice scores do not imply similar boundary accuracy, however: the mean HD95 is 41.7 and 42.0\,mm for \ac{LIIF} and \ac{CLIF}, against 17.8\,mm for \ac{DDS}. At 60 views, \ac{DDS} gives the best Dice for all structures except the vertebrae, and the lowest HD95 (5.9\,mm, against 38.5\,mm for \ac{LIIF} and 15.4\,mm for \ac{CLIF}).

Small soft-tissue structures are the main failure case. At 30 views, the pancreas Dice is zero for all three methods. At 60 views, \ac{DDS} reaches 0.50, whereas \ac{LIIF} and \ac{CLIF} remain at zero. \ac{CLIF} also has a low gallbladder Dice of 0.35 at 30 views. The pancreas is small, has a low contrast with the adjacent bowel and vessels, and is mainly delineated by thin fat planes, which the frozen \ac{AE} tends to smooth. Good overlap for large structures therefore does not ensure that small structures are segmented successfully.

The \ac{CT} reconstructions of \ac{CoReAE} also show an absolute soft-tissue \ac{HU} bias of about 22--24\,HU, against 1.3--2.4\,HU for \ac{DDS}, and a higher in-organ \ac{MAE} (40--42\,HU, against 22--26\,HU). The \ac{MAE} of \ac{DDS} shows that a small bias does not ensure low local intensity errors. Overall, the task-based results support the preservation of large organs and bones, but show larger boundary errors, failures on small structures, and a lower quantitative intensity fidelity than \ac{DDS}. On \ac{MRI}, brain structures are preserved almost as well as with \ac{DDS} (mean Dice 0.94 vs.\ 0.96 at $\times4$ and 0.92 vs.\ 0.94 at $\times8$, HD95 of 1.0--1.3\,mm for all methods).

\end{document}